\documentclass[a4paper,12pt]{article}
\pdfoutput=1
\usepackage{epsfig}
\usepackage{amssymb}
\usepackage{hyperref}

\usepackage{bookmark}

\usepackage{setspace}

\usepackage{amsmath}
\usepackage{amssymb}
\usepackage{graphicx}
\usepackage{geometry}
\usepackage{subfigure}
\usepackage{url}
\usepackage{slashed}

\newcommand{\ie}{{\it i.e.}}
\newcommand{\eg}{{\it e.g.}}
\newcommand{\cf}{{\it cf.\, }}

\newcommand{\be}{\begin{equation}}
\newcommand{\ee}{\end{equation}}
\newcommand{\br}{\begin{eqnarray}}
\newcommand{\bea}{\begin{eqnarray}}
\newcommand{\eea}{\end{eqnarray}}
\newcommand{\er}{\end{eqnarray}}
\newcommand{\ba}{\begin{array}}
\newcommand{\ea}{\end{array}}
\newcommand{\bi}{\begin{itemize}}
\newcommand{\ei}{\end{itemize}}
\newcommand{\bn}{\begin{enumerate}}
\newcommand{\en}{\end{enumerate}}
\newcommand{\bc}{\begin{center}}
\newcommand{\ec}{\end{center}}

\newcommand{\beq}{\begin{equation}}
\newcommand{\eeq}{\end{equation}}

\newcommand{\U}{\scriptscriptstyle U}
\newcommand{\D}{\scriptscriptstyle D}
\newcommand{\Q}{\scriptscriptstyle Q}

\newcommand{\E}{\scriptscriptstyle E}
\newcommand{\LL}{\scriptscriptstyle L}

\newcommand{\gsim}{\lower.7ex\hbox{$\;\stackrel{\textstyle>}{\sim}\;$}}
\newcommand{\lsim}{\lower.7ex\hbox{$\;\stackrel{\textstyle<}{\sim}\;$}}

\newcommand{\bs}{\begin{small}}
\newcommand{\es}{\end{small}}

\newcommand{\qui}{\bar e_{{\scriptscriptstyle U}_{\!i}}}
\newcommand{\qdi}{\bar e_{{\scriptscriptstyle D}_{\!i}}}

\newcommand{\BBq}{$B_q${\rm -}$\bar{B}_q$~}
\newcommand{\BBd}{$B_d${\rm -}$\bar{B}_d$~}
\newcommand{\BBs}{$B_s${\rm -}$\bar{B}_s$~}

\begin{document}
\pagestyle{empty}
\begin{center}
{\Large {\bf 
FCNC decays of SM fermions into a dark photon
}} \\
\vspace*{1.5cm}
{
 
 {\bf Emidio Gabrielli$^{{a,b}}$},   
{\bf Barbara Mele$^{c}$}, 
{\bf Martti Raidal$^{{b,d}}$}, and
{\bf Elena Venturini$^{{e}}$}
}\\

\vspace{0.5cm}
{\it
 (a) Dipart. di Fisica Teorica, Universit\`a di 
Trieste, Strada Costiera 11, I-34151 Trieste, Italy and 
INFN, Sezione di Trieste, Via Valerio 2, I-34127 Trieste, Italy}  
\\[1mm]
{\it
 (b) NICPB, R\"avala 10, 10143 Tallinn, Estonia}  \\[1mm]
 {\it
 (c) INFN, Sezione di Roma, c/o Dipart. di Fisica, Universit\`a di Roma ``La Sapienza", \\ Piazzale Aldo Moro 2, I-00185 Rome, Italy} \\[1mm]
 {\it (d) Institute of Physics, University of Tartu, Ravila 14c, 50411 Tartu, Estonia}\\[1mm]
 {\it (e) SISSA/ISAS and INFN, Via Bonomea 265, I-34136 Trieste, Italy}

\vspace*{2cm}{\bf ABSTRACT}
\end{center}

\vspace{0.3cm}
We analyze a new class of FCNC processes,   
the $f \to f^{\prime} \, \bar{\gamma}$ decays of  a fermion $f$  into a lighter (same-charge) fermion 
$f^{\prime}$ plus  a  {\it massless} neutral vector boson, a {\it dark photon} $\bar{\gamma}$.
 A massless dark photon does not   interact  at tree level with observable
fields, and the $f \!\to\! f^{\prime} \, \bar{\gamma}$ decay presents a characteristic  signature where the final fermion $f^{\prime}$ is balanced by a {\it massless invisible} system. 
Models recently proposed to explain the  exponential spread in the standard-model Yukawa couplings  can indeed foresee an extra unbroken {\it dark} $U(1)$ gauge group, and the  possibility to couple  on-shell dark photons to standard-model fermions via one-loop   magnetic-dipole kind of FCNC  interactions.
The latter are suppressed by the characteristic scale related to the mass of heavy messengers, connecting the standard model particles to the dark sector.
We compute the corresponding decay rates for the top, bottom,  and charm  decays ($t\to c\, \bar{\gamma},u\, \bar{\gamma}$, $\;b\to s\, \bar{\gamma},d\, \bar{\gamma}$, and  $c\to u \bar{\gamma}$), and for the charged-lepton decays ($\tau \to \mu\, \bar{\gamma}, e\, \bar{\gamma}$, and  $\mu \to e \bar{\gamma}$)
in terms of model parameters. We find that large branching ratios for both  quark and lepton decays are allowed in case the messenger masses are in the discovery range of the LHC. Implications of  these new decay channels
at present and future collider experiments  are briefly discussed.
\vspace*{5mm}

\noindent

\vfill\eject

\pagestyle{plain}


\section{Introduction}

One of the most intriguing aspects of the standard model (SM) is the nontrivial structure of the flavor sector, which is encoded in the corresponding structure of the Higgs-boson Yukawa couplings. The latter seem not to be originating from any global or gauge symmetry, and resemble  effective couplings rather than fundamental ones, their  eigenvalues  spanning over almost 6 orders of magnitude for charged fermions, and much more in case neutrinos are Dirac particles. The Cabibbo-Kobayashi-Maskawa (CKM) mixing matrix in the 
quark sector of weak charged currents (and the analogous one in the leptonic sector) adds further mystery to the origin and  structure of flavor. 

The recent discovery of the Higgs boson \cite{Aad:2012tfa}
has strengthened our confidence in the Higgs mechanism \cite{Englert:1964et}, 
and in the existence of its Yukawa couplings to fermions, necessary for
the fermion mass generation mechanism in the SM framework. All the observed Higgs properties seem to be in 
good agreement with the SM predictions~\cite{Khachatryan:2016vau}, although there is still  large
room for potential new physics (NP) contributions.
In this respect, the present experimental situation does not help, yet, to 
clarify whether the Yukawa couplings are fundamental or low-energy effective couplings, leaving 
space for new conjectures about the true  origin of flavor.

In case the Yukawa couplings are not fundamental, an interesting possibility is to conjecture that the chiral symmetry breaking (ChSB) and flavor structure originate from 
a dark sector and is communicated  to the SM by some kind of messenger fields~\cite{Gabrielli:2013jka,Ma:2014rua}.
The latter are by definition fields that couple both to the SM and dark-sector fields at  tree level. 
Then, due to the messenger interactions, the Yukawa couplings can be generated at one loop as effective low-energy couplings.

In this paper, we focus on the recent proposal in~\cite{Gabrielli:2013jka}, aiming at solving  the flavor hierarchy problem by explaining  the exponential spread in the Yukawa couplings at low 
energy.
For each SM fermion, this model predicts the existence of a massive fermion 
partner in the dark sector, singlet  under the SM gauge group (dubbed  {\it dark fermion} in the following), 
and a set of scalar messenger fields carrying the same SM quantum
numbers of squarks and slepton in supersymmetric models.
The Yukawa couplings $Y_f$ (where $f$ is a flavor index) are required to be vanishing
at tree level by imposing a discrete Higgs ($H$) parity,  $H\to -H$.
Then, via the spontaneous breaking of this symmetry, Yukawa couplings can be generated  at one loop. In particular, they can be induced by universal trilinear 
interactions that mix SM fields, dark fermions and messenger fields.
Due to chirality, the resulting Yukawa couplings turn out to be proportional to dark fermion masses $M_{F_f}$,
\bea
Y_f \sim \frac{M_{F_f}}{\Lambda_{\rm eff}},
\label{Yuk}
\eea
where  $\Lambda_{\rm eff}$ is an (almost) flavor-universal effective scale.
As a consequence, the observed SM Yukawa hierarchy just reflects
the structure of the dark fermion spectrum\footnote{A similar scenario with radiatively generated Yukawa couplings, and  a 
 $Y_f$ pattern as in  Eq.(\ref{Yuk}) has been 
proposed in \cite{Ma:2014rua}, although the latter does not include a discussion of the  dynamics responsible for the  
dark-fermion mass spectrum needed to give rise to the SM Yukawa hierarchy.}.

However, this conjecture alone is not sufficient to naturally solve the SM flavor hierarchy problem. A new dynamical mechanism is needed to explain the required pattern of  
dark-fermion masses. In \cite{Gabrielli:2013jka}, a  nonperturbative mechanism 
has been proposed to generate exponentially spread dark-fermion masses. It requires the existence
of an exact $U(1)_F$ gauge symmetry in the dark sector, and dark fermions 
$F_f$ 
charged under $U(1)_F$ with $\bar e_f$ quantum charges [in units of 
the fundamental $U(1)_F$ charge $\bar{e}$].
 In particular, this mechanism, based on a  Nambu-Jona-Lasinio  approach \cite{Nambu:1961tp},  predicts  
exponentially spread masses $M_{F_f}$ for  dark fermions according to the law~\cite{Gabrielli:2013jka}
\bea
M_{F_f} = \Lambda \exp{\left(-\frac{\gamma}{\bar{\alpha} \,\bar e^2_f}\right)}\, ,
\label{mass}
\eea
where $\bar{\alpha}= \bar e^2/(4\pi)$
is the $U(1)_F$ fine structure constant,  and $\gamma$ is connected to an anomalous dimension. The $\Lambda$  scale is  associated 
to the Lee-Wick term for the $U(1)_F$ gauge sector 
\cite{Lee:1971ix,Grinstein:2007mp}, which is responsible for triggering spontaneous ChSB,  and generating Dirac fermion masses \cite{Gabrielli:2013jka}.

The nonperturbative origin of the spectrum in Eq.~(\ref{mass}) as a function of $\bar{\alpha}$, is shown by the $1/\bar{\alpha}$ dependence in the exponent.
Then, by assuming order-${\cal O}(1)$
nonuniversality   among the $U(1)_F$ dark-fermion charges $\bar e_f$,   a wide 
exponential spread among fermion masses can be easily generated. Then Eq.~(\ref{mass}),
along with  Eq.(\ref{Yuk}), can provide the theoretical basis for a natural solution  to the SM flavor hierarchy problem.

A  peculiar aspect of this  model is the existence of a 
 dark photon  associated to the unbroken $U(1)_F$ gauge field, which, being massless,
  does not couple at tree level to SM fields  \cite{Holdom}. Dark-photon couplings 
to the  SM fields can instead  arise at one loop by means of higher-order operators, which are suppressed by the characteristic  messenger mass scale.

In this framework, a new interesting phenomenology is expected that can be testable at the LHC \cite{Gabrielli:2014oya,Biswas:2016jsh} and at future colliders \cite{Biswas:2015sha}. For instance,
 Higgs  effective couplings  to photon ($\gamma$) and dark 
photon ($\bar{\gamma}$), or to two dark photons, can arise at one loop due to the exchange of messenger and dark-fermion fields in \cite{Gabrielli:2014oya}. 
 These effective  couplings can lead to exotic signatures, such as the one associated to  
 the $H\to \gamma \bar{\gamma}$ decay, where the dark photon is observed in a detector
 as a massless invisible system.  The LHC has an excellent potential to  observe such decay for realistic branching ratios (BR's), in particular in the run~2~ \cite{Gabrielli:2014oya,Biswas:2016jsh}. 
Implications of  the Higgs effective couplings to  dark photons have also been 
 analyzed in $e^+e^-$ collisions~\cite{Biswas:2015sha}.

In this paper we will focus on the dark-sector flavor structure  
needed to generate the CKM matrix in a theoretical framework based on the model in~\cite{Gabrielli:2013jka}.
We will show that the required structure can potentially induce new exotic 
flavor-changing-neutral-current (FCNC) processes in the quark and lepton sectors. In particular, one foresees  a new class of FCNC decay channels,  namely  the fermion decays to a lighter fermion of the same electric charge accompanied by a massless (invisible) dark photon, 
\bea
f \to f^ {\prime}~  \bar{\gamma} \, .
\label{fdecay}
\eea

We will first analyze  the phenomenological implications of such FCNC 
decays in the top-quark, bottom-quark, and charm-quark sectors, by studying the $t\to c\, \bar{\gamma},u\, \bar{\gamma}$, $\;b\to s\, \bar{\gamma},d\, \bar{\gamma}$, and  $c\to u \bar{\gamma}$ decay channels, respectively. We will then  
extend the analysis to the leptonic sector, including the 
tau and muon decays 
$\tau \to \mu\, \bar{\gamma}, e\, \bar{\gamma}$, and  $\mu \to e \bar{\gamma}$.
In particular, we will compute different BR's and discuss their corresponding upper bounds coming from  present  phenomenological and theoretical constraints.

As mentioned above,  massless dark photons are decoupled at  tree level from  SM fields, and their  production at colliders manifests  as  missing energy $\slashed E$ and momentum $\slashed p$ in the detector,
satisfying the kinematical {\it neutrinolike} constrain $\slashed E^2\!\!-\slashed p^2=0$.
As a consequence, the FCNC $f \!\to\! f^{\prime} \, \bar{\gamma}$ decay  is characterized
by an exotic experimental signature, where the final same-charge fermion $f^{\prime}$ is balanced in a detector by an invisible system 
with vanishing invariant mass. 
 In the $f$ rest frame, neglecting radiative effects,  $f^{\prime}$ is monochromatic with energy $E_{f^{\prime}}\simeq m_f/2$, which is  a very distinctive feature that would crucially discriminate $f \!\to\! f^{\prime} \, \bar{\gamma}$ backgrounds, 
where the missing momentum is associated either to  the mismeasurement of hadronic objects or to the presence of nonmonochromatic neutrinos in the final states (as occurs in the $\mu$ or  $\tau$ decays). 
 Altogether a $f \!\to\! f^{\prime} \, \bar{\gamma}$ decay would show up experimentally by  an excellent characterization.

The plan of the paper is the following. In Sec. 2, we present the theoretical framework,  and
provide the relevant Feynman rules for the computation of the FCNC $f \!\to\! f^{\prime} \, \bar{\gamma}$ decay amplitudes. In Sec. 3, we give the analytic expressions for the amplitude of a generic $f \!\to\! f^{\prime} \, \bar{\gamma}$ decay,  and  corresponding BR. In Sec. 4, 5, 6, 7, and 8, we will analyze the 
phenomenological implications  for the FCNC decays in the top-quark, bottom-quark,
charm-quark, $\tau$ and $\mu$  sectors, respectively. Our conclusions will be given in Sec. 9. 
\section{Theoretical framework}
In this section we summarize the main aspects of the flavor model in~\cite{Gabrielli:2013jka}, providing the relevant interaction terms  
 for the FCNC $f \!\to\! f^{\prime} \, \bar{\gamma}$ decays
in the Lagrangian, and corresponding notation. More details on the model can be found in~\cite{Gabrielli:2013jka,Biswas:2015sha}.

As mentioned, the model extends the SM theory in order to  generate radiatively Yukawa couplings at one loop, assuming vanishing tree-level Yukawa couplings.
The corresponding total Lagrangian is made up of three sectors 
\bea
{\cal L}&= {\cal L}^{ Y=0}_{SM} + {\cal L}_{DS}+ {\cal L}_{MS},
\label{totlagr}
\eea
where  ${\cal L}^{Y=0}_{SM}$ is the SM Lagrangian for vanishing
tree-level Yukawa couplings, 
${\cal L}_{DS}$ is the dark-sector (DS) Lagrangian, containing the dark-fermion interactions
with the $U(1)_F$ dark-photon gauge field, 
and
${\cal L}_{MS}$ 
describes the messenger sector with its couplings to both SM and dark fields.
The ${\cal L}_{MS}$  interactions  also 
communicate the ChSB and flavor structure of the dark sector to the  observable SM sector, through the generation of  Yukawa couplings at one loop.

\subsection{The dark-quark sector}
We start by recalling the ${\cal L}_{DS}$ Lagrangian  related to the 
dark fermions associated to quarks (which we call {\it dark quarks})
and their interactions with the $U(1)_F$ gauge sector, including the
mechanism to generate exponentially spread fermion masses.
Its generalization to the leptonic sector will then be straightforward. 

For each SM quark $q^{\U_i,\D_i}$ (with $i$ a family index),  a quark replica  $Q^{\U_i,\D_i}$ is assumed in the dark sector, which is singlet under SM gauge interactions, and 
charged under an exact $U(1)_F$ gauge symmetry.
The corresponding Lagrangian is given by
\bea
{\cal L}_{DS}&=& 
i\sum_i \left( 
\bar{Q}^{\U_i}{\cal D}_{\mu}\gamma^{\mu} Q^{\U_i}+\bar{Q}^{\D_i}{\cal D}_{\mu}\gamma^{\mu} Q^{\D_i}\right)
\nonumber \\
&-&
\frac{1}{4} F_{\mu\nu} F^{\mu\nu} + \frac{1}{2\Lambda^2}  \partial^{\mu} F_{\mu\alpha} \partial_{\nu} F^{\nu\alpha},
\label{LagDS}
\eea
where ${\cal D}_{\mu}=\partial_{\mu}+i g \hat{Q} A_{\mu}$
is the usual covariant derivative associated to  
the $U(1)_F$ dark-photon $A_{\mu}$ gauge field, with $\hat{Q}$ the corresponding charge operator acting on the $Q^{\U_i}$ and  $Q^{\D_i}$ quark  
fields, and $F_{\mu\alpha}$  the  $U(1)_F$ field-strength tensor. The higher-derivative last  term in Eq.(\ref{LagDS}) is the so-called Lee-Wick term, where $\Lambda$ is the associated energy scale. 

As shown in \cite{Gabrielli:2007cp}, because of  the Lee-Wick term, which  implies  a massive spin-1 ghost particle in the spectrum, chiral 
symmetry turns out to be spontaneously broken, and dark fermions acquire mass nonperturbatively. In particular, by following the Nambu-Jona-Lasinio approach, one can show that
a Dirac quark  mass $M_{Q_f}$,  solution of the fermion mass-gap equation  corresponding to the true vacuum of the theory, exists in the weakly coupled regime in the form~\cite{Gabrielli:2007cp}
\bea
M_{Q_f}=\Lambda \exp\left\{-\frac{2\pi}{3\bar{\alpha}(\Lambda) \bar e_f^2} +\frac{1}{4}\right\}\,,
\label{mgap2}
\eea
where $\bar e_f$ stands for  the $U(1)_F$ charge eigenvalue of a generic  dark quark   of  flavor $f$, $Q_f$, in unit of the fundamental charge $\bar{e}$, and $\bar{\alpha}(\Lambda)$ 
is the effective fine structure constant (associated to $\bar{e}$) evaluated 
at the scale $\Lambda$. As already stressed, 
 this solution is truly nonperturbative (as shown by the $\bar{\alpha}$ dependence in the exponent), and  is associated 
to the true (nonperturbative) vacuum of the theory.
For $N_F$ dark quarks with $\bar e_f$ charges ($f=1,\dots N_F$), 
an exponentially spread $M_{Q_f}$ spectrum  can be generated
by Eq.~(\ref{mgap2}), just by requiring nonuniversality among the corresponding 
 $\bar e_f$ charges. Indeed, since the  $M_{Q_f}$ hierarchy in Eq.~(\ref{mgap2}) will reflect
 into the actual SM fermion Yukawa hierarchy (as discussed in the following), it turns out that,  for an integer sequence of 
$\bar e_f$ charges (and extending the present analysis to include the leptonic sector), one can easily fit most of the SM fermion mass spectrum~\cite{Gabrielli:2013jka}.

\subsection{The messenger sector and the generation of Yukawa couplings}
The  ${\cal L}_{MS}$ Lagrangian in Eq.~(\ref{totlagr}) contains messenger scalar fields,
and can be split in two terms
\bea
{\cal L}_{MS}&=&{\cal L}^{\rm 0}_{MS}+{\cal L}^{\rm I}_{MS}\, .
\label{MSMS}
\eea
${\cal L}^{\rm 0}_{MS}$ includes the kinetic term for the 
messenger fields interacting with the SM gauge fields, while  ${\cal L}^{\rm I}_{MS}$ provides the messenger interactions with the SM fermions, the dark fermions, and the Higgs boson, which are responsible for generating Yukawa couplings radiatively. 

The SM quark gauge quantum numbers fix the minimal matter content
needed for the colored messenger scalar sector, which is given by
\begin{itemize}
\item $2N$ complex scalar $SU(2)_L$ doublets: $\hat{S}_L^{\U_i}$ and $\hat{S}_L^{\D_i}$,
\item $2N$ complex scalar  $SU(2)_L$ singlets: $S_R^{\U_i}$ and $S_R^{\D_i}$,
\item one real $SU(2)_L\times U(1)_Y$ singlet: $S_0$,
\end{itemize} 
where
$\hat{S}_{\LL}^{\U_i,\D_i}=\left(\begin{array}{c}S^{\U_i,\D_i}_{\LL,1}\\S^{\U_i,\D_i}_{\LL,2}
\end{array}\right)$, 
 and $i=1,\dots,N$ ($N=3$) stands for a family index.
The $\hat{S}_{\LL}^{\U_i,\D_i}$, $S_{R}^{\U_i,\D_i}$ scalar fields 
have the SM quark quantum numbers, where the $L,R$ labels  identify the 
messengers coupled to the $L,R$ chirality components of the associated SM quarks  (just as occurs  in the case of squark fields in supersymmetric theories).
They have minimal gauge-invariant couplings to  electroweak (EW) gauge bosons and  gluons.  A minimal flavor violation hypothesis would require the Lagrangian in Eq.~(\ref{MSMS}) to be invariant under $SU(N_F)$, where $N_F$ is the number of
flavors. More generally, for any family index $i$, we can reduce the messenger mass sector to four different  
universal  mass  terms corresponding to the up/down and $L/R$ components of  the $\hat{S}_{L,R}^{\U_i}$ and $\hat{S}_{L,R}^{\D_i}$ sectors, as in minimal supersymmetric models. Notice that a more minimal hypothesis of  a  common scalar mass for the $L$ and $R$ scalar sectors is also phenomenologically acceptable.

We do not report here the expression for the interaction Lagrangian of the messenger  fields 
with the SM gauge bosons, which follows from the universal properties of gauge interactions. 
Notice that each messenger field is  also charged under $U(1)_F$, and carries the same $U(1)_F$ charge of the associated dark fermion. In other words,
$U(1)_F$ charges identify the flavor state.
A summary of relevant quantum numbers for all new   fermion  and scalar fields in the quark sector can be found in Table~\ref{tab1}.
\begin{table} \begin{center}    
\vspace{2 cm}
\begin{tabular}{|c||c|c|c|c|c|}
\hline 
Fields 
& Spin
& $SU(2)_L$ 
& $U(1)_Y$
& $SU(3)_c$
& $U(1)_F$
\\ \hline 
$\hat{S}_L^{\D_i}$
& 0
& 1/2
& 1/3
& 3
& -$\qdi$
\\ \hline
$\hat{S}_L^{\U_i}$
& 0
& 1/2
& 1/3
& 3
& -$\qui$
\\ \hline
$S_R^{\D_i}$
& 0
& 0
& -2/3
& 3
& -$\qdi$
\\ \hline
$S_R^{\U_i}$
& 0
& 0
& 4/3
& 3
& -$\qui$
\\ \hline
$Q^{\D_i}$
& 1/2
& 0
& 0
& 0
& $\qdi$
\\ \hline
$Q^{\U_i}$
& 1/2
& 0
& 0
& 0
& $\qui$
\\ \hline
$S_0$
& 0
& 0
& 0
& 0
& 0
\\ \hline \end{tabular} 
\vspace{1 cm}
\caption[]{
Spin and gauge quantum numbers for the strongly interacting messenger fields and 
corresponding dark quarks.   
$U(1)_F$ is the dark-photon gauge symmetry  in the dark sector.
}
\label{tab1}
\end{center} \end{table}

The   ${\cal L}^I_{MS}$ Lagrangian, which describes
the messenger  interactions with quarks and SM Higgs boson,
is particularly relevant for the SM flavor structure. The minimal
content of the universal interactions needed  to generate radiative (diagonal)  Yukawa couplings is
\bea
{\cal L}^I_{MS} &=&\Big\{
g_L   \left( \sum_{i=1}^{N}\left[\bar{q}^i_L Q_R^{\U_i}\right] \hat{S}^{\U_i}_{L} +
\sum_{i=1}^{N}\left[\bar{q}^i_L Q_R^{\D_i}\right] \hat{S}^{D_i}_{L}\right)
\nonumber\\
&+&
g_R \left(\sum_{i=1}^{N}\left[\bar{\scriptstyle U}^i_R Q_L^{\U_i}\right] S^{\U_i}_{R} +
\sum_{i=1}^{N} \left[\bar{\scriptstyle D}^i_R Q_L^{\D_i}\right] S^{\D_i}_{R}\right) 
\nonumber\\
&+& \;
\lambda_S S_0 \sum_{i=1}^{N}\left(\tilde{H}^{\dag} S^{\U_i}_L S^{\U_i\dag}_R+ H^{\dag} S^{\D_i}_L S^{\D_i\dag}_R \right) 
\,+\, H.c. \; \Big\}+ \; V(S_0)\,,
\label{LagMS}
\eea 
where contractions with color indices are understood. The  
$S_0$ field is a real singlet scalar, and its potential  $V(S_0)$ is invariant under the 
$S_0\!\to\!\! -S_0$ parity symmetry.
 The $g_L$ and $g_R$ constants  are  flavor-universal free parameters, whose values  can be in the perturbative regime $g_{L,R}\lsim 1$. We will assume in general 
$g_L\neq g_R$, although one could impose a higher degree of universality  by 
assuming $g_L=g_R$, with no loss of generality in the prediction 
of Yukawa couplings. In Eq.~(\ref{LagMS}),  $q^i_L$, ${\scriptstyle U}^i_R$, and ${\scriptstyle D}^i_R$ 
stand for SM quark fields, and  $H$ is the SM Higgs doublet, with
$\tilde{H}=i\sigma_2 H^{\star}$. 
One can then prevent Yukawa couplings at  tree level by imposing a combined parity symmetry under $H\to\! -H$ and $S_0\to\! -S_0$.

On the other hand, 
as shown in  \cite{Gabrielli:2013jka}, after spontaneous symmetry breaking (SSB) of the $H\to\! -H$ and $S_0\to\! -S_0$ parity symmetry  by  a nonvanishing vacuum expectation value (VEV) $\langle S_0\rangle \equiv \mu_S/\lambda_S$, the Yukawa couplings can be radiatively generated at one loop via  virtual exchange of messengers and dark fermions.
As a result, the
effective  Yukawa coupling associated to the quark of flavor $f$ turns out to be proportional  to 
the corresponding dark-quark  mass $M_{Q_f}$.
In particular, one obtains 
\cite{Gabrielli:2013jka}
\bea
Y_f&=&Y_0(x_f) \exp{\left(-\frac{2\pi}{3\bar{\alpha}(\Lambda) \bar e_f^2}\right)}\, ,
\label{Yeff}
\eea
where  the dark-quark  mass $M_{Q_f}$  has been replaced by Eq.(\ref{mgap2}),
 the one-loop $Y_0(x_f)$  function is given by
\bea
Y_0(x_f)&=&\left(\frac{g_L g_R }{16 \pi^2 }\right)
\left(\frac{\mu_S\Lambda}{\bar{m}^2}\right) C_0(x_f)\, ,
\label{Yuk0}
\eea
$\bar{m}^2$ is  the mean square mass of the messengers running in the loop, 
$x_f=M_{Q_f}^2/\bar{m}^2$, and 
\bea
C_0(x)=\frac{1-x\left(1-\log{x}\right)}{(1-x)^2}\, .
\label{C0}
\eea 
Equation (\ref{Yuk0}) is obtained
in the approximation of degenerate messenger masses  for  generic 
$SU(2)_L$ doublet $S_L$ and singlet $S_R$ fields, and in the limit of
small mixing parameter $\xi=\Delta/\bar{m}^2$, 
with $\Delta=\mu_S v$, and $v$ the Higgs VEV. 

As  from  Eqs.(\ref{Yeff}) and (\ref{Yuk0}), the top-quark Yukawa coupling can be  large and ${\cal O}(1)$,  keeping at the same time the dimensionless couplings $g_L, g_R$ small and within the perturbative regime. Indeed the Yukawa coupling turns out to be proportional to the singlet-field ($S$) VEV  ($\mu_S/\lambda_S$), and is generated only after the spontaneous breaking of the $Z_2$ symmetry. This is a general property, which is independent from the particular symmetry forbidding Yukawa couplings at the tree level. Then, a ${\cal O}(1)$ Yukawa coupling  can be achieved by choosing the $\mu_S$ scale  larger than the characteristic $\bar{m}$-mass scale  running in the loop, while keeping all other dimensionless couplings small and in the perturbative range.

In order to extend  the above 
results to larger $\xi$ mixing values, one can use the mass-eigenstate basis  for  messengers. 
Notice that, after the EW symmetry breaking, terms in the third row of the Lagrangian in Eq.(\ref{LagMS}) generate a mixing term $\Delta$ between the $SU(2)_L$ messenger doublet $S_L$, and the corresponding singlet $S_R$. The 
corresponding  Lagrangian for generic $S_{L,R}$ fields  is
\bea
{\cal L}^0_{S}&=& \partial_{\mu} \hat{S}^{\dag} \partial^{\mu}\hat{S} - \hat{S}^{\dag} \hat{M}^2_S \hat{S},
\eea
where  $\hat{S}=(S_L,S_R),$ and the  mass term involves the mass matrix 
\begin{equation}
\hat{M}^2_S = \left (
\begin{array}{cc}
m^2_L & \Delta \\
\Delta & m^2_R 
\end{array}
\right),
\label{M2}
\end{equation}
with $\Delta=\mu_S v$ parametrizing the left-right ($LR$) scalar mixing. 
The $\hat{M}^2_S$ matrix in Eq.~(\ref{M2}) can be diagonalized by the unitary matrix
\begin{equation}
U = \left (
\begin{array}{cc}
\cos{\theta} & \sin{\theta} \\
-\sin{\theta} & \cos{\theta} 
\end{array}
\right),
\label{U2}
\end{equation}
 with  
$\tan{2\theta}=\frac{2\Delta}{m_L^2-m_R^2}$.
Then, the eigenvalues of the diagonal  matrix $\hat{M}^{2 \,\rm diag}_S=U \hat{M}^2_S \,U^{\dag}$ are given by 
\bea
m^2_{\pm}=\frac{1}{2}\left(m^2_L+m_R^2 \pm
\left[(m^2_L-m^2_R)^2+4\,\Delta^2\right]^{1/2}
\right)\, , 
\eea
where, for degenerate $LR$ scenarios (namely for $m^2_L=m^2_R=\bar{m}^2$), the $U$ matrix elements
simplify to $U(i,i)=1/\sqrt{2}, \;U(1,2)=\!-U(2,1)=1/\sqrt{2}$, with square mass  
eigenvalues 
\bea
m^2_{\pm}=\bar{m}^2(1\pm \xi)\,, \;\;\;\;\;  \;\;\;\;\; \xi=\Delta/\bar{m}^2 \, .
\label{meigen}
\eea
Note that, in order to prevent tachyonic solutions, one should impose $\xi \le1$ with $\Delta>0$.
Then, we computed the generalization of the $Y_0$ expression  in Eq.(\ref{Yuk0}) 
   as a function of $\xi$, in the degenerate $LR$ scenario, which turns out to be 
   \bea
Y_0(x_f,\xi)&=&\left(\frac{g_L g_R }{16 \pi^2 }\right)
\left(\frac{\xi \Lambda}{v}\right) f_1(x_f,\xi) \, , 
\label{Yukexact}
\eea
where
\bea
f_1(x,\xi)&=&\frac{1}{2}\left[
C_0(\frac{x}{1-\xi})\frac{1}{1-\xi}+C_0(\frac{x}{1+\xi})\frac{1}{1+\xi}\right]\, ,
\label{f1}
\eea
 and $C_0(x)$ is defined by Eq.(\ref{C0}). Notice that $C_0(1)=1/2$, and, for small $x\ll 1$,  $C_0(x)\simeq 1+{\cal O}(x)$.
 Indeed, at fixed values of $\bar{m}$ and 
$\Lambda$, all Yukawa couplings must vanish for vanishing mixing $\xi\to 0$, 
since they are proportional to the  VEV  of the singlet  field $S$, $\mu_S$ (\cf  Eq.~(\ref{Yuk0})).

\subsection{The flavor structure and the CKM matrix}
Although predicting exponentially spread Yukawa couplings and providing a natural solution to the flavor hierarchy problem,  the {\it minimal} interaction Lagrangian ${\cal L}^I_{MS} $ in Eq.(\ref{LagMS}) does not account for the observed CKM mixing matrix of weak interactions. Indeed, 
the radiatively generated Yukawa couplings  turn out to be diagonal in the  weak-current interaction basis for the quark fields.
 Yukawa off diagonal terms are needed to generate the
CKM, and, in order to preserve the $U(1)_F$ gauge invariance, the universal flavor structure of the messenger interaction in Eq.(\ref{LagMS}) should be generalized as follows:
\bea
\tilde {\cal L}^I_{MS} &=&\Big\{
g_L\left( \sum_{i,j=1}^{N}\left[\bar{q}^i_L (X_L^U)_{ij} Q_R^{\U_j}\right] \hat{S}^{\U_j}_{L} +
\sum_{i,j=1}^{N}\left[\bar{q}^i_L  (X_L^D)_{ij} Q_R^{\D_j}\right] \hat{S}^{D_j}_{L}\right)
\nonumber\\
&+&
g_R\left(\sum_{i,j=1}^{N}\left[\bar{\scriptstyle U}^i_R (X_R^U)_{ij} Q_L^{\U_j}\right] S^{\U_j}_{R} +
\sum_{i,j=1}^{N}\left[\bar{\scriptstyle D}^i_R  (X_L^D)_{ij} Q_L^{\D_j}\right] S^{\D_j}_{R}\right)
\nonumber\\
&+& \;
\lambda_S S_0 \sum_{i=1}^{N}\left(\tilde{H}^{\dag} S^{\U_i}_L S^{\U_i\dag}_R+ H^{\dag} S^{\D_i}_L S^{\D_i\dag}_R\right) \,+\, H.c. \; \Big\}+ \; V(S_0)\, ,
 \label{LagMS-tilde}
\eea 
where $X_{L,R}^{U,D}$ are generic (not necessarily unitary) matrices. Notice
that the $U(1)_F$ gauge invariance and nonuniversality of $U(1)_F$ charges  
require the family index  labeling dark fermions and scalar messengers to be the same.
Then, in the weak-current basis for quark fields, 
the Yukawa couplings generated radiatively  follow the pattern
\bea
Y^{U,D}_{ij}\sim \left( X^{U,D~\dag}_L\cdot \hat{Y}^{U,D}\cdot X^{U,D}_R\right)_{ij}\, ,
\label{Ygeneral}
\eea
where the $\,\cdot \,$ symbol stands for a matrix product, and 
$\hat{Y}^{U,D}={\rm diag}[Y^{U,D}_1,Y^{U,D}_2,Y^{U,D}_3]$, with 
$Y_i^{U,D}$ ($i=1,2,3$) standing for the  Yukawa couplings  in Eq.~(\ref{Yeff}) for the {\it up} and {\it down} sectors.
Following the usual SM approach, the Yukawa matrix in Eq.~(\ref{Ygeneral}) can be diagonalized by a biunitary rotation $V^{U,D}_{L,R}$, namely 
\bea
{\rm diag}[Y^{U,D}]=V^{U,D~\dag}_{L,R}\cdot Y^{U,D}\cdot V^{U,D}_{L,R}\, ,
\eea
hence giving rise to the CKM matrix $K=V^{U~\dag}_L\cdot V^{D}_L$.

The observed structure of the  CKM matrix requires $X_{L,R}^{U,D}$ to have off diagonal entries smaller than the diagonal ones, 
with the latter almost proportional to the  unity  matrix ${\bf 1}$ in the family space. This suggests the following ansatz for the $X_{L,R}^{U,D}$ matrices
\bea
X_{L,R}^{U,D}\sim {\bf 1} + \Delta_{L,R}^{U,D}\, ,
\label{X}
\eea
where  the matrices $|\Delta_{L,R}^{U,D}|\ll 1 $ collect  
 diagonal and off diagonal terms in a less hierarchical structure\footnote{
We suggest a possible renormalization mechanism for generating a flavor structure of the $X_{L,R}^{U,D}$ matrices as required by Eq.~(\ref{X}), assuming universal tree-level couplings like in Eq.~(\ref{LagMS}). This requires  new heavy  (either scalar or vector) fields in the dark sector, which are
SM gauge singlets, and are charged under $U(1)_F$ 
with charges $Q_{ij}=\bar e_i-\bar e_j$ ($i,j=1,2,3$). Gauge invariant couplings of these new fields to both dark fermions and messenger scalars can be formed. One-loop corrections  
to the vertices of the universal interactions in Eq.~(\ref{LagMS}),  induced by these new interactions in the dark sector, can then generate the desired off diagonal transitions that can be reabsorbed in the matrix elements $\Delta_{ij}$. Being $\Delta_{ij}$ generated at  higher orders in perturbation theory, the 
hierarchy  shown in  Eq.~(\ref{X}) is automatically satisfied.
We will not consider this possibility here, and will assume 
the most general structure for the $X$ matrices, no matter what mechanism has  generated them.}.

After rotating the quark fields to the basis of mass eigenstates (entailing diagonal Yukawa couplings), the interaction terms in Eq.(\ref{LagMS-tilde}) can be transformed by 
replacing the $X_{L,R}^{U,D}$  matrices according to 
\bea
X^{U}_{L,R} \to \rho_{L,R},~~~~~~X^{D}_{L,R} \to \eta_{L,R} \, ,
\eea
where
\bea
\rho_{L,R}&\equiv& V^{U~\dag}_{L,R}\cdot X_{L,R}^{U}\, , \\
\noindent
\eta_{L,R}&\equiv& V^{D~\dag}_{L,R}\cdot X_{L,R}^{D}\, .
\label{rho-eta-matr}
\eea

If $X^{U,D}_L$ are unitary matrices, then 
 $V^{U,D}_{L,R}=(X^{U,D}_L)^{-1}$,   the $\rho_{L,R},\, \eta_{L,R}$ matrices
will just be equal to ${\bf 1}$, and  the CKM matrix will be $K=X_L^U\cdot X_L^{D\dag}$.
However,   $X^{U,D}_L$ matrices do not need to be unitary 
(or proportional to a unitary matrix), since they do not arise from 
unitary transformations (like, \eg, in the CKM-matrix case). In
 general,  $\rho_{L,R}$ and $\eta_{L,R}$ will then have nonvanishing 
off diagonal entries. This has nontrivial consequences, since off diagonal terms in the  $\rho_{L,R}$ and $\eta_{L,R}$ matrices 
can induce FCNC interactions  {\it at one loop} in the observable quark and lepton sectors.

Among the  FCNC processes induced by these new interactions, 
there is a new class of FCNC one-loop decays, that is  SM-fermion decays into a massless dark photon, via the channels $q \to q^{\prime} \bar{\gamma}$ ($\ell \to  \ell^{\prime} \bar{\gamma})$, where $q^{\prime}$ $ (\ell^{\prime})$ is a lighter quark (lepton) with same charge as $q $ $(\ell)$. 

\begin{figure}
\begin{center}
\includegraphics[width=0.9\textwidth]{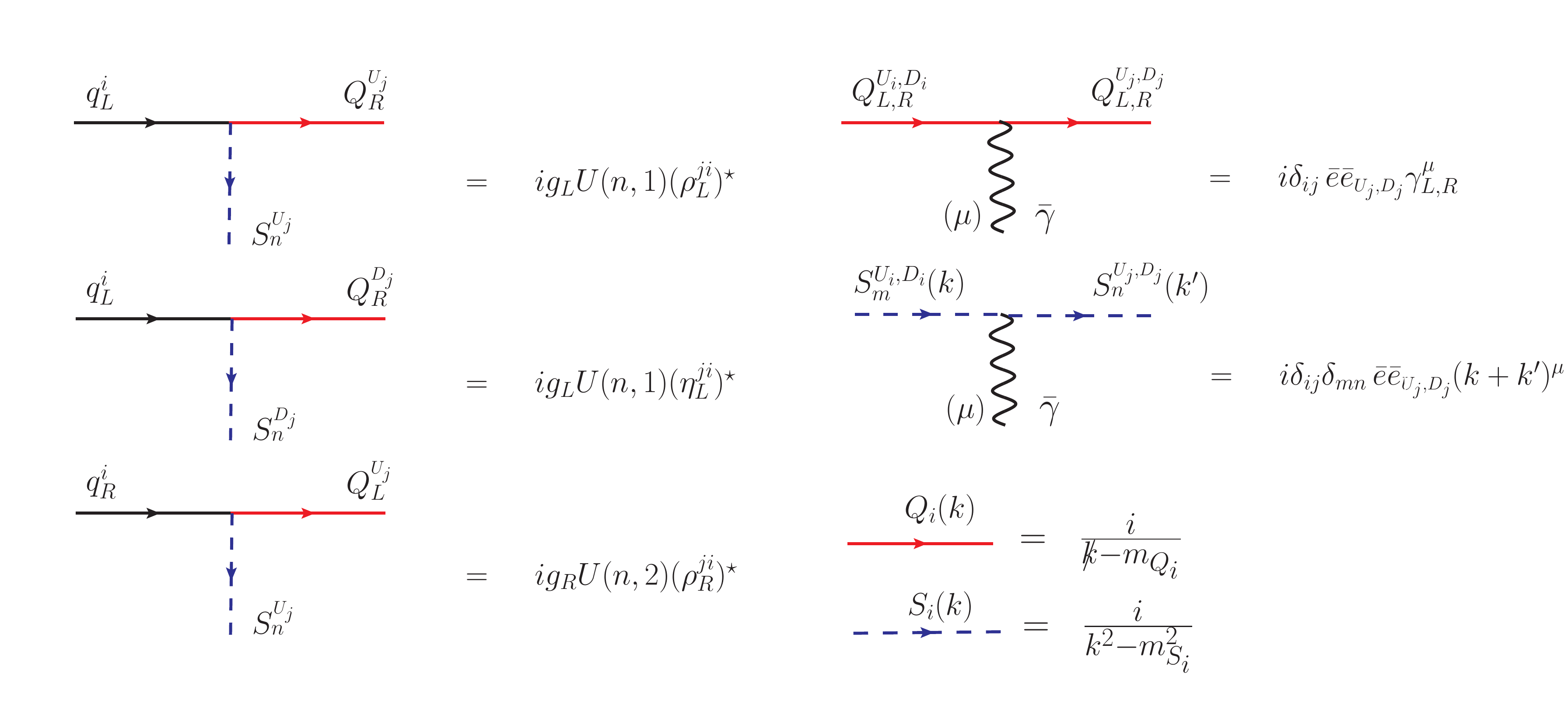}
\caption{Feynman rules for interaction vertices and propagators entering 
the computation of the one-loop $q\to q^{\prime} \bar{\gamma}$ decay amplitude.
The symbols $q_{L/R}^i$ and $Q_{L/R}^{U_i,D_i}$ stand for the quark and dark-quark  fields, respectively, with $L/R$ denoting the left-/right-handed chirality projections. $S_n^{U_j}$ and $S_n^{D_j}$ stand for the  mass eigenstates ($n=1,2$) in the {\it up} and {\it down} messenger  sectors, respectively, while $\bar{\gamma}$ is the dark-photon field.}
\label{Fig-FR}
\end{center}
\end{figure}

In the next section, we will compute the relevant amplitudes and corresponding
decay widths for this new class of processes, as well as the NP contribution to the 
 $q\to q^{\prime} \gamma$ and $\ell \to  \ell^{\prime} {\gamma}$ decays into a SM photon.
The Feynman rules relevant for the computation of the $q\to q^{\prime} \bar{\gamma}$ decay amplitude (with straightforward extension to the leptonic sector)
 can be found in Fig.~\ref{Fig-FR}. 

\section{The $q\to q^{\prime}\, \bar{\gamma}$ amplitude and decay width}
For a generic quark $q^i$, with $q=U,D$, we consider  the FCNC decay process  
\bea
q^i(p)\to q^j(p^{\prime})\; \bar{\gamma}(k)
\eea
where the indices $i, j$ run over  quark families with $i>j$, 
and 
$p,p^{\prime}$, and $k$ indicate the particle four-momenta. A generalization to the 
leptonic sector is straightforward. 
This process is induced at  one loop by the Lagrangian  in Eq.~(\ref{LagMS-tilde}) for quarks (and by its leptonic generalization for lepton decays).

\begin{figure}
\begin{center}
\vspace{-0.4cm}
\includegraphics[width=0.6\textwidth]{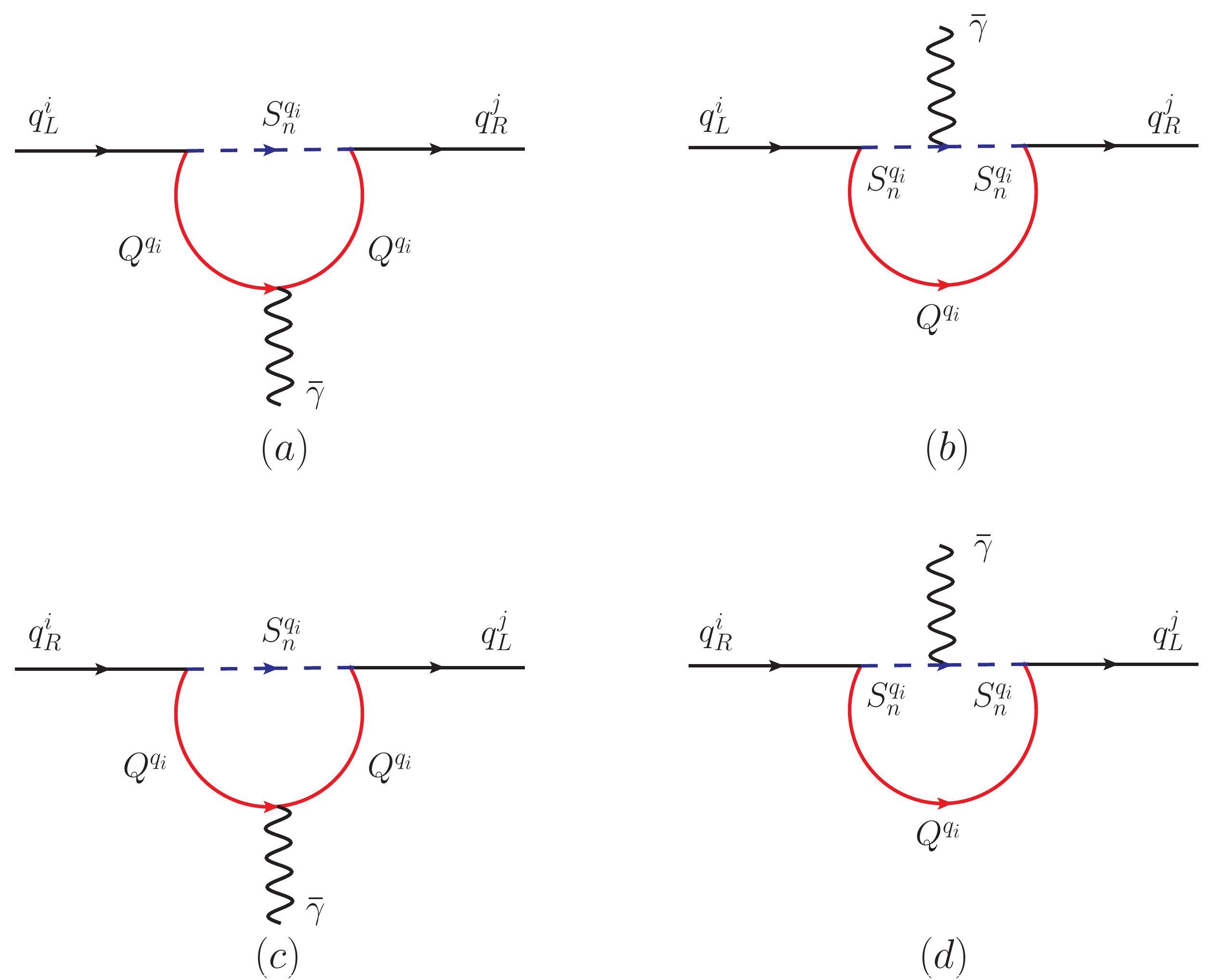}
\caption{Feynman diagrams (a)--(d) contributing to the FCNC decay
$q^i\to q^ j \bar{\gamma}$ with $q=U,D$ and $i > j$, 
where $q^{i,j}_{L,R}$ are the initial ($i$), final  ($j$) quarks, 
with $L/R$ indicating the left/right chirality projections, $Q^{q_i}$ 
and $S_n^{q_i}$ the corresponding dark quarks and messenger fields, 
respectively, with the latter in the basis of mass eigenstates ($n=1,2$),
while $\bar{\gamma}$ stands for the dark-photon line.}
\label{Fig-FD}
\end{center}
\end{figure}

The Feynman diagrams contributing to the $q^i\to q^j\, \bar{\gamma}$ process are given in Fig.~\ref{Fig-FD}. 
There are no self-energy contributions to the $q^ i\to q^ j \,\bar{\gamma}$ process, since the dark photon does not couple to SM fermions at tree level. 
The  messengers running in the loop are much heavier than the external fermion states (also in  case of the top-quark decay), and we  can safely neglect  terms of order ${\cal O}(m_{q_i}^2/\bar{m}^2)$, where $m_{q_i}$ are the external-quark masses. However, we will retain the
leading contributions induced by the initial SM fermion mass, or, equivalently, by 
its associated Yukawa coupling, and neglect the contributions of the final quark mass.

The total  amplitude in momentum space receives two independent gauge-invariant contributions
\bea
M(q^i \to q^j\, \bar{\gamma}) &=&
M(q^i_L \to q^j_R \, \bar{\gamma})\, +\, 
 M(q^i_R \to q^j_L \, \bar{\gamma})\, , 
\eea
where 
$q^{i}_{L/R}$ are 
chirality eigenstates in the $q=U,D$ sectors. The two contributions  can be parametrized as follows
\bea
M(q^i_L \to q^j_R \, \bar{\gamma}) &=& \frac{1}{(\Lambda^q_L)_{ij}}
[\bar{u}_R^{q_j} \sigma_{\alpha\mu} u_L^{q_i} ] k^{\mu} \bar{\epsilon}^{\alpha}\, , \nonumber
\\
M(q^i_R \to q^j_L \, \bar{\gamma}) &=& \frac{1}{(\Lambda^q_R)_{ij}}
[\bar{u}_L^{q_j} \sigma_{\alpha\mu} u^{q_i}_R ] k^{\mu} \bar{\epsilon}^{\alpha}\, ,
\label{amp}
\eea
where $\sigma_{\mu\nu}\equiv \frac{1}{2}[\gamma_{\mu},\gamma_{\nu}]\,$  ($[a,b]$ 
standing for the {\it a} and {\it b} matrix commutator),  $u_{L/R}\equiv \frac{1}{2}(1\mp\gamma_5) u$, and $u^{q_i}$ and $u^{q_j}$ 
correspond respectively to the $q^i$ and $q^j$ on shell bispinors in momentum space,  $\bar{\epsilon}^{\alpha}$ being the dark-photon polarization vector. Gauge invariance 
requires  $k_{\mu}\bar{\epsilon}^{\mu}=0$ for  on shell dark photons, which makes  
the contribution proportional to the $\gamma^{\mu}_{L,R}$ matrices vanish for 
 on shell massless (\ie, for $k^2=0$)  dark photons.
As a consistency check, we have controlled that this condition is satisfied at one loop.

Then, the effective scales associated to the matrix elements $(\Lambda^q_{L,R})_{ij}$   can be derived by matching Eq.~(\ref{amp}) with  the  computation of the transition amplitude, based on the Feynman diagrams in 
Fig.~\ref{Fig-FD}.
We assume massless  
final fermions, which will be a proper approximation 
for the processes considered in the following.

The contribution to the magnetic-dipole type of operator $[\bar{u}_R^{q_j} \sigma_{\alpha\mu} u_L^{q_i} ]$  (which is finite and $U(1)_F$ gauge invariant) involves different chiralities in the
external $u^{q_i}$ and $\bar{u}^{q_j}$ states. There are  two different sources for the  chirality flip. One corresponds to the mass insertion of a virtual dark quark through its propagator, and the second arises from the external quark  
masses, after the on shell conditions on $u^{q_i}$ or $\bar{u}^{q_j}$ are applied.  Since we are assuming massless final fermions, only initial fermions contribute to the latter. 

Finally, after some algebraic manipulations, the  
$\Lambda^{\U}_{L,R}$ scales relative to 
 a generic FCNC transition $q^i\to q^j$, with  $q$ in the {\it up} fermion sector, and $i,j$
 ($i >j$)  running over  three fermion generations, become  
\bea
\dfrac{1}{(\Lambda^{\U}_L)_{ij}}\!\!&=&\!\!\dfrac{\bar{e}\, m_{{\U}_i}}{\overline{m}^2_{\U}} \left[
\bar{e}_i^{\U}\dfrac{ 
\rho_{R}^{ji}}{\rho_{R}^{ii}}F_{LR}(x^{\U}_i,\xi_{\U}) -  \dfrac{g_R^2}{16\pi^2}\sum_{k=1}^{3}\bar{e}^{\U}_k  \rho_{R}^{jk} \rho_{R} ^{ki} F_{RR}(x_k^{\U},\xi_{\U})
\right]
\nonumber
\\
\dfrac{1}{(\Lambda^{\U}_R)_{ij}}\!\!&=&\!\!\dfrac{\bar{e}\, m_{{\U}_i}}{\overline{m}^2_{\U}} \left[\bar{e}_i^{\U} \dfrac{\rho_{L}^{ji}}{ \rho_{L} ^{ii}}F_{RL}(x^{\U}_i,\xi_{\U}) - \dfrac{g_L^2}{16\pi^2}  \sum_{k=1}^3\Big(  \bar{e}_k^{\U} \rho_{L} ^{jk} \rho_{L}^{ki} F_{LL}(x_k^{\U},\xi_{\U}) \right.
\nonumber\\
&+& \left. \left(\dfrac{\overline{m}^2_{\U}}{\overline{m}^2_{\D}}\right)
\bar{e}_k^{\D}\eta_{L} ^{jk} \eta_{L}^{ki} F_{LL}(x_k^{\D},\xi_{\D})\Big)\right]  ,
\label{LambdaU}
\eea
being $\bar{e}$  the charge unit for dark-photon interactions, $\bar{e}_k^{q}$ their  eigenvalues, $m_{\U_i,\D_i}$ the initial-quark masses, 
$M_{Q^{\U,\D}_i}$ the corresponding dark-quark masses, 
$x_i^{\U,\D}\equiv M_{Q^{\U,\D}_i}^2/\bar{m}_{\U,\D}^2$,  and 
 $\bar{m}_{\U,\D}\; ,\;$ $\xi_{\U,\D}$, respectively, the common average mass 
and mixing parameter in the {\it up},{\it down} messenger sectors. 

For $\Lambda^{\D}_{L,R}$ in the {\it down} quark sector, we obtain  instead
\bea
\dfrac{1}{(\Lambda^{\D}_L)_{ij}}&=&\dfrac{\bar{e}m_{{\D}_i}}{\overline{m}^2_{\D}} \left[ \bar{e}_i^{\D}\dfrac{ 
\eta_{R}^{ji}}{\eta_{R}^{ii}}F_{LR}(x^{\D}_i,\xi_{\D}) - \dfrac{g_R^2}{16\pi^2} \sum_{k=1}^{3}\bar{e}^{\D}_k  \eta_{R}^{jk} \eta_{R} ^{ki} F_{RR}(x_k^{\D},\xi_{\D})
\right]
\nonumber
\\
\dfrac{1}{(\Lambda^{\D}_R)_{ij}}&=&\dfrac{\bar{e}m_{{\D}_i}}{\overline{m}^2_{\D}} \left[\bar{e}_i^{\D} \dfrac{\eta_{L}^{ji}}{ \eta_{L}^{ii}}F_{RL}(x^{\D}_i,\xi_{\D}) -  \dfrac{g_L^2}{16\pi^2} \sum_{k=1}^3\Big(\bar{e}_k^{\D} \eta_{L} ^{jk} \eta_{L}^{ki} F_{LL}(x_k^{\D},\xi_{\D})
\right.
\nonumber\\
&+&\left. \left(\dfrac{\overline{m}^2_{\D}}{\overline{m}^2_{\U}}\right)
\bar{e}_k^{\U}\rho_{L}^{jk} \rho_{L}^{ki} F_{LL}(x_k^{\U},\xi_{\U})\Big) \right]\,  .
\label{LambdaD}
\eea
The first terms in the right-hand side of 
Eqs.~(\ref{LambdaU})--(\ref{LambdaD}) 
for the effective $\Lambda^{\D,\U}_{L,R}$  scales 
are independent from the $g_{L,R}$ couplings, since this dependence  
 has been reabsorbed in the corresponding SM Yukawa couplings, by using  
 Eq.~(\ref{Yukexact}).
\\
Furthermore,
the loop functions appearing in Eqs.~(\ref{LambdaU})--(\ref{LambdaD})  satisfy 
the conditions 
$F_{RR}(x,\xi)=F_{LL}(x,\xi)$, and $F_{RL}(x,\xi)=F_{LR}(x,\xi)$, where
\bea
F_{LL}(x,\xi)
&=&\dfrac{1}{8} \left[ \dfrac{x^2-(\xi-1)^2+2x(\xi-1) \log(\frac{x}{1-\xi})}{(x-1+\xi)^3}+\Big\{\xi \to -\xi\Big\}\right], \\
F_{LR}(x,\xi)&=&\dfrac{f_{2}(x,\xi)}{f_1(x,\xi)}\, ,
\label{FLR}
\eea
defining
\bea
f_{2}(x,\xi)&=&\dfrac{1}{2\,\xi}\left[ 
\dfrac{1-x+\xi+(1+\xi) \log\left( \frac{x}{1+\xi}\right) }{(1-x+\xi)^2}
-\Big\{\xi\to -\xi\Big\}\right] \, ,
\label{f2}
\eea
and for $f_1(x,\xi)$  given by Eq.(\ref{f1}).

Some comments on  
Eqs.(\ref{LambdaU})--(\ref{LambdaD}) are in order. 
Terms proportional to  $F_{LR}$ and $F_{RL}$ arise from the 
chirality flip induced by the virtual-dark-fermion mass insertion.
Terms proportional to $F_{LL}$ and $F_{RR}$  come instead from the chirality flip induced by the initial-quark mass $m_{U_i,D_i}$, after applying on shell
relations on external momenta $\slash\!\!\! p\, u^f_{L,R}(p)=m_f u^f_{R,L}(p)$, for a generic fermion $f$ of mass $m_f$. In the present model, all contributions turn out to be proportional to the initial
quark mass. This is because the dark-quark 
mass insertion has been reabsorbed in the corresponding quark mass, by using  the model prediction  for the one-loop effective Yukawa coupling in Eq.~(\ref{Yuk0}). 
However, for $\bar{m}_{\U}$ and $\bar{m}_{\D}$ of the same order, 
terms proportional to  $F_{LL/RR}$  are subleading 
with respect to the ones proportional to $F_{LR/RL}$, due to the suppression of the loop factors 
${(g_{L,R}^2}/{16\pi^2)}$
in Eqs.~(\ref{LambdaU})-(\ref{LambdaD}).

Finally, we report  some useful analytical expressions for  $F_{LL}(x,\xi)$ and $F_{LR}(x,\xi)$  in the limit of  small and large values of the mixing parameter $\xi$. For $\xi \ll 1$  one gets
\bea
&&\lim_{\xi\to 0} F_{LR}(x,\xi)=
\frac{2(1-x)+(1+x)\ln{x}}{(x-1)(1-x+\ln{x})}\, ,~~~
\lim_{\xi\to 0} F_{LL}(x,\xi)\,=\,
\frac{x^2 -1 -2x\ln{x}}{4(x-1)^3}\, ,\\ \noindent
&&\lim_{x\to 1}\lim_{\xi\to 0}
F_{LR}(x,\xi)=-1/3\, ~~~~~~~~~~~~~~~~~~~~~~
\lim_{x\to 1}\lim_{\xi\to 0}F_{LL}(x,\xi)=1/12\, , 
\eea
while, for large mixing $\xi \sim 1$, we get\footnote{In order to avoid stable messenger particles
 in the spectrum, for a generic quark sector $q$, the corresponding mixing
parameter $\xi$ should be bounded by $0<\xi < 1-x$, where $x=m_Q^2/\bar{m}^2$, and $m_Q$ is the  associated dark-fermion mass (see next section). Then, the logarithmic term $\ln{(1-\xi)}$, appearing in the $F_{LR}$ denominator  in the large $\xi\to 1$ expansion 
[see Eq.~(\ref{FLRlim})], will be  bounded by $\ln{(1-\xi)}<\ln{x}$. Since $x$ is nonvanishing (being dark fermions heavier than 
  the corresponding SM fermions),  
$ F_{LR}(x,\xi)$ and $ F_{LL}(x,\xi)$ will  not develop any singularity in the allowed $x$ and $\xi$ ranges.} 
\bea
&&
\!\!\!\!\!\!\!\!\!\!\!\!\!\!\!\!\!\!\!\!\!\!\!\!
F_{LR}(x,\xi)
\simeq \frac{x(2+\ln{4})-4-2x\ln{x}}{4-6x+x^2(2+\ln{2})
+(x-2)^2\ln{(1-\xi)}-2(2-2x+x^2)\ln{x}} +{\cal O}(1-\xi)\, ,
\label{FLRlim}
\\
&&
\;\;\;\;\;
\lim_{\xi\to 1} F_{LL}(x,\xi)=
\frac{4(x-1)-3x^ 2+x^ 3-2x^ 2\ln{\frac{x}{2}}}
{4(x-2)^3x}\, .
\eea

Since messenger masses are expected to be quite heavy \cite{Gabrielli:2013jka}, the $q^i\to q^j\, \bar{\gamma}$ decay process can actually be described by  an effective Lagrangian approach.  The relevant effective density Lagrangian 
${\cal L}_{\rm eff}$ contains two leading gauge-invariant operators of dimension 5,  that is  the  FC magnetic-dipole operators given by 
\bea
{\cal L}_{\rm eff}=\sum_{q=U,D}\sum_{i,j=1}^3\left(
\frac{1}{2(\Lambda^q_L)_{ij}} \Big[\bar{q}^j_R(x) \sigma_{\mu\nu} \bar{F}^{\mu\nu}(x) q^ {i}_L(x)\Big]
+ 
\frac{1}{2(\Lambda^q_R)_{ij}} \Big[\bar{q}^j_L(x)\sigma_{\mu\nu} \bar{F}^{\mu\nu}(x) q^i_R(x)\Big]\right) \!\!,
\label{Leff}
\eea 
where $i>j$,
 $\bar{F}^{\mu\nu}(x)$ is the dark-photon $U(1)_F$ field-strength tensor, 
$q^i(x)$  and $q^j(x)$  are the initial and final quark fields,  and $\Lambda_{L,R}^{\U}$ and $\Lambda_{L,R}^{\D}$ are given in Eqs.(\ref{LambdaU}) and (\ref{LambdaD}), respectively.

Using the effective Lagrangian in Eq.(\ref{Leff}), 
the total width  for $q^i\to q^j\, \bar{\gamma}$  is (neglecting the final quark mass)
\bea
\Gamma(q^i\to q^j \bar{\gamma}) &=& \frac{m_{q_i}^3}{16 \pi^3}\left(
\frac{1}{(\Lambda_L^q)_{ij}^2}+\frac{1}{(\Lambda_R^q)_{ij}^2}
\right)\, .
\label{width}
\eea
Notice that, due to the chiral suppression of the initial  quark masses $m_{q_i}$
entering in the $\Lambda_{L,R}^ q$ scales [see  Eqs.(\ref{LambdaU})--(\ref{LambdaD})], the  width  turns out to be proportional to the 
fifth power of the decaying quark mass $m_{q_i}$, suppressed by the fourth power of the corresponding average messenger mass $\bar{m}_q$, according to the expression 
\bea
\Gamma(q^i\to q^j \bar{\gamma})
&\sim& \frac{m_{q_i}^5}{16 \pi^3\bar{m}_q^4}\times {\rm (loop\, functions)}\, .
\label{width-top}
\eea
In the following discussion, the relevant independent parameters will be $\bar m_q$ (which controls the order of magnitude of the decay width), the mixing parameter $\xi_q$, (which, at large values~$\sim1$, pushes the smallest $\bar m_q$ eigenvalues   of the messengers running in the loop toward the lowest values [\cf Eq.~(\ref{meigen})], hence enhancing the decay amplitudes), and $x_i^{\U,\D}$
(which sets the dark-fermion mass scale with respect to the messenger one).

Furthermore, it will be convenient to define a universal-flavor  (UF) scenario, where one has {\it up-down } flavor universality in   the mass  sector of the colored messenger fields (\ie, $\bar{m}^2_{\U}=\bar{m}^2_{\D}\equiv \bar{m}^2$). The latter is the most symmetric and predictive framework that one can envisage in the present model.
We also define a  {\it nonuniversal} flavor (NUF) scenario, where one relaxes the 
{\it up} and {\it down}  flavor universality in the messenger sector,
 and assumes a universal $\bar{m}^2_{\D}$ mass 
 in the {\it down} sector which is independent from the universal 
$\bar{m}^2_{\U}$ mass in the {\it up} sector. 
\section{The $t\to (c,u) \, \bar{\gamma}$ decays}
In this section we analyze the FCNC decay of the top quark
\bea
t\to q \, \bar{\gamma}\, ,
\eea
where in the final state there can be either a $c$ or a $u$ quark.
Using Eq.~(\ref{width-top}),
the corresponding BR, in the massless final-quark limit,    
can be parametrized in terms of the tree-level  BR($t\to W b$), as follows
\bea
{\rm BR}(t\to q \, \bar{\gamma})&=&\frac{{\rm BR}(t\to W b)}{\sqrt{2} G_F |V_{tb}|^2\rho(x_W)}\left(
\frac{1}{(\Lambda^{tq}_L)^2}+\frac{1}{(\Lambda^{tq}_R)^2}\right)
\, ,
\eea
where  $\rho(x)=\left(1-x\right)^2\left(1+2x\right)$, 
$\Lambda^{tu}_{L,R}\equiv (\Lambda^{\U}_{L,R})_{31}$, 
$\Lambda^{tc}_{L,R}\equiv (\Lambda^{\U}_{L,R})_{32}$, 
$x_W=\frac{M_W^2}{m_t^2}$, being $M_W$ and $m_t$ the $W^{\pm}$ and top-quark mass, respectively. The  relevant $\Lambda^{tu}_{L,R}$ and $\Lambda^{tc}_{L,R}$ expressions are in  Eq.~(\ref{LambdaU}).

Assuming a universal average messenger mass $\bar{m}_{\U}=\bar{m}_{\D}=\bar m,$
 the mass-scale dependence of ${\rm BR}(t\to q \, \bar{\gamma})$  
turns out to be  
\bea
{\rm BR}(t\to q \, \bar{\gamma})&\propto& \frac{m_t^2}{\bar{m}^4 G_F}\, .
\label{scale-dep}
\eea
 The lower allowed value of the average messenger mass 
$\bar{m}$ is constrained by dark-matter (DM)  and vacuum-stability bounds, and, as a consequence, the $1/\bar{m}^4$ term in Eq.(\ref{scale-dep}) strongly suppresses the $t\to q \, \bar{\gamma}$ decay. 
In particular, we will  prevent  {\it stable} colored and EW messenger particles in the spectrum, 
 which would conflict with DM constraints, hence 
allowing messenger decays into dark fermions according to the interaction Lagrangian
in Eq.~(\ref{LagMS-tilde}). In the following, by DM constraints we indicate the requirements that the mass spectrum is such that all 
the messenger decays are kinematically allowed.

\subsection{DM and vacuum stability constraints for $t\to q \bar{\gamma}$}
We now  discuss the relevant theoretical bounds in the scalar messenger sector,  and, in the following subsection, the corresponding  upper bounds on   BR$(t\to q \, \bar{\gamma})$.
  We will assume, for the moment, {\it up-down } flavor universality (\ie, the UF scenario
  defined above). 
By using  Eqs.(\ref{Yeff}) and (\ref{Yukexact}) for the radiatively generated 
Yukawa couplings, we obtain the following prediction for the generic mass $M_{Q_i}$ of 
the dark fermion associated to the SM quark $q_i$, as a function of the  quark mass $m_i$,
\bea
M_{Q_i}&=&m_i\left(\frac{\,16\pi^2}{g_Lg_R }\right)\frac{1}{\xi f_1(x_i,\xi)}\, ,
\label{MQi}
\eea
 where $x_i=M_{Q_i}^2/\bar{m}^2$ and 
 $\xi$ is the universal mixing parameter in the colored messenger sector. Note that,  the quark masses $m_i$ as well as the running coupling constants $g_L,g_R$, appearing in Eqs.(\ref{MQi}), (\ref{bound3}), (\ref{boundU}), and (\ref{boundD}), are evaluated at the messenger mass scale $\mu\sim \bar{m}$.

Being $m^2_{\pm}=\bar{m}^2(1\pm\xi)$  the eigenvalues of the 
 {\it up-down } degenerate messenger mass spectrum, in order to avoid {\it stable} messengers,
 the lightest messenger mass ${m}_{-}$ must be larger 
than the mass  of the heaviest dark fermion, that is  $M_{Q_t}$,  associated to the top-quark~\cite{Gabrielli:2013jka},
\bea
{m}_{-} \ge M_{Q_t}\, .
\label{DMcondition}
\eea
On the other hand, the vacuum stability condition requires  $\xi\le 1$, in order to avoid either tachyons in the spectrum or color/charge--breaking minima through
the generation of nonvanishing  VEV in the messenger scalar sector ~\cite{Gabrielli:2013jka}.
Because of the $U(1)_F$ gauge invariance in the dark sector,  
Eq.(\ref{DMcondition}) is sufficient to avoid stability for all messenger fields, and  to guarantee that all dark fermions are stable particles. By using Eqs.~(\ref{meigen})
and (\ref{MQi}),  
Eq.(\ref{DMcondition}) can be rephrased into the following lower bound
on the average messenger mass in the colored messenger sector 
\bea
\bar{m}^2\ge \left(\frac{16 \pi^2}{g_Lg_R}\right)^2 
\frac{m_t^2}{\xi^2 f_1^2(x_t,\xi)(1-\xi)}\, ,
\label{bound2}
\eea
where $m_t$ is the  top-quark mass. Notice that also  the rhs of 
Eq.~(\ref{bound2}) depends  on 
$\bar{m}$ through the ratio $x_t=M_{Q_t}^2/\bar{m}^2$ entering  the loop function $f_1(x,\xi)$ defined in Eq.(\ref{f1}). At fixed $\xi$, 
the lowest $\bar{m}$ bound  corresponds to equality in Eq.(\ref{DMcondition})  and can be  obtained by replacing $x_t\to 1-\xi$ inside  $f_1(x_t,\xi)$ in Eq.(\ref{bound2}).
The lowest $\bar{m}$ minimum in Eq.(\ref{bound2}) is then a pure function of $\xi$, namely
\bea
\bar{m}&\ge&m_t  \left(\frac{16 \pi^2}{g_Lg_R}\right) F(\xi)
\,  ,
\label{bound3}
\eea
where  $F(x)$ is given by
\bea
F(x)&=&\frac{8x\sqrt{1-x}}{2x+(1-x)^2
\log{\left(\frac{1-x}{1+x}\right)}}\, .
\eea
For $x\ll 1$,  the formula above simplifies to $F(x)\simeq 2/x+1/3 +{\cal O}(x)$, 
while, for  $x\simeq 1$, one obtains
$F(x)\simeq 4\sqrt{1-x}+{\cal O}((1-x)^{3/2})$.

By relaxing the full flavor  universality in the messenger sector,  and 
 restricting mass degeneracy  to  the {\it up} and {\it down} messenger sectors separately, the above bounds in Eq.(\ref{bound3}) can be generalized as follows:
\bea
\bar{m}_{\U}&\ge&m_t  \left(\frac{16 \pi^2}{g_Lg_R}\right) F(\xi_{\U})
\,  ,
\label{boundU}
\eea
\bea
\bar{m}_{\D}&\ge&m_b  \left(\frac{16 \pi^2}{g_Lg_R}\right) F(\xi_{\D})
\, ,
\label{boundD}
\eea
where $\bar{m}_{\U(\D)}$ and $\xi_{\U(\D)}$ refer to  the {\it up} ({\it down}) sector.
Notice that in the rhs of Eq.(\ref{boundD}) the bottom-quark mass $m_b$ replaces  $m_t$, since
we are now assuming different average messenger masses (\ie, $\bar{m}^2_{\U}\neq \bar{m}^2_{\D}$) for the {\it up}  and {\it down} sectors.
A generalization of the above bounds to the leptonic sector is straightforward.

Accordingly, for $ m_t=173.2$ GeV  and a bottom-quark pole mass $ m_b=4.78$~GeV~\cite{Agashe:2014kda}, we find in the large  $\xi_{\U,\D}$ regime
\bea
\bar{m}_{\U} &\ge&\frac{(110 ~{\rm TeV})~K_{t}(\bar{m})}{g_Lg_R}\sqrt{1-\xi_U}\, ,
\label{mUbound} \\
\bar{m}_{\D} &\ge& \frac{(3 ~{\rm TeV})~K_{b}(\bar{m})}{g_Lg_R}\sqrt{1-\xi_D}\, ,
\label{mDbound}
\eea
where the factors $K_{t,b}(\bar{m})<1$,   defined by 
$m_{t,b}(\bar{m})=K_{t,b}(\bar{m}) \, {m}_{t,b}$, 
contain  the renormalization
effects 
connecting the top and bottom running masses  $m_{t,b}(\bar{m})$ at the scale $\bar{m}$ with  their pole masses ${m}_{t,b}$.

On the other hand, for  $\xi \ll 1$ the lower bounds on
$\bar{m}_{\U,\D}$  in Eqs.~(\ref{boundU}) and (\ref{boundD}) are stronger, due to the enhancement factor $F(\xi)\sim 1/\xi$ at small $\xi$. The singular behavior for $\xi\ll 1$ is a consequence of the vanishing of Yukawa couplings for $\xi\to 0$ at fixed dark fermion masses $M_{Q_i}$ (\cf Eq.~(\ref{Yukexact})). Hence, 
 large $M_{Q_i}$ values  are needed to compensate the latter suppression in Eq.~(\ref{Yukexact}),  
 and even larger  $\bar{m}$ values  due to the DM constraints in Eq.~(\ref{DMcondition}).
Note that, if we assume flavor universality in the 
{\it up} and {\it down} sector for messenger fields, then the strongest bound on $\bar{m}$ in Eq.~(\ref{boundU}) applies.
In the UF scenario, since the $t\to q \, \bar{\gamma}$ width   scales as 
$1/\bar{m}^4$  (\cf Eq.~(\ref{scale-dep})), the corresponding BR($t\to q \, \bar{\gamma}$) will be severely constrained, especially for small $\xi$ mixing.

Since, for sufficiently large mixing, the limits in Eqs.~(\ref{mUbound})-(\ref{mDbound}) might go below the messenger mass  
bounds  arising from their nonobservation in direct pair production at the LHC , we will distinguish
in our study the $\xi$ ranges that correspond to lower mass limits that could be in conflict with the LHC results.
Actually, although the present model shows features that are similar to the SUSY phenomenology, the actual LHC mass bounds depend nontrivially on the model parameters. Dedicated LHC analyzes will be needed in order to set robust bounds on the corresponding particle and parameter spectra.
Then, in our analysis we will just assume a few {\it tentative} mass bounds the could be derived 
for messenger searches at the LHC and set the corresponding maximal $\xi$ mixing value
not to overcome these tentative bounds. In particular, in the following we will assume that the LHC
presently excludes pair production of {\it colored} messengers lighter than 1 TeV
and  of {\it colorless} (EW) messengers lighter than 300 GeV. 

In the numerical analysis of the following sections, since we aim at a  simplified LO analysis, we will not include the QCD running of relevant couplings and masses. Hence, the numerical behavior reported in all tables and figures will correspond to setting all quark masses to their pole mass values.


\subsection{Upper bounds on ${\rm BR}(t\to q \, \bar{\gamma})$}
A rough  
estimate of the upper bounds on   ${\rm BR}(t\to q \, \bar{\gamma})$,
versus the relevant free parameters of the model, needs a few 
 working assumptions. In 
the UF scenario (that is for $\bar{m}^2_{\U} = \bar{m}^2_{\D} \equiv \bar{m}^2$ and  
$\xi_{\U}=\xi_{\D}\equiv \xi$), we can see that in the 
rhs of the two equations entering Eq.~(\ref{LambdaU}) 
for $\Lambda^{\U}_{L,R}$ [or  equivalently in Eq.~(\ref{LambdaD}) for $\Lambda^{\D}_{L,R}$] 
the first terms in parenthesis 
are dominant over the second ones, being the latter suppressed by the loop factor $g_{R/L}^2/16 \pi^2$. Since  $F_{LR/RL}$ and $F_{LL/RR}$
 are almost of the same order,  
we can safely neglect  the contribution of the second terms 
in Eq.(\ref{LambdaU}). In order to further simplify the analysis, one can also 
assume universality between the $L/R$ quark couplings to 
dark fermions (\ie, 
$g_L=g_R$), and  the $\rho_{L,R}$ matrix elements (\ie, $(\rho_L)_{ji}=(\rho_R)_{ji}$). 

Under the UF assumption and neglecting $g_{R/L}^2/16 \pi^2$ terms, disregarding   overall factors from couplings, the generic $q\to q^{\prime} \, \bar{\gamma}$  
width   depends on three fundamental 
parameters, \ie, the average messenger mass $\bar{m}$, the 
mixing parameters $\xi_{\U}$, and  $x_3^{\U}$, 
 satisfying the conditions
in Eqs.~(\ref{bound3}).
Then, since  BR$(t\to q \, \bar{\gamma})\sim 1/\bar{m}^4$, 
the largest allowed BR upper bound corresponds to  the equality condition in
 Eq.~(\ref{mUbound}). Analogous conclusions hold for the FCNC decay
in dark photon in the {\it down}-quark sector.

In Table~\ref{tab2}, we report  the results for the maximum allowed 
${\rm BR}(t\to q\bar{\gamma})$,
 satisfying the vacuum stability bounds and DM constraints, 
versus the mixing parameter $\xi=\xi_{\U}$. 
 The results assume $U(1)_F$ charges and other multiplicative couplings normalized 
 to~1.  In particular, in Table~\ref{tab2}, we set  
  $g_{L,R}=1$, and 
 $\bar{e}\,\bar{e}^{\U}_3 \; =\rho^{33,13,23}_{L,R}=1$, with all other elements of flavor matrices set to zero.
In the last two columns we report  
$\bar{m}^{\rm min}$, the minimum  $\bar{m}$ allowed by DM constraints, and the minimum ${m}_{-}^{\rm min}$ of the corresponding lowest messenger mass eigenvalue,  as defined in Eq.~(\ref{meigen}).
\begin{table} \begin{center}    
\begin{tabular}{|c||c|c|c|}
\hline 
$\xi$ 
& ${\rm BR}^{\rm max}(t\to q\, \bar{\gamma})$
& $\bar{m}^{\rm min}[{\rm TeV}]$
& ${m}_{-}^{\rm min}[{\rm TeV}]$
\\ \hline 
0.1
& $5.6\times 10^{-15}$
& 554
& 526
\\ \hline
0.2
& $1.0\times 10^{-13}$
& 279
& 249
\\ \hline
0.3
& $6.0\times 10^{-13}$
& 185
& 155
\\ \hline
0.5
& $7.5\times 10^{-12}$
& 107
& ~\,75

\\ \hline
0.7
& $7.0\times 10^{-11}$
& ~\,67
& ~\,37

\\ \hline
0.8
& ~$\!2.5\times 10^{-10}$
& ~\,52
& ~\,23
\\ \hline
0.9
& $\!\!1.6\times 10^{-9}$
& ~\,35
& ~\,11
\\ \hline
~\,0.95
& $\!\!8.3\times 10^{-9}$
& ~\,\,25
& ~~~~~\,5.5
\\ \hline
~\,0.99
& $\!\!2.6\times 10^{-7}$
& ~\,11
& ~~~~~\,\,1.1
\\ \hline
\end{tabular} 
\caption[]{ Maximum values of
${\rm BR}(t\to q\, \bar{\gamma})$ in the UF scenario allowed 
by vacuum stability and DM constraints,
corresponding to the minimum allowed average messenger mass  $\bar{m}^{\rm min}$,
and to the lightest {\it up-down} universal messenger mass eigenvalue 
${m}_{-}^{\rm min}=\bar{m}^{\rm min}\sqrt{1-\xi}$ versus the mixing parameter $\xi$. Results are in unit of couplings, that is they assume 
 $\bar{e}\,\bar{e}^{\U}_3 \;=g_{L,R}=\rho^{33,13,23}_{L,R}=1$, with all other elements of flavor matrices set to zero.}
\label{tab2}
\end{center} 
\end{table}

The resulting allowed  
${\rm BR}(t\to q\, \bar{\gamma})$ values  get tiny for small mixing $\xi_{U}$,  but 
might approach detectability at future colliders   in case one  assumes a quite large mixing (which is typical of natural theories~\cite{Biswas:2016jsh}). Indeed, for  
$\xi_{U}=0.95$ one can achieve a BR, in unit of couplings, of the order of $10^{-8}$, which can go up to values  $\sim 10^{-7}$ for 
$\xi_{U}=0.99$. These bounds are effective for 
couplings of the order ${\cal O}(1)$, and in the more realistic case of perturbative smaller couplings
they could be even more severe.
 On the other hand, there are theoretical arguments  suggesting  the values  $\bar{\alpha}\sim (0.05 - 0.2)$, while large $\xi$ mixings and $g_{L,R}\sim{\cal O}(1)$ are favored in order to avoid large corrections to the Higgs-boson mass~\cite{Gabrielli:2013jka, Biswas:2015sha}. 
Therefore, the effect of a more realistic  coupling-constant normalization in the present scenario can induce  
a suppression of order $(10^{-1}\!\!-\!\!10^{-2})$ on the BR  upper bounds in Table \ref{tab2}, modulo possible small values of $\rho^{13,23}_{L,R} $.

We now  relax the 
{\it up} and {\it down}  flavor universality in the messenger sector,
 and assume a universal $\bar{m}^2_{\D}$ mass 
 in the {\it down} sector independent from the universal 
$\bar{m}^2_{\U}$ mass in the {\it up} sector (the NUF scenario defined above).
Then, the DM constraints on $\bar{m}_{\D}$ are less severe according to 
Eq.~(\ref{boundD}), and  
allow  lighter messenger masses in the {\it down} sector, which would in turn permit a 
larger  ${\rm BR}(t\to q\, \bar{\gamma})$.  Indeed, $\bar{m}_{\D}$ enters  the $\Lambda_R$ scale  in Eq.(\ref{LambdaU}) which 
receives contributions from both the {\it down} and {\it up} messenger sectors.

In Table~\ref{tab3} we show the maximum  
${\rm BR}(t\to q\, \bar{\gamma})$ allowed  in the NUF scenario, versus $\xi_{\D}$,  computed  using  $\bar{m}_{\D}$ given by the equality in Eq.(\ref{boundD}).
We have neglected the contributions induced by the $1/\Lambda^{tq}_{L}$ scale  (which are suppressed by terms  $\sim 1/\bar{m}^2_{\U}$), and retained only the $F_{LL}$ contribution in $1/\Lambda^{tq}_R$.
We remind that the $F_{LL}$ term comes from the chirality flip contribution to the FC magnetic-dipole operator induced by the external states, and thus  is suppressed with respect
to  other contributions by a loop factor  $\sim g_L^2/(16\pi^2)$. 
\begin{table} \begin{center}    
\begin{tabular}{|c||c|c|c|}
\hline 
$\xi_{\D}$ 
& ${\rm BR}^{\rm max}(t\to q\, \bar{\gamma})$
& $\bar{m}_{\D}^{\rm min}[{\rm TeV}]$
& ${m}_{{\D}_{-}}^{\rm min}[{\rm TeV}]$
\\ \hline 
0.1
& $1.2\times 10^{-14}$
& 15
& 14
\\ \hline
0.2
& $2.1\times 10^{-13}$
& ~~~~7.7
& ~~~~6.9
\\ \hline
0.3
& $1.3\times 10^{-12}$
& ~~~~5.1
& ~~~~4.3
\\ \hline
0.5
& $1.8\times 10^{-11}$
& ~~~~2.9
& ~~~~2.1
\\ \hline
0.7
& $2.4\times 10^{-10}$
& ~~~~1.9
& ~~~~1.0
\\ \hline
(0.8)
& $\!\!1.3\times 10^{-9}$
& ~~~~1.4
& \,~~~~~0.64
\\ \hline
(0.9)
& $\!\!2.0\times 10^{-8}$
& \,~~~~~0.97
& \,~~~~~0.31
\\ \hline
~\,(0.95)
& $\!\!3.1\times 10^{-7}$
& \,~~~~~0.68
& \,~~~~~0.15
\\ \hline
~\,(0.99)
& $\!\!1.8\times 10^{-4}$
& \,~~~~~0.30
& \,\,\,~~~~0.03
\\ \hline
\end{tabular} 
\caption[]{ Maximum values of
${\rm BR}(t\to q\, \bar{\gamma})$ in the NUF scenario allowed 
by vacuum stability and DM constraints,
corresponding to the minimum allowed average messenger mass  $\bar{m}_{\D}^{\rm min}$,
and to the lightest {\it down}  messenger mass eigenvalue 
${m}_{\D_{-}}^{\rm min}= \bar{m}_{\D}^{\rm min}\sqrt{1-\xi_{\D}}$,   versus the mixing parameter $\xi_{\D}$. Results are in unit of couplings, that is they assume 
 $\bar{e}\,\bar{e}^{\D}_3 \;=g_{L,R}=\eta^{33,13,23}_{L,R}=1$, with all other elements of flavor matrices set to zero.
Values of $\xi_{\D}$ in parenthesis might be excluded by direct searches of colored scalar particles.}
\label{tab3}
\end{center} \end{table}
Despite the 
suppression factor $1/(16\pi^2)$, the upper bounds on the 
${\rm BR}(t\to q\, \bar{\gamma})$ in Table~\ref{tab3} are more relaxed than the 
UF-scenario ones  in  Table~\ref{tab2}, since  $\bar{m}_{\D}$ can be much lower than $\bar{m}_{\U}$  in the NUF scenario.
Values  $\xi_{\D} > 0.7$ (shown in parenthesis  in Table~\ref{tab3}) 
might be excluded by  direct searches
of colored scalar particles at the LHC,  since they correspond to light messenger 
masses in the {\it down} sector below 1 TeV.  Anyway, a dedicated search able to  substantiate the latter statement (which depends on model-dependent features) has not yet  been performed at the LHC.

\begin{figure}
\vspace{-2.5cm}
\begin{center}
\hspace{-5.cm}
\includegraphics[width=0.75\textwidth]{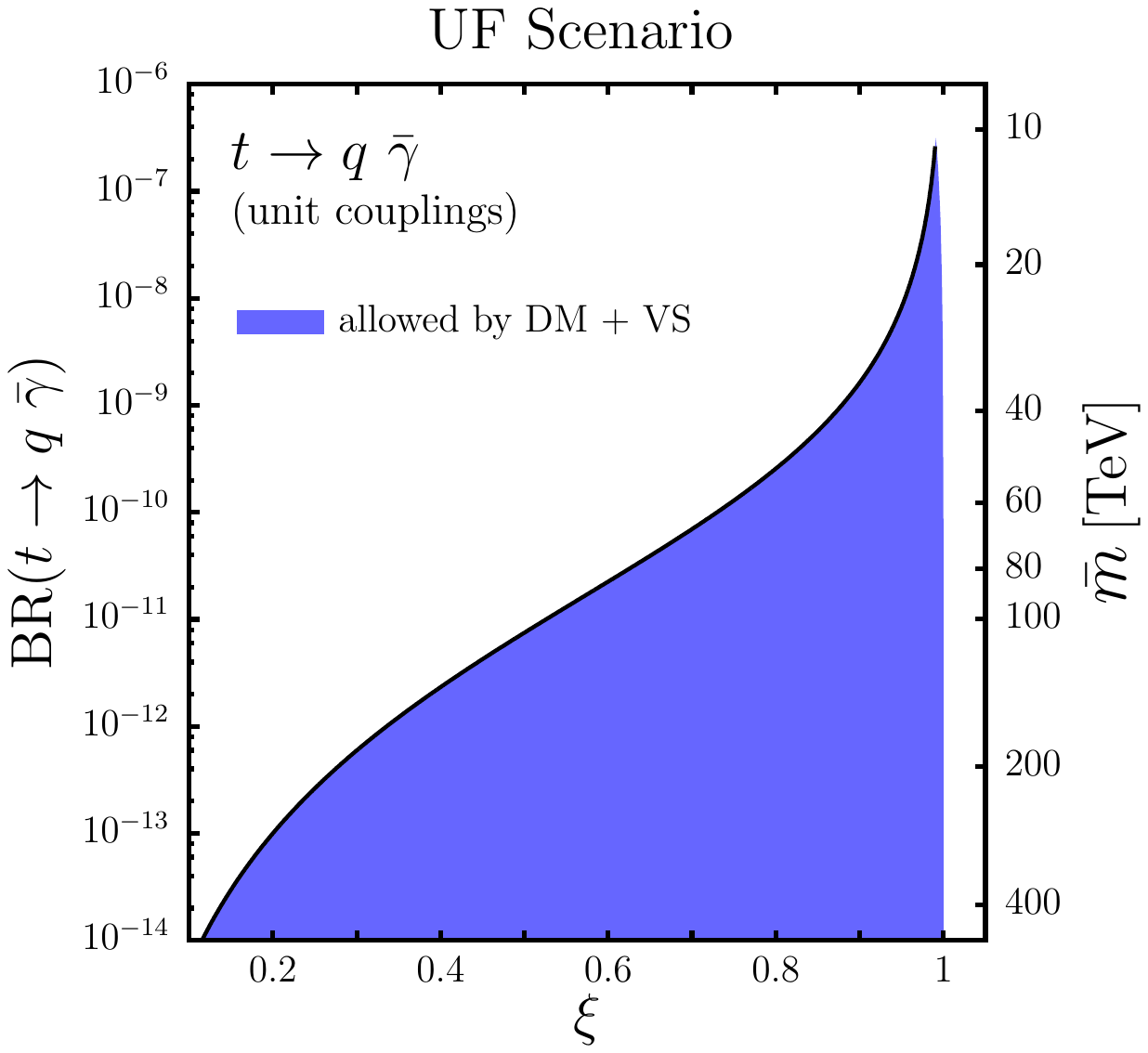}
\hspace{-4.cm}
\includegraphics[width=0.75\textwidth]{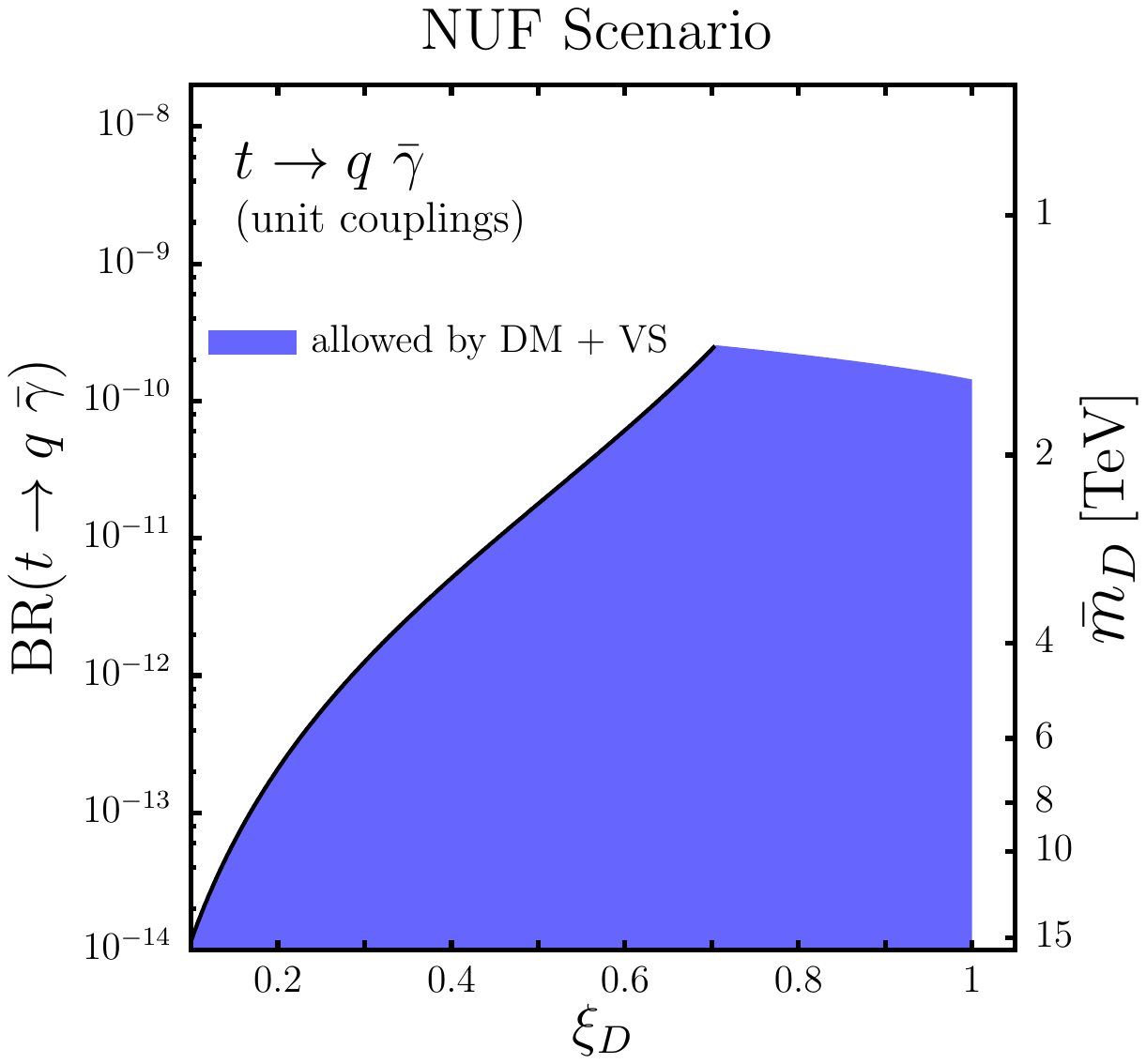}
\hspace{-5cm}
\vspace{-7.cm}
\caption{Allowed regions (colored areas) by DM and vacuum stability (VS)  constraints
for  ${\rm BR}(t\to q\, \bar{\gamma})$ and for the average messenger mass scales 
$\bar{m}$ and $\bar{m}_{\D}$
versus the corresponding mixing $\xi$ and $\xi_{\D}$, in the UF  (left) and NUF  (right) scenarios, respectively.}
\label{fig-BR-top}
\end{center}
\end{figure}

We summarize the above results in Fig.~\ref{fig-BR-top}, where  we show the regions of ${\rm BR}(t\to q\, \bar{\gamma})$ and relevant average messenger mass ($\bar{m}$ and 
$\bar{m}_{\D}$ for the UF and NUF scenarios respectively)  allowed by the DM and vacuum stability constraints versus the mixing parameters $\xi$ and $\xi_D$, in the UF and NUF scenarios, respectively. 
Notice that, at fixed mixing, the black bold upper line in the blue region gives, on the left
vertical axis, the upper bound on ${\rm BR}(t\to q\, \bar{\gamma})$, and, on the right vertical axis, the corresponding lower $\bar{m}_{\D}$ value.
The upper bound for   $\xi_{\D} > 0.7$ in the left plot is ruled  
 by  direct searches
of colored scalar particles at the LHC,  since it corresponds to light messenger 
masses in the {\it down} sector of 1 TeV.

We now discuss the constraints coming from possible dark-fermion and messenger  
contributions to the FCNC decays $t\to q\gamma$,
where the dark photon is replaced by a SM photon in the
final state. In the SM this channel
receives the main contribution from $W$ and $b$-quark loops, whose amplitude, 
due to the Glashow-Iliopoulos-Maiani (GIM) mechanism \cite{Glashow:1970gm},  is suppressed by terms  $\sim V_{ts}m_b^2/M_W^2$ (where $V_{ts}$ is the CKM matrix element), which makes the corresponding 
decay rate quite small. 

The SM values of  BR($t\to c\, \gamma$)   and BR($t\to u\, \gamma$) are  a few $10^{-14}$  and a few $10^{-16}$, respectively~\cite{AguilarSaavedra:2002ns}.
However, in the present framework, $t\to q \gamma$ would receive extra contributions  from  loops of messengers and dark fermions, involving the same flavor structures entering  the $t\to q \bar{\gamma}$ amplitude (see Appendix for details). We will then assume that these further   contributions to the $t\to q \gamma$ amplitude are dominant with respect to the SM one, and apply the present experimental constraints on  BR($t\to q \gamma$) to indirectly constrain the $t\to q \bar{\gamma}$ decay rate.

 Analytical results for the extra $t\to q \gamma$ amplitude 
are reported in the Appendix, by 
  retaining only the dominant contributions proportional to the 
dark-fermion masses. By applying the same  approximation to the $t\to q\, \bar{\gamma}$ amplitude, we get a simplified relation that connects the two BR's by the following expression:
\bea
{\rm BR}(t\to q \,\bar{\gamma})
&=&
\frac{\bar{\alpha}}{\alpha}\left(
\frac{\bar e_3^{\U}\, f_2(x_3^{\U},\xi_{\U})}
{e_{\U}\, {\bar{f}_2(x_3^{\U},\xi_{\U}})}\right)^2
{\rm BR}(t\to q \,\gamma)\, ,
\label{BRratios}
\eea
where $\alpha=1/137$   
is the electromagnetic (EM) fine structure constant,
$e_{\U}=2/3$ is the top-quark EM charge, 
 ${f}_{2}(x,y)$ is given in Eq.~(\ref{f2}), 
and   $\bar{f}_{2}(x,y)$ is derived in the Appendix. Notice that, in 
Eq.~(\ref{BRratios}), the factor connecting the two BR's does not depend on the flavor matrices, since the latter are the same for the dominant contributions to the two processes, and approximately cancel out in the BR ratio.
Then, neglecting the SM contributions, theoretical  BR($t\to q\, \gamma$) upper bounds versus the relevant model parameters  can be 
 obtained from  Tables~\ref{tab2} and \ref{tab3}, by means of  Eq.~(\ref{BRratios}).

Conversely,  the LHC present  constraints on ${\rm BR}(t\to q\, \gamma)$ can set  indirect experimental upper bounds on ${\rm BR}(t\to q\, \bar{\gamma})$ versus  
$x_3^{\U}$ and $\xi_{\U}$, by means of Eq.(\ref{BRratios}). The present  ${\rm BR}(t\to q\, \gamma)$ upper limits at 95\% C.L., reported by the CMS collaboration, are~\cite{Khachatryan:2015att}
\bea
{\rm BR}^{\rm exp}(t\to u\, \gamma) &<& 1.3\times 10^{-4}\\
{\rm BR}^{\rm exp}(t\to c\, \gamma) &<& 1.7\times 10^{-3}\, .
\label{CMSbounds}
\eea
Actually, the stringent DM constraints  in Eqs.~(\ref{boundU})--(\ref{boundD})
 set quite strong upper limits on 
BR($t\to q \bar{\gamma}$), 
and push  
a possible NP contribution to $t\to q \gamma$ 
in this scenario  well below the present experimental sensitivity to this channel. 
On the contrary, if we relax DM constraints, and assume that  NP contributions completely saturate the  $ {\rm BR}^{\rm exp}(t\to q\, \gamma)$ experimental 
limits in Eq.~(\ref{CMSbounds}), we can derive indirect experimental 
 ${\rm BR}(t\to q \bar{\gamma})$ upper bounds versus 
$\bar{\alpha}$, $\xi_{\U}$, and $x_3^{\U}$. For instance, in the UF scenario,
assuming $(\bar e^{\U}_3)^2\bar{\alpha}\sim 0.1$ as a reference value  for the relevant combination of $U(1)_F$ couplings, as indicated by  {\it naturalness} arguments (see Appendix in \cite{Biswas:2015sha}), we get the following 
upper bounds on ${\rm BR}(t\to q\, \bar{\gamma})$, for representative $\xi_{\U}$ and $x_3^{\U}$ values:
\begin{itemize} 
\item  for $\xi_{\U}=0.1$, and $x_3^{\U}=0.8$ (small-mixing regime)
\bea
{\rm BR}^{(t\to u \gamma)}(t\to u\, \bar{\gamma}) &<& 1.8\times 10^{-2}
\left(\frac{\bar{\alpha}}{0.1}\right)
\, \\
{\rm BR}^{(t\to c \gamma)}(t\to c\, \bar{\gamma}) &<& 2.3\times 10^{-1}
\left(\frac{\bar{\alpha}}{0.1}\right)
\,
\eea
\item  for $\xi_{\U}=0.8$, and $x_3^{\U}=0.1$ (large-mixing regime)
\bea
{\rm BR}^{(t\to u \gamma)}(t\to u\, \bar{\gamma}) &<& 3.4\times 10^{-2}
\left(\frac{\bar{\alpha}}{0.1}\right)
\, \\
{\rm BR}^{(t\to c \gamma)}(t\to c\, \bar{\gamma}) &<& 4.4\times 10^{-1}
\left(\frac{\bar{\alpha}}{0.1}\right)
\, .
\eea
\end{itemize}
The resulting upper bounds are much weaker than the  ones in Tables~\ref{tab2} and \ref{tab3} set by DM 
constraints\footnote{ 
Notice that the ${\rm BR}^{(t\to q \gamma)}(t\to q \bar{\gamma})$ upper bounds derived from the present 
experimental ${\rm BR}(t\to q{\gamma})$ constraints increases by decreasing the $x_3^{\U}$, thanks to the $\log{x_3^{\U}}$ enhancement of 
 ${\rm BR}(t\to q \bar{\gamma})$ with respect to  ${\rm BR}(t\to q \gamma)$. In the $t\to q \bar{\gamma}$ amplitude, the $\log{x_3^{\U}}$
term in the loop function $f_2(x_3^{\U},\xi_{\U})$ defined in 
Eq.~(\ref{f2})
 is due to an infrared effect in 
 the diagrams where a dark photon is radiated from internal dark-fermion lines. Indeed, at small $x$, the $f_2(x,\xi)$ function behaves as 
$f_2(x,\xi)\simeq \frac{2\xi}{1-\xi}(1+\log{x-\log(1-\xi))}+{\cal O}(x)$ .
The $\log x\,$ term is absent in the corresponding $t\to q \gamma$ loop function $\bar  f_2(x,\xi)$ Eq.~(\ref{effeduebarra}), since  dark fermions are not charged
under EM interactions.}.
 Note that such large values of the upper bounds overwhelm the possibility of having 
 extra top decay channels allowed by the present measurement of BR($t\to Wb$)~\cite{Agashe:2014kda}.

In conclusion, by imposing vacuum stability and DM constraints, 
we expect that   allowed ${\rm BR}(t\to q\, \bar{\gamma})$  values 
 do not exceed  $ \sim (10^{-8}\!\!-\!\!10^{-7})$, which are barely close to the HL-LHC maximum experimental sensitivity on rare top-quark processes, but might be well inside the exploration domain of a future hadron collider at 100 TeV~\cite{Arkani-Hamed:2015vfh}. However, larger ${\rm BR}(t\to q\, \bar{\gamma})$ values,  up to  $(10^{-5}\!\!-\!\!10^{-4})$, could  in principle be achieved, provided the LHC constraints on colored scalar particle production can be avoided in 
 case of  messengers that are lighter than 1 TeV (\cf Table~\ref{tab3}).
On the other hand, in case one can evade both DM constraints and LHC direct
bounds on colored scalar production, the expected  ${\rm BR}(t\to q\, \bar{\gamma})$
range is essentially just limited by the present accuracy on the measurement of BR($t\to Wb$).

\section{The $b\to (s,d) \, \bar{\gamma}$ decays}
Here we analyze the FCNC decay $b\to q \, \bar{\gamma}$, with $q=s,d$.
Its total width is given by Eq.(\ref{width}), with $i=3$ and $j=2,1$ for the $q=s,d $ transitions, respectively. The corresponding BR can conventionally be expressed in terms of  ${\rm BR}^{\rm exp}(B\to X_c \bar{\nu} e)=(10.65\pm 0.16)\%$, the 
world-average measurement of the $B$-meson semileptonic BR~\cite{Agashe:2014kda}-\cite{average-SL}. To this aim, the tree-level semileptonic $b\to c\, e \bar{\nu}$ decay  width, $\Gamma^b_0$, can be expressed through 
\bea
\Gamma^b_0&=& \frac{G_F^2 m_b^5 |V_{cb}|^2}{192 \pi^3} f_1(z_{cb})\, ,
\label{Gammab0}
\eea
where $f_1(x)=1-8 x +8 x^3-x^4-12 x^2 \log{x}$, with $z_{cb}=m_c^2/m_b^2$, and
$V_{cb}$ is the relevant CKM matrix element. 
Then, one has
\bea
{\rm BR}(b\to q \bar{\gamma})&=& \frac{12\, {\rm BR}^{\rm exp}(B\to X_c \bar{\nu} e)} {G_F^2 |V_{cb}|^2 m_b^2 f_1(z_{cb})}\left(\frac{1}{(\Lambda_L^{bq})^2}
                 +\frac{1}{(\Lambda_R^{bq})^2}
\right)\, ,
\label{BRbsDP}
\eea
with $q=s,d$.  The expressions needed for   
$\Lambda^{bs}_{L,R}\equiv (\Lambda^{\D}_{L,R})_{32}$,  and $\Lambda^{bd}_{L,R}\equiv (\Lambda^{\D}_{L,R})_{31}$  can be found in Eq.(\ref{LambdaD}).
Note that the ${\rm BR}(b\to q \bar{\gamma})$ dominant $m_b$ dependence cancels out in 
$1/(\Lambda^{bq}_{L,R})^2$, since the Yukawa couplings are generated radiatively. 
For our numerical analysis, we use  
the central values of the $c$-quark and $b$-quark 
pole masses, 
$m_c=1.67$ GeV and $m_b=4.78$ GeV,  respectively, and the $V_{cb}$ central value $V_{cb}=(42.46 \pm 0.88)\times 10^{-3}$, extracted from the $B$ semileptonic BR reported above~\cite{Agashe:2014kda}--\cite{average-SL}.
\subsection{DM and vacuum stability constraints for $b\to q \bar{\gamma}$}
Following the same approach adopted for the top-quark decays described in the previous section, we now 
 present  the theoretical ${\rm BR}(b\to q \, \bar{\gamma})$ upper bounds.  
 We neglect the second term in the square brackets in Eq.~(\ref{LambdaD}), which is 
 of order $\sim g_{L,R}^2/(16 \pi^2)$.  
Contrary to the top-quark case, we can neglect the latter term  in the NUF
scenario as well, since  no enhancement is expected in the corresponding 
contributions
in the 
$b$-quark case, not even in the one proportional to the $\rho_L$ matrix elements in Eq.~(\ref{LambdaD})  (the latter being suppressed by  $1/\bar{m}^2_{\U}$, which is typically smaller than 
$1/\bar{m}^2_{\D}$).

In order to simplify the analysis, 
we set all couplings (including the flavor matrix elements $\rho_{L,R}$) to 1, and consider only the dependence on the average messenger mass and 
corresponding mixing parameter in the {\it down} messenger sector. 
We consider first the UF scenario in which $\bar{m}^2_{\D}=\bar{m}^2_{\U}$, and  $\xi_{\D}=\xi_{\U}$. We also assume  
symmetric left-right couplings,  $g_L=g_R$, and left-right flavor matrices, $\rho_L=\rho_R$.  Then, ${\rm BR}(b\to q\, \bar{\gamma})$ gets its maximum for the minimum allowed $\bar{m}_{\D}$. According to Eq.~(\ref{mUbound}), for large
$\xi_{\U}$, this corresponds to 
$\bar{m}^{\rm min}_{\D}=110 \,{\rm TeV}\sqrt{1-\xi_{\D}}$ for $g_{L,R}=1$, and $K_{b}(\bar{m})=1$.

Notice that the relevant dark-fermion mass entering the FCNC $b$ decays is the heaviest dark fermion associated to the {\it down} sector, which appears
through the $x^{\D}_3$ dependence of the loop functions.
In the UF hypothesis, we can relate the  $x^{\U}_3$ and  $x^{\D}_3$ variables by assuming that the dark fermion masses are approximately a rescaled version of the  SM fermion masses. This is a realistic 
approximation since the $C_0(x)$ loop function in Eq.(\ref{C0}) has a weak $x$ dependence  in the range $0<x< 1$. Then,
 for  assessing  BR$(b\to q \bar{\gamma})$ upper bounds in the UF scenario, we will assume the following approximated 
relation: 
\bea
x^{\D}_3 \simeq x^{\U}_3 \frac{m_b^2}{m_t^2}\, .
\label{x3D}
\eea

In Table~\ref{tab4},  the  ${\rm BR}(b\to q \,\bar{\gamma})$ upper bounds   induced by vacuum stability and DM constraints  
in the UF  scenario are presented, as a function of the mixing parameter
$\xi_{\D}$. These results hold for unit couplings. For arbitrary couplings, the results in Table~\ref{tab4} must be multiplied by the product 
$(\bar{e}\,\bar{e}^{\D}_3\eta_L^{j3}/\eta_L^{33})^2$, or analogously $(\bar{e}\,\bar{e}^{\D}_3\eta_R^{j3}/\eta_R^{33})^2$, with $j=1,2$.

\begin{table} \begin{center}    
\begin{tabular}{|c||c|c|c|}
\hline 
$\xi$ 
& ${\rm BR}^{\rm max}(b\to q\, \bar{\gamma})$
& $\bar{m}^{\rm min}[{\rm TeV}]$
& ${m}_{-}^{\rm min}[{\rm TeV}]$
\\ \hline 
0.1
& $7.5\times 10^{-9}$
& 554
& 526
\\ \hline
0.2
& $1.2\times 10^{-7}$
& 279
& 249
\\ \hline
0.3
& $6.5\times 10^{-7}$
& 185
& 155
\\ \hline
0.5
& $6.2\times 10^{-6}$
& 107
& ~\,75
\\ \hline
0.8
& $1.2\times 10^{-4}$
& ~\,52
& ~\,23
\\ \hline
0.9
& $6.0\times 10^{-4}$
& ~\,35
& ~\,11
\\ \hline
~\,0.95
& $2.6\times 10^{-3}$
& ~\,25
& ~~~~~\,5.5
\\ \hline
~\,0.99
& $6.7\times 10^{-2}$
& ~\,11
& ~~~~~\,1.1
\\ \hline
\end{tabular} 
\caption[]{Maximum allowed  ${\rm BR}(b\to q\, \bar{\gamma})$
 after applying vacuum stability and DM constraints, corresponding to the minimum allowed average mass $\bar{m}^{\rm min}$,  
and to the lightest universal messenger mass eigenvalue 
${m}_{-}^{\rm min}=\bar{m}^{\rm min}\sqrt{1-\xi}$, versus  the mixing parameter $\xi$, in the UF scenario.
Results are in unit of couplings, that is they assume 
 $\bar{e}\,\bar{e}^{\U}_3\;=g_{L,R}=\eta^{33,13,23}_{L,R}=1$, with all other flavor matrix elements set to zero.}
\label{tab4}
\end{center} 
\end{table}
On the other hand, in the NUF scenario, lower $\bar{m}_{\D}$ values  
are allowed by vacuum stability and  DM constraints, and quite larger  
${\rm BR}(b\to q\, \bar{\gamma})$ values can be reached. The  NUF scenario results 
 versus  $\xi_{\D}$ are presented in Table~\ref{tab5}.
\begin{table} \begin{center}    
\begin{tabular}{|c||c|c|c|}
\hline 
$\xi_{\D}$ 
& ${\rm BR}^{\rm max}(b\to q\, \bar{\gamma})$
& $\bar{m}_{\D}^{\rm min}[{\rm TeV}]$
&${m}_{{\D}_{-}}^{\rm min}[{\rm TeV}]$
\\ \hline 
0.1
& $5.8\times 10^{-5}$
& 15
& 14
\\ \hline
0.2
& $1.1\times 10^{-3}$
& ~~~~7.7
& ~~~~6.9
\\ \hline
0.3
& $6.3\times 10^{-3}$
& ~~~~5.1
& ~~~~4.3
\\ \hline
0.5
& $7.9\times 10^{-2}$
& ~~~~2.9
& ~~~~2.1
\\ \hline
(0.6)
& $2.4\times 10^{-1}$
& ~~~~2.3
& ~~~~1.5
\\ \hline
\end{tabular} 
\caption[]{Results as in Table~\ref{tab4} but for
the NUF scenario, where we assume $\bar{e}\,\bar{e}^{\D}_3 \;=g_{L,R}=
\eta^{33,13,23}_{L,R}=1$, with all other flavor matrix elements set to zero.
The range $\xi_{\D} \ge 0.6$  might be excluded by the condition 
BR$(b\to s ~{\rm X_{\rm inv}})< {\cal O}(10\%)$, where  ${\rm X}_{\rm inv}$ stands for inclusive invisible particles (see text). }
\label{tab5}
\end{center} 
\end{table}
One can see that particularly large ${\rm BR}(b\to q\, \bar{\gamma})$ values  are allowed  in  case of large $\xi_{\D}$ mixing, 
that are possibly well inside the discovery range of  future B factories and FCC-ee.
On the other hand, an experimental bound BR$(b\to s ~{\rm X_{\rm inv}})< {\cal O}(10\%)$~\cite{Ciuchini:1996vw} (where  ${\rm X}_{\rm inv}$ stands for the inclusive invisible channel\footnote{In present experimental analysis,  kinematical distributions are according to SM, where  $X_{\rm inv}$ is given by  $\nu\bar{\nu}$ pairs. For the possibility to constrain  nonstandard final states with $X_{\rm inv}$, see \cite{Kamenik:2011vy}.}) might exclude the range  $\xi_{\D} \gsim 0.6$, when  all  relevant couplings are set to 1.

\subsection{BR($b\to s \,\bar \gamma$) upper bounds   from the BR($b\to s\, \gamma$) measurement}
We now consider the experimental constraints coming from the measurement of the
 $b\to s\, \gamma$ decay rate into a photon, and in particular the bounds on 
NP contributions to BR($b\to s\, \gamma$). The $b\to s\, \gamma$ process is known with high precision in the 
SM, with a next-to-next-to-leading-order (NNLO) accuracy  in QCD (see \eg \, \cite{Buras:2011we} for a complete review on the subject). The most updated SM theoretical prediction provides the value~\cite{Misiak:2015xwa}
\bea
{\rm BR}(B\to X_S \, \gamma)&=&(3.36\pm 0.23)\times 10^{-4}\, .
\label{bsgSM}
\eea
The effective low-energy Hamiltonian for the ${\Delta B=1}$ transitions, describing the $b\to s\, \gamma$ decay, is given by
\bea
H^{\Delta B=1}_{eff}=-\frac{4 G_F}{\sqrt{2}} V^{\star}_{32}
V^{\star}_{33} \sum_{i=1}^8 C_i(\mu_b) Q_i(\mu_b)\, ,
\eea
where the complete basis of operators $Q_i$ in the SM can be found 
\eg \, in \cite{Buchalla:1995vs}. The Wilson coefficients 
$C_i(\mu_b)$ are evaluated at the low-energy scale 
$\mu_b\sim {\cal O}(m_b)$ and  have been computed at the 
NNLO in QCD~\cite{Buras:2011we}. 
The $Q_7$ and $Q_8$ operators (conventionally, the magnetic-dipole and
chromagnetic-dipole operators, respectively) are the main 
 operators receiving  contributions from NP, as occurs in our scenario, and are defined as
\bea
Q_7&=&\frac{e}{16\pi^2}m_b(\bar{s}_L \sigma^{\mu\nu} b_R) F_{\mu\nu}
\nonumber
\\
Q_8&=&\frac{g_S}{16\pi^2}m_b(\bar{s}_L \sigma^{\mu\nu}T^a b_R) G^a_{\mu\nu}\, ,
\label{Q7Q8}
\eea 
where 
$F_{\mu\nu}$, $G^a_{\mu\nu}$ are the EM and QCD field strengths, respectively, with 
$a=1,...,8$ running on the adjoint representation of the QCD $SU(3)_c$ 
group.

The present NP scenario will give a contribution at  one loop to 
the Wilson coefficients of the $Q_7$ and $Q_8$ 
operators at the $M_W$ scale, namely to 
$C_7(M_W)$ and $C_8(M_W)$, respectively. 
The   corresponding $b\to s \,\gamma$ and $b\to s \,g$ decay amplitudes   induced by these operators (with $g$ standing for a gluon) can be found in the Appendix. However, the present model induces also contributions to two new local operators $\tilde{Q}_7$ and $\tilde{Q}_8$, which are defined by assuming an opposite chirality structure in Eq.~(\ref{Q7Q8})~\cite{Gabrielli:2000hz}. We will refer to $\tilde{C}_7(M_W)$ and $\tilde{C}_8(M_W)$ as the  Wilson coefficients  corresponding  to $\tilde{Q}_7$ and $\tilde{Q}_8$ at the $M_W$ scale.

NP effects in $b\to s \, \gamma$ can be parametrized in a model-independent way by  introducing the  $R_{7,8}$ and 
$\tilde{R}_{7,8}$ parameters defined
at the EW scale as
\bea
R_{7,8}\equiv \frac{C^{\rm NP}_{7,8}(M_W)}{C^{\rm SM}_{7,8}(M_W)}
,~~~
\tilde{R}_{7,8}\equiv \frac{\tilde{C}^{NP}_{7,8}(M_W)}{C^{\rm SM}_{7,8}(M_W)} , 
\label{R78}
\eea
where $C_{7,8}^{\rm NP}$ include the pure NP contribution. The Wilson coefficients above are meant to be 
evaluated at the leading order (LO). We are now considering their effect on 
${\rm BR}(B\to X_s \gamma)$ evaluated at the next-to-leading order (NLO)~\cite{Chetyrkin:1996vx}, where 
nonperturbative $1/m_b$ \cite{Voloshin:1996gw} and $1/m_c$ \cite{Falk:1993dh}
corrections have been included. Although the $b\to s \, \gamma$ rate is  known at the NNLO~\cite{Misiak:2015xwa}, the  LO accuracy for NP effects is sufficient for the purposes of the present analysis. Indeed, we restricted to  a 1-loop matching, while a true NLO accuracy in the NP effects would require  a (nontrivial to perform) 2-loop matching.

By inserting the $R_{7,8}$ and $\tilde{R}_{7,8}$ definition  
in the final ${\rm BR}(B\to X_s \gamma)$ expression, as 
 in \cite{Chetyrkin:1996vx}, one obtains \cite{Gabrielli:2000hz}
\bea
{\rm BR}(B\to X_S \, \gamma)&=& (3.36 \pm 0.26)\times 10^{-4}\left(
1+0.622 R_7+0.090(R_7^2+\tilde{R}_7^2)
\right.\nonumber\\
&+&\left.
0.066 R_8+0.019(R_7R_8+\tilde{R}_7\tilde{R}_8)+0.002(R_8^2+\tilde{R}_8^2)\right)
\, ,
\label{BRparam}
\eea
where, with respect to \cite{Gabrielli:2000hz}, we rescaled the SM central value
by the most updated result at the NNLO accuracy \cite{Misiak:2015xwa}, and kept 
the (1-$\sigma$) SM uncertainty.

The experimental measurements of the CP- and 
isospin-averaged ${\rm BR}({B}\to X_s \,\gamma)$ by CLEO \cite{Chen:2001fja}, 
Belle \cite{Abe:2001hk}, and BABAR \cite{Aubert:2007my}
lead to the combined value \cite{Amhis:2014hma}
\bea
{\rm BR}^{\rm exp}({B}\to X_S \,\gamma)&=&(3.43\pm0.21\pm 0.07)\, \times \, 10^{-4}
\label{BRexp}
\eea

\begin{figure}
\vspace{-2.5cm}
\begin{center}
\includegraphics[width=0.7\textwidth]{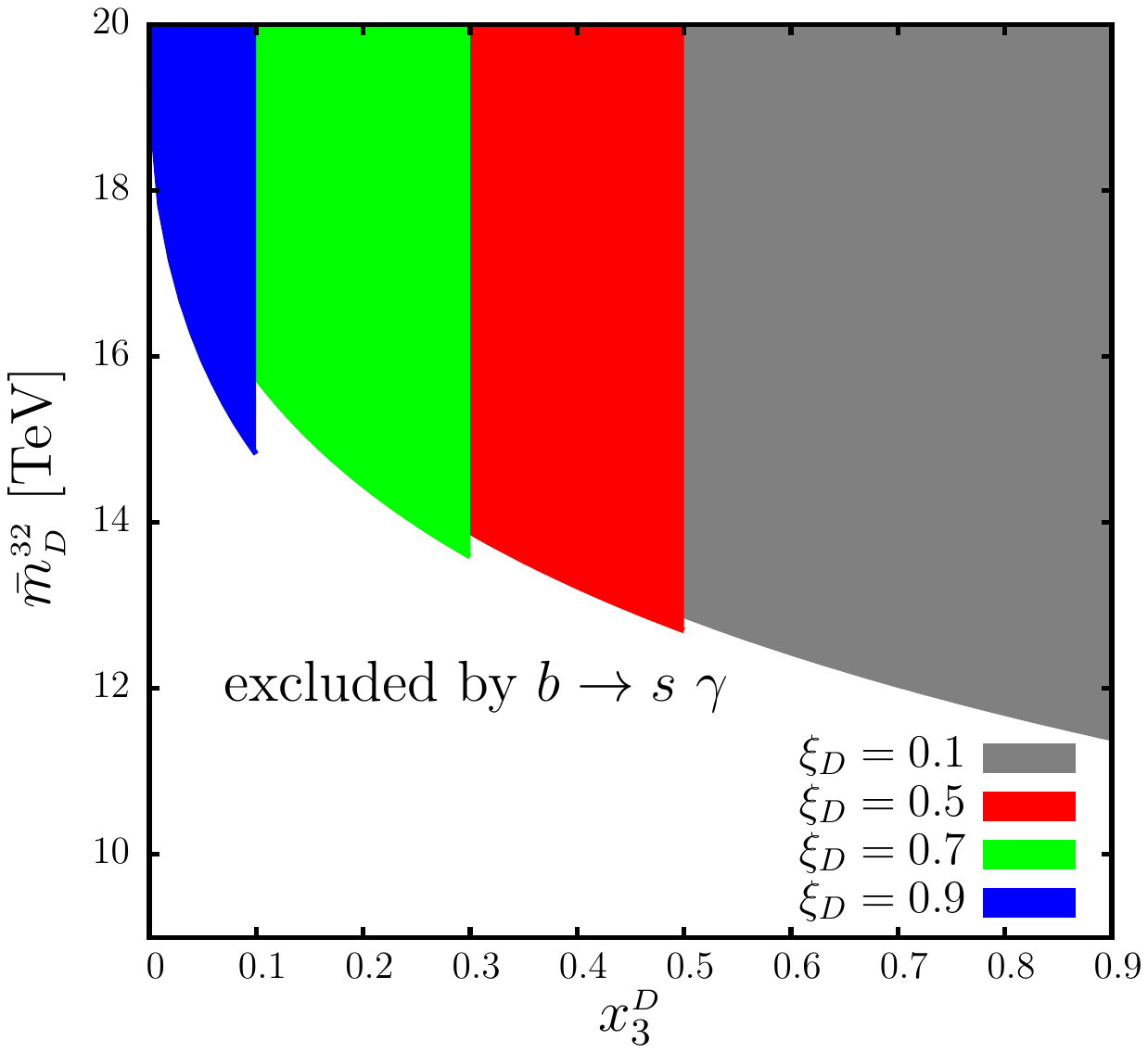}
\vspace{-7.5cm}
\caption{Regions allowed by $b\to s \,\gamma$ data 
at 95\% C.L. (represented by superimposed colored areas), 
for the effective messenger mass scale 
$\bar{m}^{32}_{\D}$ defined in Eq.~(\ref{MD}), as a function of  $x_3^{\D}$ and for different  values of the 
mixing  parameter $\xi_{\D}$. Regions $x_3^{\D}>1-\xi_{\D}$ are excluded by DM constraints.
}
\label{fig2}
\end{center}
\end{figure}

In order to constrain the  contributions induced by the present NP scenario, 
we will make a few  simplifying assumptions. As  can be seen from the coefficients multiplying the $R_i$ and $R_iR_j$ terms in the right-hand side of 
Eq.~(\ref{BRparam}), the linear term in $R_7$ has the dominant weight. 
Since in the present scenario  $R_{7,8}$ and $\tilde{R}_{7,8}$ 
are expected to be all of the same order,  
 to simplify the analysis we neglect all the  $R_i$ terms but the linear term in $R_7$
in the rhs of Eq.(\ref{BRparam}),  which will be a fair approximation for the purposes of the present analysis.
 Then, by requiring that the theoretical 
central value  lies inside the experimental $2$-$\sigma$ band of Eq.(\ref{BRexp}) (with a standard deviation 
$\sigma=0.22 \times  10^{-4}$  obtained by summing in quadrature the statistical 
and systematic errors), one obtains the following
upper bounds
\bea
|R_7| &\lsim & 0.139 ~~~~~{\rm for }~~~ {\rm sign}(R_7) =+1 \\ \nonumber
|R_7| &\lsim & 0.071 ~~~~~{\rm for }~~~ {\rm sign}(R_7) =-1 \, .
\label{R7bounds}
\eea
Since the  $R_7$ sign is not 
predicted in the present framework, we will impose the most conservative upper bounds on 
$|R_7|<0.139$, corresponding to the positive $R_7$ sign.
According to  the results given in the Appendix,  the $R_7$ absolute value is given by 
\bea
|R_7|=\frac{ 2\pi^2\sqrt{2}}{3G_F V_{32}^{\star}V_{33} \bar{m}^2_{\D} |C_7^{SM}(M_W)|}
\left|\frac{\eta_L^{23}}{\eta_L^{33}}\right| \bar{F}_{LR}(x_3^{\D},\xi_{\D})
\eea
where the expression for the function $\bar{F}_{LR}(x,\xi)$ can be found in Eq.~(\ref{fbarLR}) in the Appendix, and $C_7^{\rm SM}(M_W)=-0.193$  for  $m_t=170$~GeV~\cite{Buchalla:1995vs}. Then, the constraint $|R_7|<0.139$  sets
a lower bound on the  {\it effective} messenger mass 
scale $\bar{m}^{32}_{\D}$, defined as 
\bea
\bar{m}^{32}_{\D}\equiv \bar{m}_{\D}\sqrt{\left|\frac{\eta_L^{33}}{\eta_L^{23}}\right|}\, ,
\label{MD}
\eea
versus $x_3^{\D}$ and $\xi_{\D}$.
In Fig.~\ref{fig2}, we plot the   $\bar{m}^{32}_{\D}$ regions excluded at 95\% C.L.
by  $b\to s \gamma$ data, as a function of  $x_3^{\D}$, and for several values of the $\xi_{\D}$ mixing.

If we now combine the DM constraints on
$\bar{m}_{\D}=\bar{m}_{\U}$ in Eq.(\ref{bound3}),
with the ones from $b\to s\, \gamma$ in Fig.~\ref{fig2}, we can see that the latter do not allow to set any stringent upper limit 
on the flavor matrix elements $\eta_L^{23},\eta_L^{33}$, since
in this case $\bar{m}_{\U}$ would be always inside the allowed regions of 
$\bar{m}^{32}_{\D}$ in Fig.~\ref{fig2} [see Eq.~(\ref{bound3})].

On the contrary,  in the  NUF scenario, lower messenger masses
are allowed (\cf  Eq.(\ref{mDbound})), 
and strong upper bounds
on the combination $\frac{|\eta_L^{23}|}{|\eta_L^{33}|}$ arise from the 
$b\to s \,\gamma$ constraints. 
For example, combining DM and 
 $b\to s \,\gamma$ constraints we get, for  $\xi_{\D}=0.5$ , 
\bea
\Big|\frac{\eta_{\LL}^{23}}{\eta_{\LL}^{33}}\Big| &<& 7\times 10^{-2}
\left(\frac{\bar{m}_{\D}}{3\,{\rm TeV}}\right)^2 ,
~~~~~~~~{\rm if}
~~~\bar{m}_{\D}\ge 3~{\rm TeV} .
\eea

We now analyze the ${\rm BR}(b\to s\, \bar{\gamma})$ upper bounds given by the $b\to s \, \gamma$
data in Fig.~\ref{fig2}.
For simplicity,  we will assume a left-right symmetry, namely $\Lambda^{bs}_{L}=\Lambda^{bs}_{R}$.
Then, the   $1/(\bar{m}^{32}_{\D})^2$ scale, defined by Eq.~(\ref{MD}),
factorizes in   both the $b\to s \gamma$ and $b\to s \bar{\gamma}$ amplitudes. For the NP contribution  saturating the 
 $R_7<R_7^{\rm max}=0.139$ bound arising from the $b\to s\, \gamma$ measurement, we get then 
\bea
\frac{1}{\Lambda^{bs}_{\LL}} &<& 
 \left( \frac{3 m_b\, \bar{e}\bar{e}_3^{\D}\,G_F V_{32}^{\star} V_{33}}{2\pi^2\sqrt{2}}\right)
\bar{F}_{LR}(x_3^{\D},\xi_{\D})\;
R_7^{\rm max} |C_7^{\rm SM}(M_W)| \, ,
\label{Lambdabound}
\eea
which can be translated into an upper bound on  ${\rm BR}(b\to s\, \bar{\gamma})$. 
In particular, 
we obtain, for representative $\xi_{\D}$ and $x_3^{\D}$ values\footnote{Actually, these bounds are independent from the matrix elements $\eta_{L,R}^{23}$ and $\eta_{L,R}^{33}$ only if we require the left-right universality assumption $\eta_{L}^{ji}=\eta_{R}^{ji}$ or by considering the contribution of each of them at a time, since these can factorize in 
both $b\to s \gamma$ and $b\to s \bar{\gamma}$ amplitudes.}
\begin{itemize} 
\item  for $\xi_{\D}=0.1$ and $x_3^{\U}=0.8$ (small-mixing regime)
\bea
{\rm BR}^{(b\to s \gamma)}(b\to s\, \bar{\gamma}) &<& 6.9\times 10^{-3}
\left(\frac{\bar{\alpha}}{0.1}\right)
\, ,
\label{boundb1}
\eea
\item  for $\xi_{\D}=0.8$ and $x_3^{\U}=0.1$ (large-mixing regime)
\bea
{\rm BR}^{(b\to s \gamma)}(b\to s\, \bar{\gamma}) &<& 1.0\times 10^{-2}
\left(\frac{\bar{\alpha}}{0.1}\right)
\, ,
\label{boundb2}
\eea
\end{itemize}
where we have set $\bar{e}^{\D}_3=1$, and used the approximated relation for $x_3^{\D}$ in Eq.(\ref{x3D}). Typical values $\bar{\alpha}\simeq 0.1$ are natural in the present framework \cite{Biswas:2015sha}. In the  NUF scenario, where  $x_3^{\U}$ and $x_3^{\D}$ are independent
variables, we get 
\begin{itemize} 
\item  for $\xi_{\D}=0.1$ and $x_3^{\D}=0.8$ (small-mixing regime) 
\bea
{\rm BR}^{(b\to s \gamma)}(b\to s\, \bar{\gamma}) &<& 2.5\times 10^{-4}
\left(\frac{\bar{\alpha}}{0.1}\right)
\, ,
\label{boundb1NUF}
\eea
\item  for $\xi_{\D}=0.8$ and $x_3^{\D}=0.1$ (large-mixing regime)
\bea
{\rm BR}^{(b\to s \gamma)}(b\to s\, \bar{\gamma}) &<& 4.8\times 10^{-4}
\left(\frac{\bar{\alpha}}{0.1}\right)
\, .
\label{boundb2NUF}
\eea
\end{itemize}
Notice that these upper bounds 
are independent from the effective messenger scale $\bar{m}^{32}_{\D}$, since the latter 
 has been set to saturate the upper bound on $R_7$ coming 
from the $b\to s\, \gamma$ data.
\begin{figure}
\vspace{-3.0cm}
\begin{center}
\hspace{-5.cm}
\includegraphics[width=0.75\textwidth]{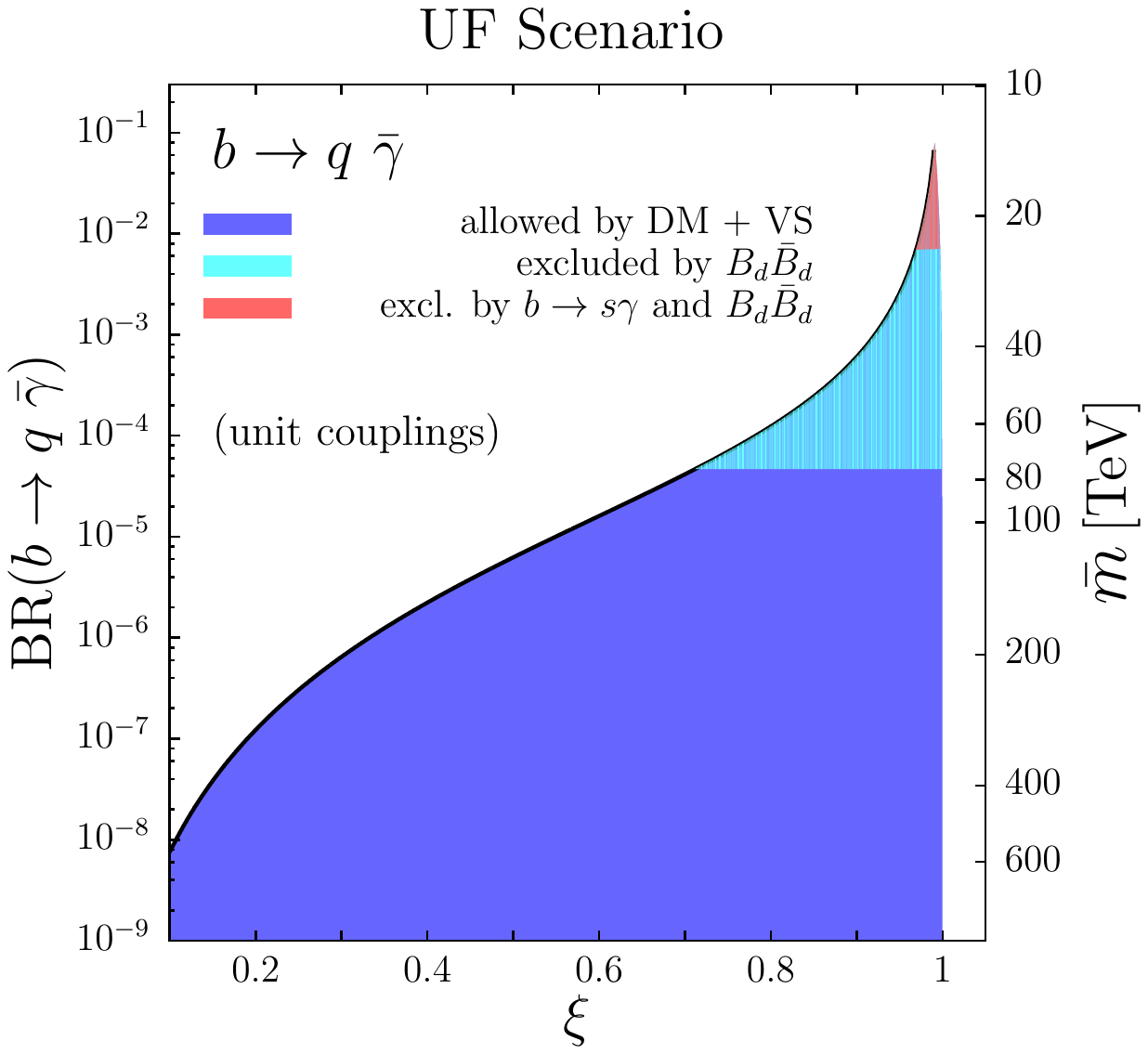}
\hspace{-4.cm}
\includegraphics[width=0.75\textwidth]{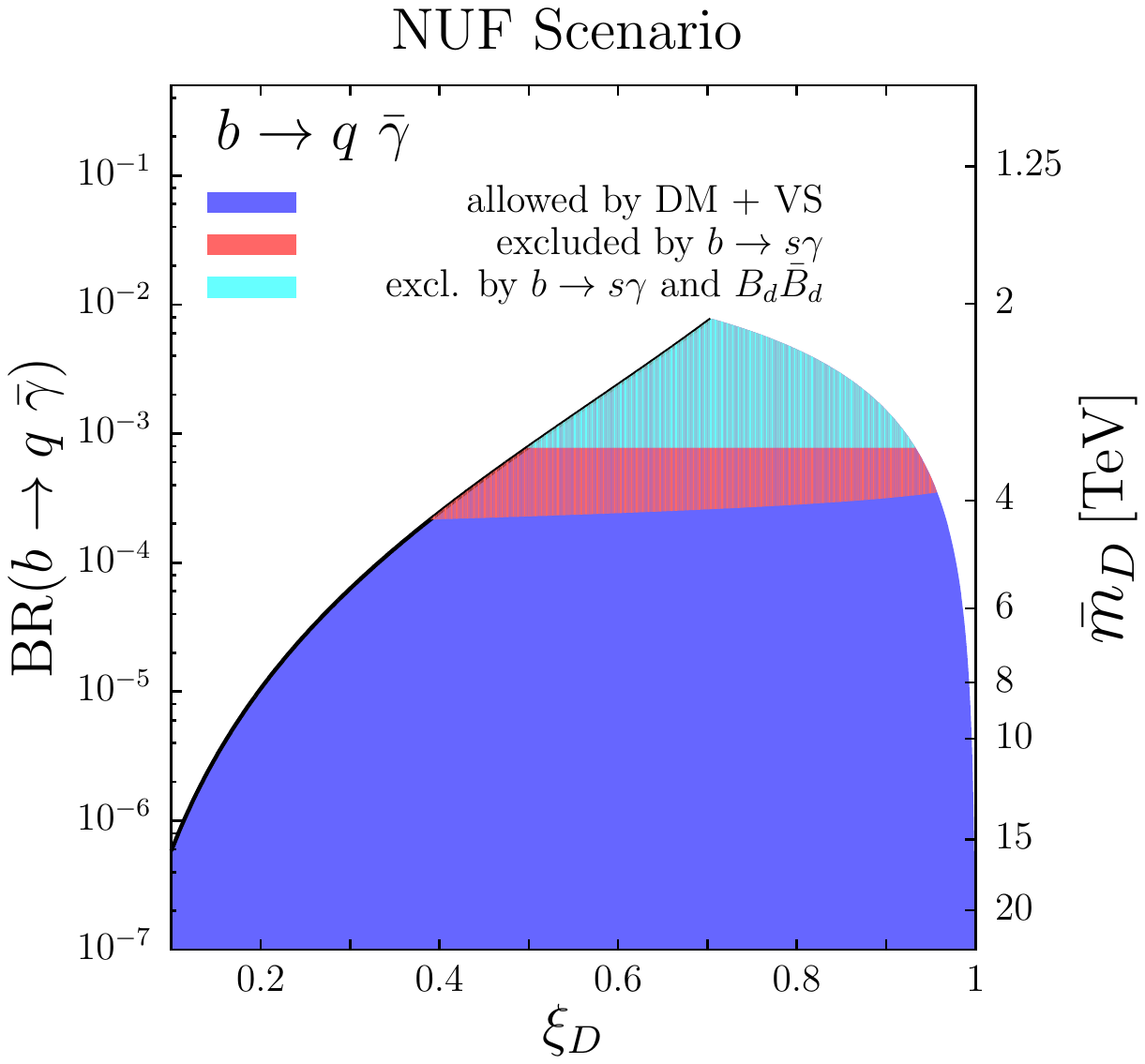}
\hspace{-5cm}
\vspace{-8.cm}
\caption{Allowed regions (dark blue colored areas) by DM and vacuum stability (VS)  constraints
for  ${\rm BR}(b\to q\, \bar{\gamma})$ and for the average messenger mass scales 
$\bar{m}$ and $\bar{m}_{\D}$,
versus the corresponding mixing $\xi$ and $\xi_{\D}$, in the UF  (left) and NUF  (right) scenarios, respectively. 
In the UF (NUF) scenario, we assume  
$\bar e \, \bar e_3^{\D}=1$,\;$ \eta^{j3}_L/\eta^{33}_L=1\,(0.1)$, with $j=1,2$. Red
 regions are excluded by the $b\to s \gamma$ constraints, and light-blue
 regions are excluded by both $b\to s \gamma$ and $B_d\bar{B}_d$ mixing
constraints.}
\label{fig-BR-bot}
\end{center}
\end{figure}
\noindent
In Fig.~\ref{fig-BR-bot}, we show the resulting ${\rm BR}(b\to q\, \bar{\gamma})$
expectations versus mixing. The blue area corresponds to the allowed ranges,
while the red area select the regions excluded by the ${\rm BR}(b\to q\,{\gamma})$
bounds. 
One can see that large values for ${\rm BR}(b\to q\, \bar{\gamma})$ are presently allowed,
both in the UF (left plot) and NUF (right plot). In particular,
for unit couplings, the UF scenario allows ${\rm BR}(b\to q\, \bar{\gamma})$'s up to ($10^{-8}-10^{-3}$), depending on the mixing value, while the NUF scenario allows up to ($10^{-6}-10^{-4}$).

\subsection{BR($b\to q \,\bar \gamma$) upper bounds  from \BBq mixing measurements}
In this section we estimate the largest effect induced by the NP contribution to the effective Hamiltonian for the $\Delta B=2$ transitions. Then we will analyze its impact on the \BBd and \BBs mixing measurements. 

The effective Hamiltonian for the $|\Delta B_s|=2$ transitions,
induced by the Lagrangian in Eq.(\ref{LagMS-tilde}),  is given by
\bea
H_{ef\!f}^{|\Delta B_s|=2}&=&\frac{1}{64 \pi^2 \bar{m}^2(1-\xi)}\left[
\sum_{i=1}^{5} C_i Q_i + \sum_{i=1}^{3}\tilde{C}_i \tilde{Q}_i \right]+{\rm H.c.}
\label{HeffB}
\eea
where the operators $Q_{1-5}$ are defined as
\bea
Q_1&=&\left(\bar{b}_L^{\alpha}\gamma_{\mu}s_L^{\alpha}\right)\left(\bar{b}_L^{\beta}\gamma^{\mu}s_L^{\beta}\right)
\nonumber \\
Q_2&=&\left(\bar{b}_R^{\alpha}s_L^{\alpha}\right)\left(\bar{b}_R^{\beta}s_L^{\beta}\right)
\nonumber \\
Q_3&=&\left(\bar{b}_R^{\alpha}s_L^{\beta}\right)\left(\bar{b}_R^{\beta}s_L^{\alpha}\right)
\nonumber \\
Q_4&=&\left(\bar{b}_R^{\alpha}s_L^{\alpha}\right)\left(\bar{b}_L^{\beta}s_R^{\beta}\right)
\nonumber \\
Q_5&=&\left(\bar{b}_R^{\alpha}s_L^{\beta}\right)\left(\bar{b}_L^{\beta}s_R^{\alpha}\right)
\label{Qbasis}
\eea
with $\tilde{Q}_i=Q_i(L\leftrightarrow R)$ and $q_{L,R}\equiv \frac{1}{2}(1\mp \gamma_5) q$. Also, $q=b,s$ stand for the $b$-quark and
$s$-quarks fields, respectively, and $\alpha,\beta$ are color indices
(sum over color indices is understood). The operator basis corresponding to the effective Hamiltonian for  $|\Delta B_d|=2$ is simply obtained by 
replacing $s$ with $d$ quark fields in  $Q_i$ and $\tilde{Q}_i$ operators in Eq.(\ref{Qbasis}).

In order to obtain the Wilson coefficients $C_i$ and  $\tilde{C}_i$, we compute the contributions at  one loop to the box diagrams  for the process $\bar{b}s\rightarrow b\bar{s}$, by neglecting  quark masses and external momenta. Since we are interested to  their dominant effect, we will work in the approximation of large mixing $\xi$, which allows us to restrict to the contribution of the Feynman diagrams in which only the two lightest scalars circulate in the loop. 
In the
left-right symmetric scenario considered here, this corresponds to consider
in the box diagram only the propagation of two degenerate messengers with mass square $m^2_{-}=\bar{m}^2(1-\xi)$. Since we are interested in constraining only the combination of flavor matrix elements $\eta_{32}$ and $\eta_{31}$ (which enter the 
$b\to s \bar{\gamma}$ and $b\to d \bar{\gamma}$ processes, respectively), in order to simplify the analysis, we
will consider  only the contribution to the \BBq mixing induced by the dark-fermion associated to the $b$-quark, namely $Q_{\D_3}$, while we assume  for the diagonal entries, $\eta_{33}=1$ and $\eta_{ii}=0$ for $i=1,2$.

By using the above approximations and performing the matching between the amplitude of $\bar{b}s\rightarrow b\bar{s}$ computed from the full theory and the one obtained by  the effective Hamiltonian in Eq.(\ref{HeffB}), we obtain the following results for the Wilson
coefficients evaluated at the messenger mass scale $\bar{m}_{-}$:
\bea
C_1&=&\frac{1}{2}C^2_{LL}\Delta_1\, , ~~~~~\tilde{C}_1\,=
\,\frac{1}{2}C^2_{RR}\Delta_1\, ,
\nonumber\\
C_2&=&\tilde{C_2}\,=\,0\, ,
\nonumber\\
C_3&=& \frac{1}{2}C^2_{RL}\Delta_2\, , ~~~~~
\tilde{C}_3\,=\, \frac{1}{2}C^2_{LR}\Delta_2\, ,~~~~~
\nonumber\\
C_4&=&-2C_{LR}C_{RL}\Delta_1\, , ~~~~~ C_5\,=\,C_{LR}C_{RL}
\Delta_2\, ,
\label{WilsonB}
\eea
where the coefficients $C_{L,R}$ are defined as
\bea
C_{LL}&=&g_L^2\eta_L^{3j}(\eta_L^{j2})^{\star}\, ,~~~~~
C_{RR}\,=\,g_R^2\eta_R^{3j}(\eta_R^{j2})^{\star}\, ,
\nonumber\\
C_{LR}&=&g_Lg_R\eta_L^{3j}(\eta_R^{j2})^{\star}\, ,~~
C_{RL}\,=\,g_Lg_R\eta_R^{3j}(\eta_L^{j2})^{\star}\, 
\label{CC}
\eea
and $j=3$ in case one considers only the exchange of the $Q_{D_3}$ dark-fermion.
As for the quantities $\Delta_{1,2}$, which parametrize the loop integrals, we get the following results for the UF and NUF scenarios
\bea
\Delta_1^{UF}&=&-\frac{1}{4}\, ,~~~~~\Delta_2^{UF}\,=\,0\, ,
\nonumber \\
\Delta_1^{NUF}&=&-\frac{1}{12}\, ,~~~~~\Delta_2^{NUF}\,=\,\frac{1}{6}\, .
\label{integrals}
\eea
In the UF scenario, the loop integrals in Eq.(\ref{integrals}) have been obtained by setting to zero the dark-fermion mass, which is well justified 
since in this case the average messenger mass is much larger than  $M_{Q_{\D_3}}$. On the other hand, in the NUF scenario, we have retained the contribution of the dark-fermion mass of third generation $M_{Q_{D_3}}$ and set it equal to the lightest messenger mass $M_{Q_{D_3}}^2\simeq \bar{m}^2(1-\xi)$, as assumed in the the NUF scenario contribution to  $BR(b\to q \bar{\gamma})$ in order to pinpoint  the largest effect.
Regarding the effective Hamiltonian for the $|\Delta B_d|=2$ transitions, the corresponding Wilson coefficients can be obtained by the $C_i$ and $\tilde{C}_i$ expressions above, by replacing in Eq.({\ref{CC}) the $\eta_{L,R}^{j2}$ matrix elements  by $\eta_{L,R}^{j1}$, with $j=3$.

The contribution to the \BBq mixing amplitude $M_{12}^q$ is given by
\bea
M_{12}^q&=&\frac{\langle B_q |H_{ef\!f}^{|\Delta B_{q}|=2}|\bar{B}_q\rangle}{2M_{B_{q}}}
\eea
where $M_{B_{q}}$ is the  neutral $B$-meson mass, with $q=d,s$.
Combining the SM with the NP contributions, one obtains for the difference
of the neutral B meson mass eigenstates system
$\Delta M_q=M_H^q-M_L^q=2|M_{12}^q|$ , where  $M^q_H$ and $M^q_L$ are the corresponding heavy and light mass eigenstates of the neutral \BBq system respectively, 
\cite{Lenz:2010gu}
\bea
\Delta M_d&=&0.502~{\rm ps}^{-1}|\Delta_d|\, ,
\nonumber\\
\Delta M_s&=&17.24~{\rm ps}^{-1}|\Delta_s| \, ,
\label{deltaM}
\eea
where $z_{t}=\frac{\bar{m}_t^2}{M_W^2}$ , and the $\Delta_q$ quantities are defined as
\bea
\Delta_q\equiv 1+\frac{M_{12}^{NP,q}}{M_{12}^{SM,q}}\, .
\eea
Above, $M_{12}^{NP,q}$ ($M_{12}^{SM,q}$) stands for the NP (SM) corresponding contribution. In the above Eq.(\ref{deltaM}), we assume  the central values reported in \cite{Lenz:2010gu}, in particular
$|V_{tb}V_{td}^{\star}|=0.0086$, $|V_{tb}V_{ts}^{\star}|=0.04$,
$f_{B_d}^2 B_{B_d}=(0.17~{\rm GeV})^2$, $f_{B_s}^2 B_{B_s}=(0.21~{\rm GeV})^2$,
and $S(\frac{\bar{m}_t^2}{M_W^2})=2.35$,
where $S(x)$ is the Inami-Lim function for the top-quark contribution from the box diagram, $\bar{m}_t$ the top-quark mass in $\bar{{\rm MS}}$ scheme evaluated at  $\bar{m}_t$ scale [$\bar{m}_t(\bar{m}_t)=0.957 m_t$], 
 $f_{B_q}$ the $B_q$ decay constants, and $B_{B_{q}}$ the bag factors related to the matrix element of the corresponding $\Delta B=2$ SM operators.

We then computed the Wilson coefficients at the low energy scale of order
${\cal O}(m_b)$, and the matrix elements of the operators appearing in Eq.(\ref{Qbasis}) at the NLO and evaluated at the same scale, by using the results of 
\cite{Becirevic:2001jj}, where the same structure for the effective Hamiltonian was considered.
Following the results of \cite{Lenz:2012az}, the present \BBq mixing measurements  imply
\bea
{\rm Re}(\Delta_d)&=&0.823^{+0.143}_{-0.095}\, ,
~~~~{\rm Re}(\Delta_s)\,=\, 0.965^{+0.133}_{-0.078}\, ,
\eea
where corresponding errors are at 1-$\sigma$ level.
Assuming a constructive NP contribution to the SM result  (and real $\eta$ matrices), where the NP contribution to $(|{\rm Re}(\Delta_d)|-1)$ is more constrained, we require $(|{\rm Re}(\Delta_q)|-1)$ to lie at the 2-$\sigma$ level in the 
following ranges
\bea
0\,\le\,|{\rm Re}(\Delta_d)|-1 \, <\, 0.109\, , ~~~~~~~
0\,\le\,|{\rm Re}(\Delta_s)|-1 \, < \, 0.231\, .
\eea
In Fig.~\ref{fig-BR-bot}, we show the effect of the  \BBq mixing constraints on  ${\rm BR}(b\to q \bar{\gamma})$ versus $\xi$, for the UF (left plot) and NUF (right plot) scenarios. The light-blue areas  are the excluded ones. We focus on  the \BBd mixing
constraints [which hold for the  ${\rm BR}(b\to d \bar{\gamma})$ case], since the   regions excluded by the \BBs mixing are always outside the  area allowed by DM and $b\to s \gamma$ constraints.
One can see that the \BBd mixing is quite effective, disfavoring  ${\rm BR}(b\to d \bar{\gamma})$ values above
$5\times 10^{-5}$ and $8\times 10^{-4}$ for the  UF and NUF scenarios,  
respectively.
\section{The $c\to u \, \bar{\gamma}$ decay}
Here we analyze the FCNC decay $c\to u \, \bar{\gamma}$, following the 
same approach as used for the heavier quarks.
The corresponding total width is given by Eq.(\ref{width}), where $i=2$ and $j=1$ for the 
$c\to u$ transition. We will express BR($c\to u \, \bar{\gamma}$) 
in terms of the inclusive decay   rate  
${\rm BR}^{\rm exp}(c\to \ell^+ X)=(0.096\pm 0.004)\%$ (with $X$ standing 
for {\it anything}) \cite{Agashe:2014kda}, by approximating  
$\Gamma(c\to \ell^+ X)$ with the   
Cabibbo-allowed tree-level $c\to s \,e^+ \nu_e$ decay width.

Then, one has
\bea
{\rm BR}(c\to u \,\bar{\gamma})&=& \frac{12\; {\rm BR}^{\rm exp}(c\to \ell^+ X) }{G_F^2 |V_{cs}|^2 m_c^2 f_1(z_{uc})}\left(\frac{1}{(\Lambda_L^{cu})^2}
                 +\frac{1}{(\Lambda_R^{cu})^2}
\right)\, ,
\label{BRcuDP}
\eea
where $V_{cs}$ is the relevant CKM matrix element, and $z_{uc}=m_u^2/m_c^2$, with $f_1(x)$ defined  by Eq.(\ref{Gammab0}). The expressions needed for   
$\Lambda^{cu}_{L,R}\equiv (\Lambda^{\U}_{L,R})_{21}$ can be found in Eq.(\ref{LambdaU}). For our numerical analysis, we use  the central value of 
$V_{cs}=0.986\pm 0.016$, extracted from the average of the $D$ leptonic and
semileptonic decays~\cite{Agashe:2014kda}.

Following the same strategy as the one described for the  top and $b$-quark cases, we report in Table~\ref{tabcharm} the results for the maximum allowed value
of ${\rm BR}(c\to u\,\bar{\gamma})$, satisfying the vacuum stability 
bounds and DM constraints versus 
the mixing parameter $\xi=\xi_{\U}=\xi_{\D}$ 
($\xi=\xi_{\D}$) in the UF (NUF) scenario.
These results assume $U(1)_F$ charges and other multiplicative couplings normalized 
to~1. In particular, in Table~\ref{tabcharm} one has 
$\bar{e}\,\bar{e}^{\U}_2 \; =g_{L,R}=\rho^{12,22}_{L,R}=\eta^{12,22}_{L,R}=1$, with all other elements of flavor matrices set to zero.
\begin{table} \begin{center}    
\begin{tabular}{|c||c|c|}
\hline 
$\xi$ 
& ${\rm BR}^{\rm max}_{\rm UF}(c\to u\, \bar{\gamma})$
& ${\rm BR}^{\rm max}_{\rm NUF}(c\to u\, \bar{\gamma})$
\\ \hline 
0.1
& $1.0\times 10^{-11}$
& $2.9\times 10^{-13}$
\\ \hline
0.2
& $1.6\times 10^{-10}$
& $4.9\times 10^{-12}$
\\ \hline
0.3
& $8.5\times 10^{-10}$
& $2.8\times 10^{-11}$
\\ \hline
0.5
& \!\!$8.1\times 10^{-9}$
& $3.8\times 10^{-10}$

\\ \hline
0.7
& \!\!$5.3\times 10^{-8}$
& \!\!$5.1\times 10^{-9}$

\\ \hline
0.8
& \!\!$1.6\times 10^{-7}$
& \!\!$2.9\times 10^{-8}$

\\ \hline
0.9
& \!\!$7.5\times 10^{-7}$
& \!\!$4.9\times 10^{-7}$

\\ \hline
~\,0.95
& \!\!$3.2\times 10^{-6}$
& \!\!$7.7\times 10^{-6}$

\\ \hline
\end{tabular} 
\caption[]{ Maximum values of
${\rm BR}(c\to u\, \bar{\gamma})$ allowed 
by vacuum stability and DM constraints versus the mixing parameter $\xi=\xi_{\U}=\xi_{\D}$ and $\xi=\xi_{\D}$, in the UF and NUF scenarios, respectively. Results are  in unit of couplings, that is they assume  
$\bar{e}\,\bar{e}^{\U}_2 \; =g_{L,R}=\rho^{12,22}_{L,R}=\eta^{12,22}_{L,R}=1$, with all other elements of flavor matrices set to zero.}
\label{tabcharm}
\end{center} 
\end{table}

Finally, in Fig.~\ref{fig-BR-charm}, we show the corresponding regions of ${\rm BR}(c\to q\, \bar{\gamma})$ values   allowed by DM and vacuum stability constraints
 versus the mixing parameter.  
 The blue area corresponds to the allowed ranges.
Experimental upper bounds on 
${\rm BR}(c\to q\,{\gamma})$ do not further constraint the blue regions in this case. 
One can see that large values for ${\rm BR}(c\to q\, \bar{\gamma})$ are presently allowed,
both in the UF (left plot) and NUF (right plot). In particular,
for unit couplings, the UF scenario allows ${\rm BR}(c\to q\, \bar{\gamma})$'s up to 
($10^{-11}\!\!-\!\!10^{-4}$), depending on the mixing value, while the NUF scenario allows up to ($10^{-13}\!\!-\!\!10^{-8}$).

\begin{figure}
\vspace{-2.5cm}
\begin{center}
\hspace{-5.cm}
\includegraphics[width=0.75\textwidth]{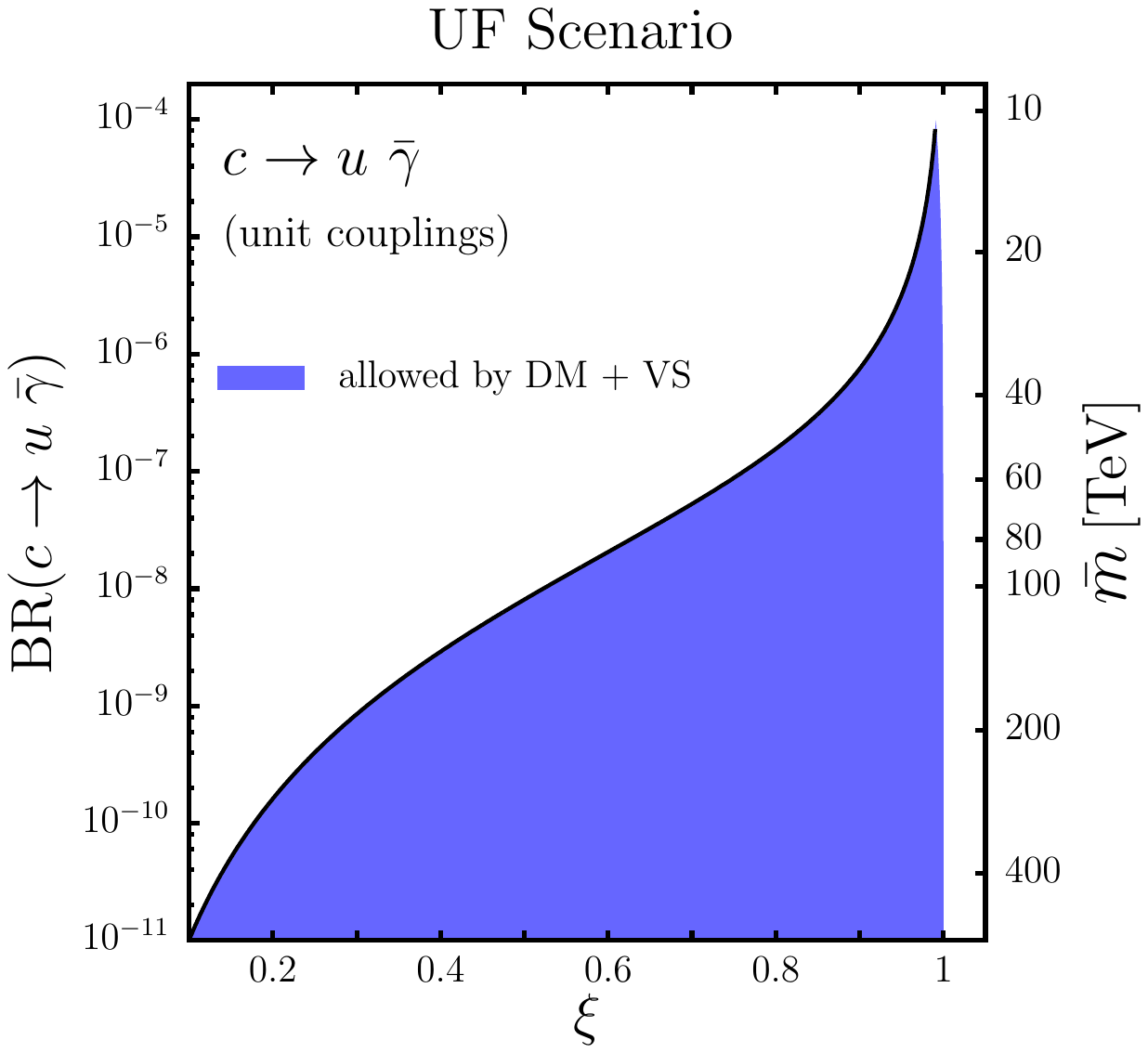}
\hspace{-4.cm}
\includegraphics[width=0.75\textwidth]{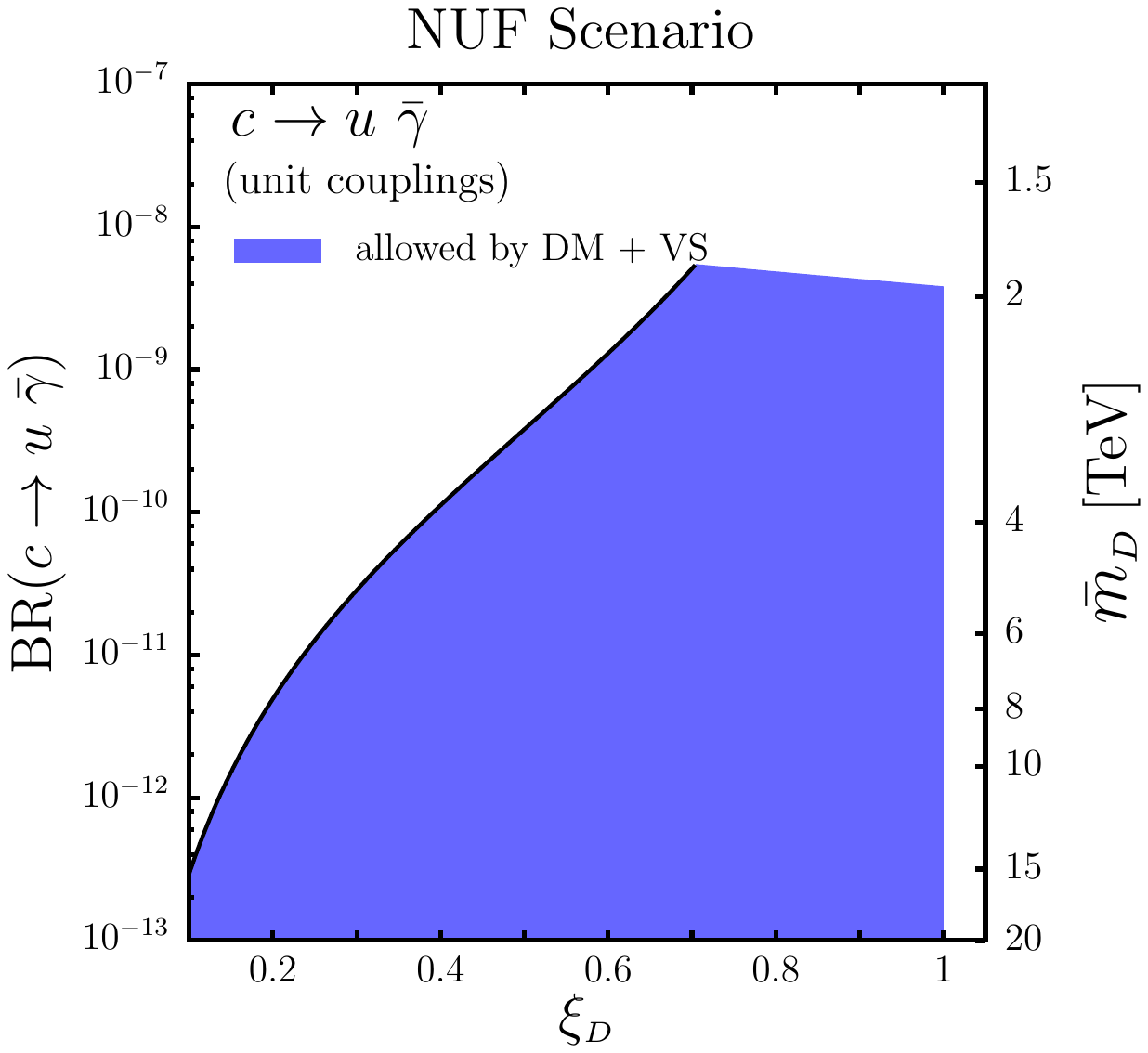}
\hspace{-5cm}
\vspace{-7.cm}
\caption{Allowed regions (colored areas) by DM and vacuum-stability (VS)  constraints
for  ${\rm BR}(c\to u\, \bar{\gamma})$ and for the average messenger mass scales 
$\bar{m}$ and $\bar{m}_{\D}$
versus the corresponding mixing $\xi$ and $\xi_{\D}$, in the UF  (left) and NUF  (right) scenarios, respectively. In the left (right) plots 
we assume $\bar{e}\,\bar{e}^{\U}_2=1$,\;$ \rho_L^{12}/ \rho^{22}_L\simeq1$
($\bar{e}\,\bar{e}^{\D}_2 =1$,\;$ \eta_L^{12}/ \eta^{22}_L\simeq1$) with all other matrix elements of flavor matrices set to zero.}
\label{fig-BR-charm}
\end{center}
\end{figure}


\section{The $\tau\to (\mu,e) \, \bar{\gamma}$ decays}
We now  consider the extension of the model described in Sec.~2 to the leptonic sector
in order to also generate effective {\it lepton} Yukawa couplings. Notice that we will not include the possibility of Majorana masses neither at tree level nor radiatively generated  for the neutrino sector, and neutrinos will be assumed to acquire 
only a Dirac mass through the SM Higgs mechanism.
Although one can also radiatively  generate Majorana masses  in this framework, we will not consider  this possibility here.

New dark fermions will be associated to the charged leptons and neutrinos, 
as occurs in the quark scenario, with a corresponding set of color singlet messenger fields,  having the same $SU(2)_L\times U(1)_Y$ quantum numbers of the ones
related to the lepton sector. Moreover, in order to 
generate the Pontecorvo-Maki-Nakagawa-Sakata (PMNS)~\cite{Maki:1962mu}
lepton mixing matrix, we will generalize the mechanism described in Sec.~2.3 
for the origin of the CKM matrix to the leptonic sector.
The induced PMNS matrix will be also unitary, since we will not include any seesaw mechanism.

The generalization to the leptonic sector of the interaction Lagrangian in 
Eq.~(\ref{LagMS-tilde})
 is straightforward,
consisting just in the substitution of  quark messenger and dark-fermion fields by the corresponding ones in the leptonic sector.
We then just provide the notation for the 
new flavor mixing matrices in the leptonic sector. In particular, 
after rotating the lepton fields to the mass-eigenstate basis, new
flavor matrices $\tilde{\rho},\tilde{\eta}$ will appear in the 
leptonic Lagrangian corresponding to Eq.~(\ref{LagMS-tilde}), where 
\bea
\rho_{L,R}&\to& \tilde{\rho}_{L,R} \nonumber\\
\eta_{L,R}&\to& \tilde{\eta}_{L,R}\, .
\label{matrixel}
\eea

In this framework, we first analyze the flavor-violating tau lepton decays
\bea
\tau \to \ell \, \bar{\gamma}\, ,
\eea
where $\ell=\mu,e$~\footnote{Flavor violating $\tau$ decays into a {\it massive} neutral vector have been considered in \cite{Heeck:2016xkh}.}. 
The corresponding decay width can be inferred by Eq.~(\ref{width}), with $i=3$ and $j=2,1$ for the $\ell=\mu,e $ transitions, respectively.  
The $\Lambda_{L,R}^{\tau \mu}$ and $\Lambda_{L,R}^{\tau e}$ expressions can be obtained from   $(\Lambda^{\D}_{L,R})_{32}$ and $(\Lambda^{\D}_{L,R})_{31}$  as defined 
in Eq.~(\ref{LambdaD}),  where the quark
masses in the down sector $m_{\D_i}$, 
the dark-quark masses $M_{\Q^{\U,\D}_i}$, and  
average messenger masses $\bar{m}_{\U,\D}$ are replaced by the corresponding
ones in the leptonic sector, namely  $m_{\E_i}$, $M_{\LL^{\U,\D}_i}$, 
$\bar{m}_{\LL}^{\U,\D}$, respectively. 
In Eq.~(\ref{LambdaD}), one then makes the replacements  
$(x^{\D}_3,\xi_{\D})\to (x^{\LL}_3,\xi_{\LL})$, where   
$x_3^{\LL}\equiv (M_{\LL_3^{\D}}/\bar{m}_{\LL}^{\D})^2$, and  $g_{L,R}\to \bar{g}_{L,R}$~, where $\bar{g}_{L,R}$ are the relevant couplings in the leptonic sector. As for   
the flavor matrices,  Eq.~(\ref{matrixel}) applies.

We can now express ${\rm BR}(\tau\to \ell \,\bar{\gamma})$ by normalizing it 
to  
${\rm BR}^{\rm exp}(\tau\to \nu_{\tau} \bar{\nu}_{\mu} \mu)=(17.41\pm 0.04)\%$~\cite{Agashe:2014kda},  assuming  the following $\tau\to \mu\, \nu_{\tau} \bar{\nu}_{\mu} $ tree-level decay width  
\bea
\Gamma^{\tau}_0&=& \frac{G_F^2 \;m_{\tau}^5}{192 \;\pi^3} f_1(z_{\mu\tau})\, ,
\label{Gamma0tau}
\eea
where the function $f_1(x)$ is defined just after Eq.~(\ref{Gammab0}), and 
$z_{\mu\tau}=m_{\mu}^2/m_{\tau}^2$. Then, one obtains
\bea
{\rm BR}(\tau\to \ell \,\bar{\gamma})&=& \frac{12\; {\rm BR}^{\rm exp}_{
\tau\to \nu_{\tau} \bar{\nu}_{\mu} \mu}} {G_F^2 \;m_{\tau}^2\; f_1(z_{\mu\tau})}\left(\frac{1}{(\Lambda_L^{\tau \ell})^2}+\frac{1}{(\Lambda_R^{\tau \ell})^2}
\right)\, .
\label{BRtauDP}
\eea

We will restrict  to the UF scenario, where the average 
messenger masses for the {\it up} and {\it down} $SU(2)_L$ messenger fields
in the leptonic sector are assumed to be the same, namely  $\bar{m}_{\LL}^{\U}=\bar{m}_{\LL}^{\D}\equiv \bar{m}_{\LL}$. Moreover, in $\Lambda_{L,R}^{\tau \ell}$ we will neglect the  terms  proportional to $\bar{g}_L^2/(16\pi^2)$ (\cf Eq.~(\ref{LambdaD})).

Regarding the constraints coming from  DM and vacuum stability, 
in the leptonic
sector the bounds in Eq.~(\ref{boundU})  reads 
\bea
\bar{m}_{\LL}&\ge&m_{\tau}  \left(\frac{16 \pi^2}{\bar{g}_L\bar{g}_R}\right) F(\xi_{\LL})
\, ,
\label{boundL}
\eea
where 
$\xi_{\LL}$ is the universal mixing parameter for the leptonic messenger masses. Then, at large $\xi_{\LL}$, one has 
\bea
\bar{m}_{\LL} \gsim 1.1\, \sqrt{1-\xi_{\LL}}~{\rm TeV}\, .
\eea
The corresponding maximum  allowed ${\rm BR}(\tau\to \ell\, \bar{\gamma})$ is reported in 
Table~\ref{tab6}, where  all relevant couplings are set  to~1.
\begin{table} \begin{center}    
\begin{tabular}{|c||c|c|c|}
\hline 
$\xi_{\LL}$ 
& ${\rm BR}^{\rm max}(\tau\to \ell \, \bar{\gamma})$
& $\bar{m}_{\LL}^{\rm min}[{\rm TeV}]$
& ${m}_{{\LL}_{-}}^{\rm min}[{\rm TeV}]$
\\ \hline 
~\,0.05
& $2.3\times 10^{-7}$
& \!11  
& \!11
\\ \hline 
0.1
& $3.8\times 10^{-6}$
& ~~~\,5.7
& ~~~\,5.4
\\ \hline
0.2
& $6.9\times 10^{-5}$
& ~~~\,2.9
& ~~~\,2.6
\\ \hline
(0.3)
& $4.1\times 10^{-4}$
& ~~~\,1.9
& ~~~\,1.6
\\ \hline
(0.4)
& $1.6\times 10^{-3}$
& ~~~\,1.4
& ~~~\,1.1
\\ \hline
(0.5)
& $5.2\times 10^{-3}$
& ~~~\,1.1
& ~~~\,0.8
\\ \hline
\end{tabular} 
\caption[]{Maximum values of ${\rm BR}(\tau\to \ell \, \bar{\gamma})$
allowed  by vacuum stability and DM constraints in the UF scenario for the leptonic sector,   corresponding to the minimum allowed average mass $\bar{m}_{\LL}^{\rm min}$, and to the lightest universal messenger mass eigenvalue ${m}_{\LL_{-}}^{\rm min}=\bar{m}^{\rm min}\sqrt{1-\xi_{\LL}}$  versus the mixing parameter $\xi_{\LL}$. Results
are in unit of couplings, that is they assume $\bar{e}\,\bar{e}^{\LL}_3\; =\bar{g}_{L,R}=
\tilde{\eta}^{33,13,23}_{L,R}=1$, with all other elements of flavor matrices set to zero. BR's corresponding to values of $\xi_{\LL}\gsim 0.2$ might be excluded at 90\% C.L.
by direct constraints on ${\rm BR}(\tau\to \ell {\gamma})$ (see text).}
\label{tab6}
\end{center} 
\end{table}

Radiative lepton-flavor violating (LFV) decays  $\tau^- \to \ell^-  \gamma$, with $\ell=\mu,e$, indirectly constrain $\tau$ decays into dark photons. 
The present experimental upper bounds at 90\% C.L. are \cite{Aubert:2009ag}
\bea
{\rm BR}(\tau^- \to e^-  \gamma)&<&  3.3 \times 10^{-8}\, ,
\nonumber
\\
{\rm BR}(\tau^- \to \mu^-  \gamma)&<&  4.4 \times 10^{-8}\, .
\label{taulim}
\eea
The SM contribution to the LFV $\tau\to  \ell\, \gamma$ decays  is  negligible, due to the GIM suppression and tiny neutrino masses, even accounting for the PMNS
matrix. However, the NP contribution could be potentially quite large.
In the present scenario the corresponding prediction is  
\bea
{\rm BR}(\tau\to \ell \,\gamma)&=& \frac{12\; {\rm BR}^{\rm exp}_{
\tau\to \nu_{\tau} \bar{\nu}_{\mu} \mu}} {G_F^2\; m_{\tau}^2 \;f_1(z_{\mu\tau})}\left(\frac{1}{(\bar{\Lambda}_L^{\tau \ell})^2}+\frac{1}{(\bar{\Lambda}_R^{\tau \ell})^2}
\right)\, ,
\label{BRtauP}
\eea
where the expressions for $\bar{\Lambda}_{L,R}^{\tau \ell}$ can be derived
from the general formulas in the Appendix, by
replacing the $\eta_{L,R}$ matrices according to Eq.(\ref{matrixel}), and the variables
$(x_3^{\D},\xi_{\D})$ by $(x_3^{\LL},\xi_{\LL})$.

As discussed above for $b$ decays, a characteristic {\it effective} messenger mass scale $1/(\bar{m}_{\LL}^{32})^2$ given by
\bea
\bar{m}_{\LL}^{32}\equiv \bar{m}_{\LL}\sqrt{\left|\frac{\tilde{\eta}_L^{33}}{
\tilde{\eta}_L^{j3}}\right|}\, ,
\label{ML}
\eea
with $j=2$ and $1$  for $\mu$ and $e$ final states, respectively,
 factorizes  in both   $\tau \to \ell  \gamma$ and $\tau \to \ell \bar{\gamma}$ BR's.
Then, the bounds in Eq.(\ref{taulim})
can be straightforwardly  converted into  lower bounds on the effective mass scale $\bar{m}_{\LL}^{32}$ 
\bea
\bar{m}_{\LL}^{32}&>&\left(\frac{96\,\pi\,\alpha\, 
{\rm BR}^{\rm exp}_{\tau\to \nu_{\tau} \bar{\nu}_{\mu} \mu}}{{\rm BR}^{\rm max}_{\ell\gamma} \;G_F^2 \;f_1(z_{\mu\tau})}\right)^{1/4}\sqrt{\bar{F}_{LR}(x_3^{\LL},\xi_{\LL})}\, ,
\label{Mtaubound}
\eea
where ${\rm BR}^{\rm max}_{\ell\gamma}\equiv 4.4\;(3.3) \times 10^{-8}$ for $\ell=\mu\;(e)$.

In Fig.~\ref{fig3} we plot the excluded regions for $\bar{m}_{\LL}^{32}$ 
corresponding to the constraint in Eq.(\ref{Mtaubound}), versus $x_3^{\LL}$ and
for some values of the mixing parameter $\xi_{\LL}$.
One can see that the constraints depend on $x_3^{\LL}$, with $\bar{m}_{\LL}^{32} \lsim (6.7$--$11.3)$ TeV  in the  
region $x_3^{\LL}< 1- \xi_{\LL}$, for $\xi_{\LL}\gsim 0.1$.

\begin{figure}
\vspace{-2.5cm}
\begin{center}
\includegraphics[width=0.7\textwidth]{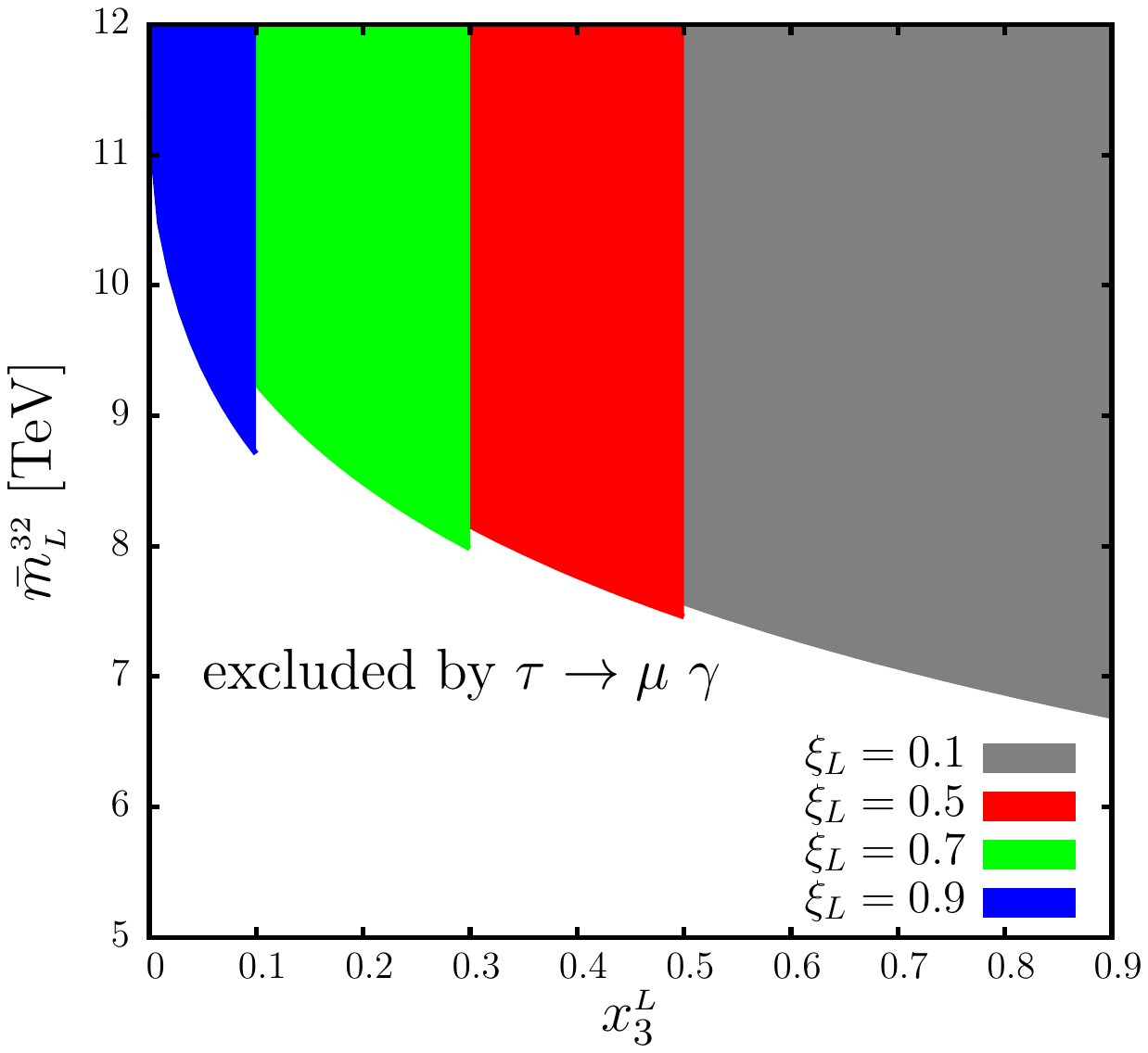}
\vspace{-7.5cm}
\caption{Regions  allowed by  constraints on BR($\tau\to \mu \gamma$)
at 90\% C.L.  (represented by superimposed colored areas), 
for the {\it effective}   messenger mass scale 
$\bar{m}_{\LL}^{32}$ defined in Eq.(\ref{ML}), as a function of  $x_3^{\LL}$ and for several values of the 
mixing $\xi_{\LL}$ parameter. Regions $x_3^{\LL}>1-\xi_{\LL}$ are excluded by DM constraints.}
\label{fig3}
\end{center}
\end{figure}
Analogous results for the constraints from $\tau\to e \gamma$ can be simply rescaled from the ones in Fig.~\ref{fig3}, by using the corresponding upper bound in Eq.(\ref{taulim}).

As we can see from the results in Fig.~\ref{fig3}, the constraints 
from  $\tau\to \mu \gamma$ or $\tau\to e \gamma$ on the effective scale 
$\bar{m}_{\LL}^{32}$ are a more relaxed than the corresponding 
ones from $b\to s \gamma$, 
for same values of $x^{\D}_3,\xi_{\D}$ and $x^{\LL}_3,\xi_{\LL}$ (see Fig.~\ref{fig2} for comparison). The reason is 
that the $b\to s \gamma$ decay gets the main contribution from the SM, and the 
constraints apply mainly on the interference between the SM and NP amplitude. On the other hand, for the $\tau \to \ell \gamma$ decay, the SM contribution is negligible and 
the constraints apply directly on the new physics contributions to the 
amplitude squared.

Now, we combine the constraints from $\tau\to \ell \gamma$ decay with the corresponding 
ones from DM and vacuum stability. 
If we compare the values of $\bar{m}_{\LL}^{\rm min}$ in Table~\ref{tab6}
with the excluded regions in Fig.~\ref{fig3}, we can see that no significant upper limits on the mixing matrices $\tilde{\eta}_{L,R}$ can be set  at small mixing, since
the lower bounds from DM constraints on the average mass $\bar{m}_{\LL}$ are always above the  regions excluded by experimental constraints on BR($\tau\to \ell \,\gamma$). On the other hand, for large mixing, the 
DM constraints are relaxed, and we obtain for example
\bea
\left|\frac{\tilde{\eta}^{23}_{\LL}}{\tilde{\eta}^{33}_{\LL}}\right| &<& 1.3\times 10^{-2} 
\left(\frac{\bar{m}_{\LL}}{870\, {\rm GeV}}\right)^2,~~~{\rm for }~~\xi_{\LL}=0.6\, ,
\nonumber \\
\left|\frac{\tilde{\eta}^{23}_{\LL}}{\tilde{\eta}^{33}_{\LL}}\right| &<& 2.7\times 10^{-3} 
\left(\frac{\bar{m}_{\LL}}{360\, {\rm GeV}}\right)^2,~~~{\rm for }~~\xi_{\LL}=0.9\, .
\eea

Finally,  we give below the upper bounds on
${\rm BR}(\tau\to \ell\,\bar{\gamma})$ which satisfy the $\tau\to \ell\, \gamma$
constraints. In particular, for  small and large mixing values we get
\begin{itemize} 
\item  for $\xi_{\LL}=0.1$ and $x_3^{\LL}=0.8$ (small-mixing regime)
\bea
{\rm BR}^{(\tau\to \mu \gamma)}(\tau\to \mu\, \bar{\gamma}) &<& 2.6\times 10^{-6}
\left(\frac{\bar{\alpha}}{0.1}\right)\, ,
\, \\
{\rm BR}^{(\tau\to e \gamma)}(\tau\to e\, \bar{\gamma}) &<& 2.0\times 10^{-6}
\left(\frac{\bar{\alpha}}{0.1}\right)\, ,
\,
\eea
\item  for $\xi_{\LL}=0.8$ and $x_3^{\LL}=0.1$
(large-mixing regime)
\bea
{\rm BR}^{(\tau\to \mu \gamma)}(\tau\to \mu\, \bar{\gamma}) &<& 5.1\times 10^{-6}
\left(\frac{\bar{\alpha}}{0.1}\right)\, ,
\, \\
{\rm BR}^{(\tau\to e \gamma)}(\tau\to e\, \bar{\gamma}) &<& 3.8\times 10^{-6}
\left(\frac{\bar{\alpha}}{0.1}\right)
\, .
\label{BRtaubounds}
\eea
\end{itemize}

\begin{figure}
\vspace{-3.cm}
\begin{center}
\hspace{-5.cm}
\includegraphics[width=0.75\textwidth]{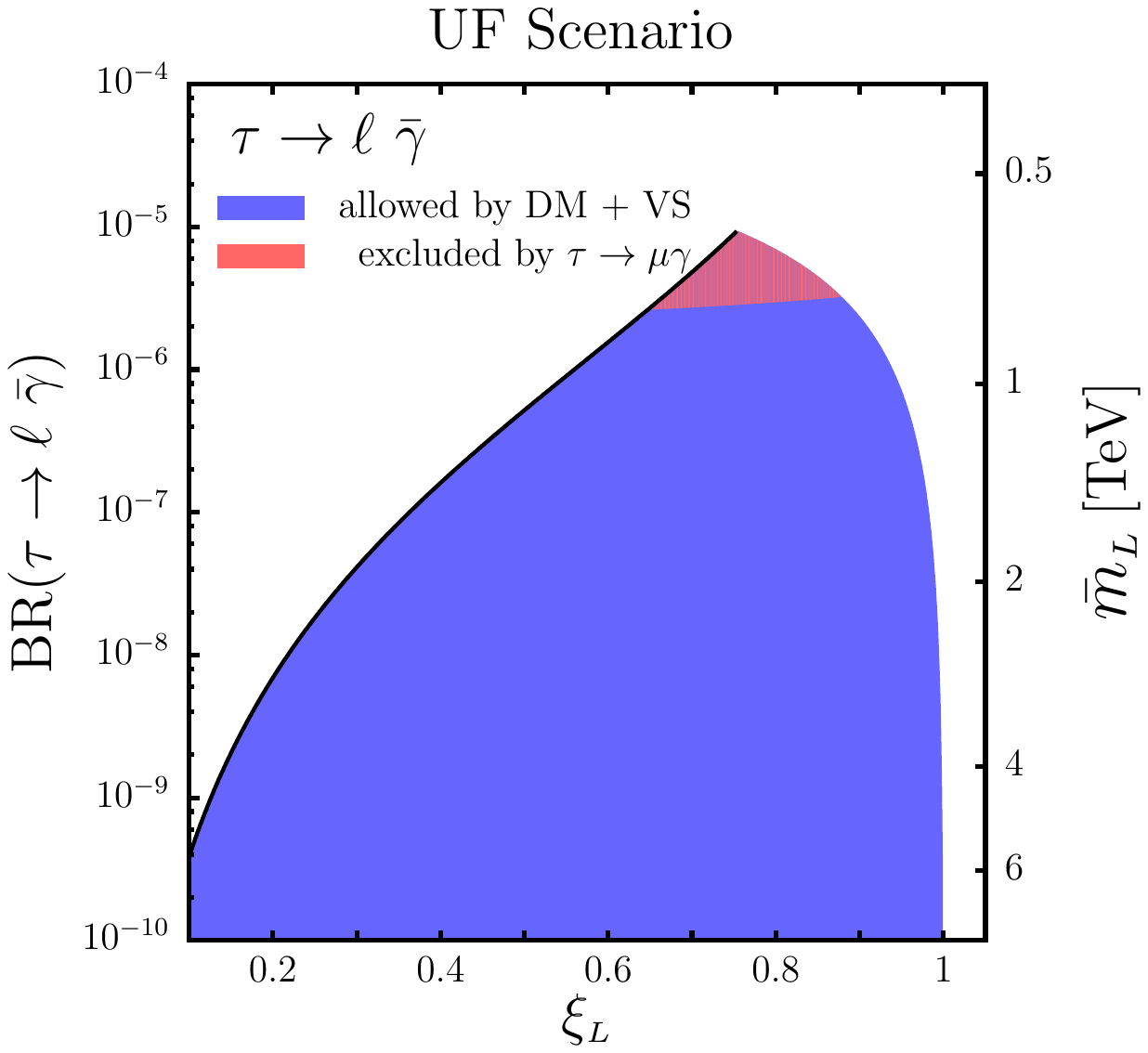}
\hspace{-4.cm}
\includegraphics[width=0.75\textwidth]{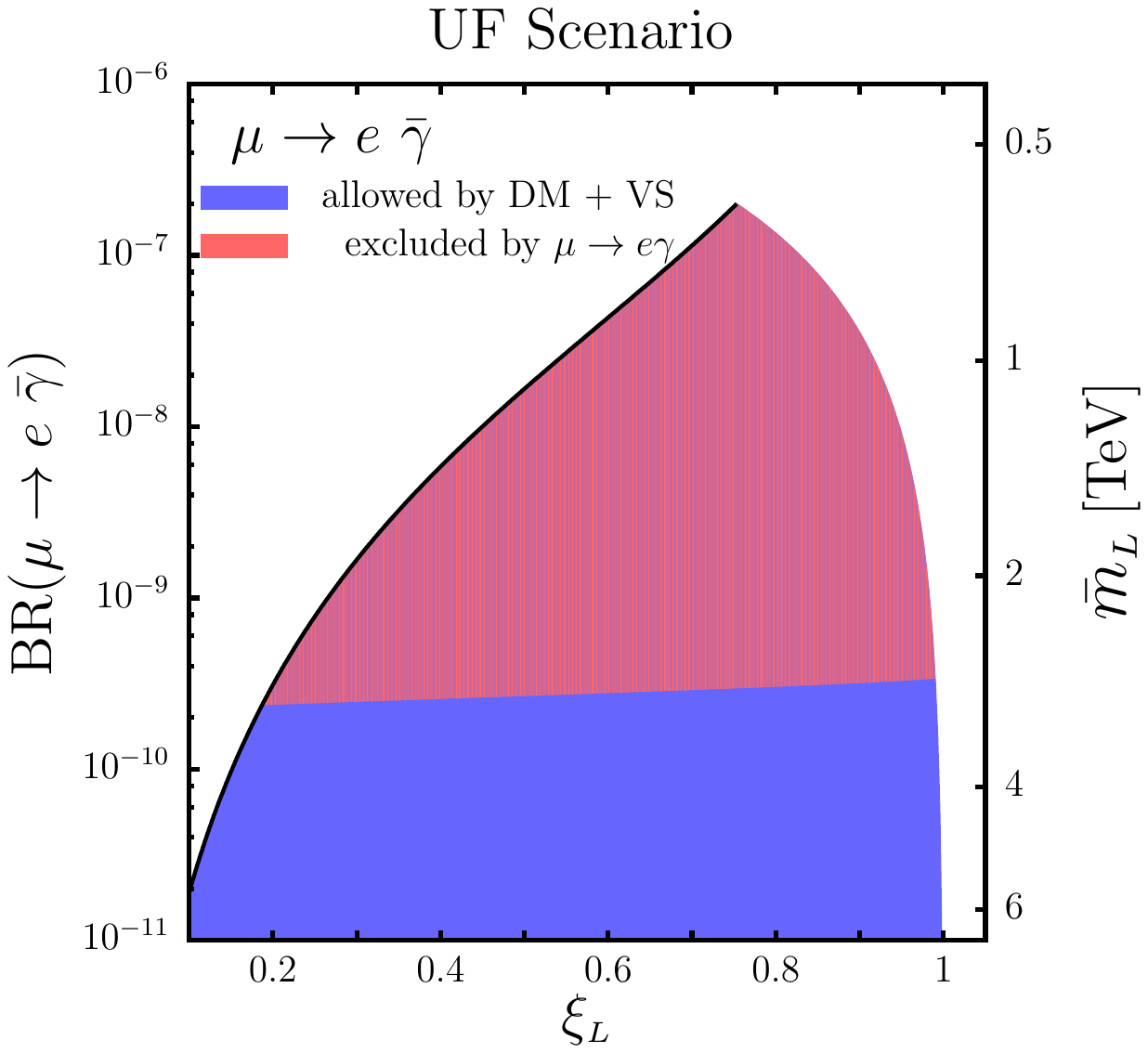}
\hspace{-5cm}
\vspace{-7.cm}
\caption{Regions allowed  by DM and vacuum stability (VS)  constraints
for  ${\rm BR}(\tau \to \ell\, \bar{\gamma})$ (left) and 
${\rm BR}(\mu \to e \, \bar{\gamma})$  (right), and 
for the average messenger mass scale 
 $\bar{m}_{\LL}$,
versus the  mixing  $\xi_{\LL}$, in the UF   scenario (blue areas).
Superimposed red areas are the subregions excluded by direct constraints on
${\rm BR}(\ell \to \ell'\, {\gamma})$. 
In the left (right) plot, we assume 
$\bar{e} \, \bar{e}_3^{\LL}=1$,\;$   \tilde\eta^{j3}_L/\tilde\eta^{33}_L=10^{-2} $
($\bar{e} \, \bar{e}_3^{\LL}=1$,\;$   \tilde\eta^{12}_L/\tilde\eta^{22}_L=10^{-4}$), with $j=1,2$.
}
\label{fig-BR-tau-mu}
\end{center}
\end{figure}

In Fig.~\ref{fig-BR-tau-mu} (left plot), we show the resulting ${\rm BR}(\tau \to \ell \, \bar{\gamma})$
expectations versus mixing,  in the UF scenario. The blue area corresponds to the allowed ranges,
while the red area selects the subregions excluded by the ${\rm BR}(\tau \to \mu \,{\gamma})$
bounds. 
One can see that,
for $\bar{e} \, \bar{e}_3^{\LL}=1$,\;$  \tilde\eta^{j3}_L/\tilde\eta^{33}_L=10^{-2} $ (with $j=1,2$),  ${\rm BR}(\tau \to \ell \, \bar{\gamma})$'s up to ($10^{-10}-10^{-6}$) are allowed, depending on mixing.
\section{The $\mu\to e\, \bar{\gamma}$ decay}
Here we analyze the radiative LFV muon decay 
\bea
\mu\to e\, \bar{\gamma}\, ,
\eea
following the  analysis done for the LFV $\tau$ decay into a dark photon. As for the  $\tau$ lepton, the corresponding BR can be parametrized in terms of the tree-level
BR($\mu\to \nu_{\mu}\bar{\nu}_{e} e$), as follows
\bea
{\rm BR}(\mu\to e \,\bar{\gamma})&=& \frac{12\; {\rm BR}^{\rm exp}_{
\mu\to \nu_{\mu} \bar{\nu}_{e} e}} {G_F^2 \;m_{\mu}^2\; f_1(z_{e\mu})}\left(\frac{1}{(\Lambda_L^{\mu e})^2}+\frac{1}{(\Lambda_R^{\mu e})^2}
\right)\, ,
\label{BRmuDP}
\eea
where  notations 
are defined in the previous section, and 
${\rm BR}^{\rm exp}(\mu\to \nu_{\mu} \bar{\nu}_{e} e)\simeq 100\%$~\cite{Agashe:2014kda}.
As  in Eq.~(\ref{ML}), we define an 
{\it effective} messenger mass $\bar{m}_{\LL}^{21}$ given by
\bea
\bar{m}_{\LL}^{21}\equiv \bar{m}_{\LL}\sqrt{\left|\frac{\tilde{\eta}_{\LL}^{22}}{
\tilde{\eta}_{\LL}^{12}}\right|}\, .
\label{MLmu}
\eea
which factorizes in the BR if we require the $L$-$R$ symmetry by assuming $\bar g_L=\bar g_R$.
The maximum allowed 
BR($\mu\to e\, \bar{\gamma}$) by DM and vacuum stability constraints are reported in Fig.~\ref{fig-BR-tau-mu} (right plot),  where we also report 
the constraints due to the 
LFV $\mu\to e \gamma$ decay. The present experimental upper bound at 90\% C.L. has been recently obtained by the MEG experiment at the Paul Scherrer Institute
~\cite{Adam:2013mnn}
\bea
{\rm BR}^{\rm exp}(\mu\to e\, \gamma) < 4.2 \times 10^{-13} \,.
\label{BRmu}
\eea
As in the $\tau$-lepton case, the SM contribution to the $\mu\to e \gamma$ decay rate is  negligible, due to the GIM suppression and tiny neutrino masses. Then, the upper 
bound in Eq.(\ref{BRmu}) can constrain the effective scale $\bar{m}_{\LL}^{21}$
defined above. In particular, one has
\bea
\bar{m}_{\LL}^{21}&>&\left(\frac{96\,\pi\,\alpha\, 
{\rm BR}^{\rm exp}_{\mu\to \nu_{\mu} \bar{\nu}_{e} e}}{{\rm BR}^{\rm max}_{\mu\gamma} \,G_F^2 \,f_1(z_{e\mu})}\right)^{1/4}\sqrt{\bar{F}_{LR}(x_2^{\LL},\xi_{\LL})}\, ,
\label{MLbound}
\eea
where ${\rm BR}^{\rm max}_{\mu\gamma}\equiv 4.2\times 10^{-13}$, and one can assume  $x_2^{\LL}\simeq (\frac{m_e}{m_{\mu}})^2 x_3^{\LL}$. Results are reported in Fig.~\ref{fig4}, where we plotted $\bar{m}_{\LL}^{21}$ versus $x_3^{\LL}$. 
\begin{figure}
\vspace{-2.5cm}
\begin{center}
\includegraphics[width=0.7\textwidth]{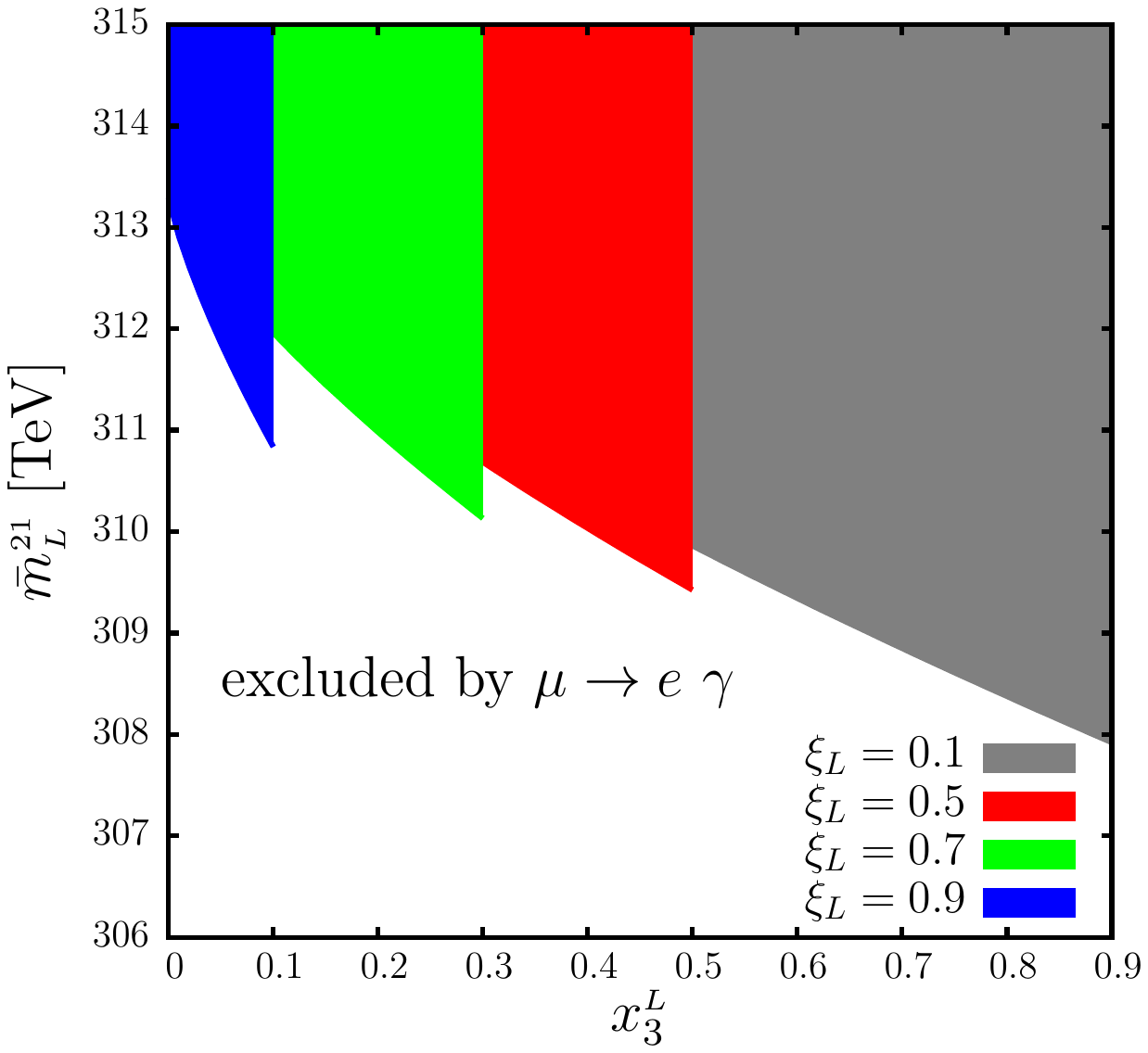}
\vspace{-7.5cm}
\caption{Regions  allowed by $\mu\to e \,\gamma$ constraints
at 90\% C.L. (represented by superimposed colored areas), 
for the {\it effective} messenger mass scale 
$\bar{m}_{\LL}^{21}$ defined in Eq.(\ref{MLmu}), as a function of  $x_3^{\LL}$ and for different values of the 
mixing $\xi_{\LL}$. Regions $x_3^{\LL}>1-\xi_{\LL}$ are excluded by DM constraints.}
\label{fig4}
\end{center}
\end{figure}
One can see that the constraints have a weak dependence on $x_3^{\LL}$, with $\bar{m}_{\LL}^{21} \lsim (308$--$313)$ TeV  in the  
region $x_3^{\LL}< 1- \xi_{\LL}$, for $\xi_{\LL}\gsim 0.1$.
Indeed, since
$x_2^{\LL}\simeq (\frac{m_e}{m_{\mu}})^2 x_3^{\LL}\ll 1$,
the $x_2^{\LL}$ dependence of BR is almost flat in the range $x_2^{\LL} \ll 1$, due to  the absence of $\log(x)$  infrared  singularities for $x\to 0$ in  $\bar{F}_{LR}(x,\xi)$.

By combining DM constraints on $\bar{m}_{\LL}$ 
with the ones from $\mu\to e\, \gamma$, 
considering the $\bar{m}_{\LL}$  lower   bound from DM and vacuum stability for a few values of $\xi_{\LL}$, we get  
\bea
\left|\frac{\tilde{\eta}^{12}_{\LL}}{\tilde{\eta}^{22}_{\LL}}\right| &<& 3.4\times 10^{-4} 
\left(\frac{\bar{m}_{\LL}}{5.7\, {\rm TeV}}\right)^2~~~,~~~{\rm for }~~\xi_{\LL}=0.1\, ,
\nonumber \\
\left|\frac{\tilde{\eta}^{12}_{\LL}}{\tilde{\eta}^{22}_{\LL}}\right| &<& 1.2\times 10^{-5} 
\left(\frac{\bar{m}_{\LL}}{1.1\, {\rm TeV}}\right)^2~~~,~~~{\rm for }~~\xi_{\LL}=0.5\, ,
\nonumber \\
\left|\frac{\tilde{\eta}^{12}_{\LL}}{\tilde{\eta}^{22}_{\LL}}\right| &<& 1.3\times 10^{-6} 
\left(\frac{\bar{m}_{\LL}}{360\, {\rm GeV}}\right)^2~~~,~~~{\rm for }~~\xi_{\LL}=0.9\, .
\eea

Finally, from the $\mu\to e\, \gamma$ constraints, we obtain the following upper bounds 
\begin{itemize} 
\item  for $\xi_{\LL}=0.1$ and $x_3^{\LL}=0.8$ (small-mixing regime)
\bea
{\rm BR}^{(\mu\to e \gamma)}(\mu\to e\, \bar{\gamma}) &<& 3.9\times 10^{-10}
\left(\frac{\bar{\alpha}}{0.1}\right)\, ,
\eea
\item  for $\xi_{\LL}=0.8$ and $x_3^{\LL}=0.1$ (large-mixing regime)
\bea
{\rm BR}^{(\mu\to e \gamma)}(\mu\to e\, \bar{\gamma}) &<& 6.2\times 10^{-10}
\left(\frac{\bar{\alpha}}{0.1}\right)
\, .
\label{BRmubounds}
\eea
\end{itemize}

In Fig.~\ref{fig-BR-tau-mu} (right plot), we show the resulting ${\rm BR}(\mu \to e \, \bar{\gamma})$
expectations versus mixing,  in the UF scenario. As before, the blue area corresponds to the allowed ranges,
while the superimposed red area selects the regions excluded by the ${\rm BR}(\mu \to e \, {\gamma})$
bounds. 
One can see that, 
for $\bar{e} \, \bar{e}_3^{\LL}=1$,\;$   \tilde\eta^{12}_L/\tilde\eta^{22}_L=10^{-4}$,  ${\rm BR}(\mu \to e \, \bar{\gamma})$'s up to ($10^{-11}-10^{-10}$) are allowed.

\section{Conclusions}
We have studied the  FCNC decays  of SM fermions    into a dark photon, $f\to f' \bar\gamma$,
 as foreseen by NP models with an extra unbroken $U(1)$ gauge group, acting on both a dark sector
and a messenger sector, whose dynamics could explain the observed Yukawa coupling hierarchy.
Model-dependent predictions for the corresponding BR's have been worked out, based
on constraints given by  DM abundance, vacuum stability, present non observation of
non-SM states at the LHC, and bounds on the related radiative $f\to f' \gamma$ decay rates.

We have found that large and possibly measurable BR's are allowed in most cases.
In particular, for typical coupling strengths, predicted BR($f\to f' \bar\gamma$)'s reach $ \sim (10^{-10}\!\!-\!\!10^{-7}$)  
for the decays of top-quark, 
 $ \sim (10^{-4}\!\!-\!\!10^{-3}$) for the $b$-quark, 
 $ \sim (10^{-8}\!\!-\!\!10^{-4}$) for the $c$-quark,
 $ \sim (10^{-10}\!\!-\!\!10^{-6}$) for the $\tau$-lepton, and 
 $ \sim (10^{-11}\!\!-\!\!10^{-10}$) for the $\mu$-lepton, 
depending on the mixing parameters and on the flavor-universality structure of the NP sector.

Most importantly, such decay channels are characterized by 
 new peculiar two-body  signatures,  where a final SM fermion is balanced by a massless invisible 
($\nu$-like) system. The latter could be looked for at present and future colliders
through dedicated searches, with high potential for either excluding large regions of the model parameter space or discovering a NP signal.

For instance,
 large FCNC $tq\bar\gamma$ couplings  might give  rise to  new  signatures associated to top-quark production in high-energy collisions.
Indeed, top-pair production at  hadron colliders could be an ideal laboratory where to search for two-body $m_t$ resonances made up of a monochromatic jet and 
$\nu$-like missing energy/momentum  
associated to the {\it undetected} dark photon in the $t\to q \bar\gamma$ final state,  
where $E_{jet}\sim E_{miss}\sim m_t/2$ in the top c.m.   system.

An even more striking signature would correspond to the partonic $qg\to t \bar\gamma$  scattering, occurring via an $s$-channel $u$-$,c$-quark exchange. In this case a single
top-quark system with unbalanced momentum would be  associated to a {\it massless} invisible system.
Such peculiar and  clean collider top-quark signatures are not present in the SM, and  possible backgrounds may arise only from particle and jet mismeasurements.
Based on  the BR upper bounds reported above, 
searches for FCNC top couplings to stable dark photons 
 might indeed be explorable at future hadron colliders, like the FCC-hh, where 
a statistics of about 10$^{12}$ top pairs could be available~\cite{Arkani-Hamed:2015vfh}.
Note that the $(t+ E_{miss})$ final states are presently considered by LHC experiments in NP searches for {\it massive} invisible systems \cite{Aad:2014wza},\cite{CMS:2016flr}.

As far as lighter flavors are concerned, the scenario looks even more promising. 
Huge and measurable values for ${\rm BR}(b\to q\, \bar{\gamma})$, where $q=s,d$, are presently allowed.
Experimentally, as in the top-quark case, the $b\to q\, \bar{\gamma}$ is characterized by a peculiar signature, where the invisible massless dark photon equally shares the initial $b$-hadron energy and momentum with an $s$- or $d$-initiated hadronic system.
While hadron colliders are not the ideal place where to reconstruct such features,
 electron-positron $B$ factories~\cite{Wang:2015kmm} can offer the clean collision environment needed to control the invisible-system kinematical characteristics. An even better control could be available at future $Z$ factories 
(as possibly foreseen at the ILC~\cite{Baer:2013cma}, the FCC-ee~\cite{Gomez-Ceballos:2013zzn}, and  the CEPC~\cite{CEPC-SPPCStudyGroup:2015csa}, running at the $Z$ peak), where the large boost of the $b$ hadrons
could help in disentangling the invisible dark photon with high accuracy.

Similar features are shared by potentially measurable charm, tau, and muon decay
rates into a dark photon, which can also be naturally scrutinized in $e^+e^-$ collisions with large integrated luminosities. In particular, at the  FCC-ee running on the $Z$ peak, clean samples of ${\cal O}(10^{11}$-$10^{12}$) heavy-quark and lepton pairs of each given flavor from $Z\to f\bar f$ could be available~\cite{fccee}, that, in absence of major systematics, could be sensitive
to BR's into dark photons down to ${\cal O}(10^{-10}$). Dedicated studies will be needed to accurately assess the actual 
sensitivity of present and future experiments to the FCNC and LFV  fermion decay channels into a stable dark photon, naturally predicted in the  theoretical NP framework considered in the present analysis.
\section*{Acknowledgements}
EG would like to thank the CERN Theory Division for the kind hospitality during the preparation of this work. EV thanks the Theoretical Physics Department of 
the University of Trieste where part of this work has been done.
This work was supported by the ERC Grant No. IUT23-6 and by the EU through the ERDF CoE program.

\section*{Appendix}
Here we present the analytical expressions for the NP contributions to the
generic  FCNC radiative  decay amplitude corresponding to the process 
\bea 
f^i\to f^j\, \gamma \, ,
\eea
where $\gamma$ stands for a SM photon, and the indices $i,j$ 
($i>j$, with $i=3$ for the 
heaviest generation) both 
run
on the fermion families either in the {\it up} or in the {\it down} $SU(2)_L$ sector. The   
Feynman diagrams contributing 
to this process are given by the diagrams 
(b) and (d) in Fig.~\ref{Fig-FD}, plus the usual flavor-changing self-energy 
(FCSE) 
contributions, that we do not show here. The FCSE graphs are required by  gauge invariance, although  not contributing 
to the $f^i\to f^j\, \gamma$ decay amplitude for an on shell photon, being 
the latter proportional to a flavor-changing magnetic-dipole operator.
The $f^i\to f^j\, \gamma$ amplitude,  for different $L/R$ chirality states, has the same structure as  Eq.(\ref{amp}) for the $f^i\to f^j\,\bar\gamma$ amplitude, namely
\bea
M(f^i_L \to f^j_R \, \gamma) &=& \frac{1}{\bar{\Lambda}^f_L}
[\bar{u}_R^j \sigma_{\alpha\mu} u_L^i ] q^{\mu} \epsilon^{\alpha} \, ,\nonumber
\\
M(f^i_R \to f^j_L \, \bar{\gamma}) &=& \frac{1}{\bar{\Lambda}^f_R} 
[\bar{u}_L^j \sigma_{\alpha\mu} u^i_R ] q^{\mu} \epsilon^{\alpha}\, ,
\label{ampgg}
\eea
where $\epsilon^{\alpha}$ is the photon polarization vector. In the low
energy approximation, 
the mass scales $\bar{\Lambda}^f_{L,R}$ do not depend on external momenta and
can be worked out by matching the amplitude in Eq.~(\ref{ampgg}) with the 
result of the full computation in the low energy limit.
We neglect terms suppressed by loop factors and 
 provide  only the contributions proportional to the product 
$g_Lg_R$. Then, similarly to  Eqs.~(\ref{LambdaU})--(\ref{LambdaD}) (with a different loop function), we obtain for the {\it up} quark sector
\bea
\dfrac{1}{(\bar{\Lambda}^{\U}_L)_{ij}}&=&\dfrac{m^{\U}_i}{\overline{m}^2_{\U}} \left(e\,
e_i^{\U}\dfrac{ 
\rho_{R}^{ji}}{\rho_{R}^{ii}}\bar{F}_{LR}(x^{\U}_i,\xi_{\U})\right) \, ,
\\
\dfrac{1}{(\bar{\Lambda}^{\U}_R)_{ij}}&=&\dfrac{m^{\U}_i}{\overline{m}^2_{\U}} \left(e\, e_i^{\U} \dfrac{\rho_{L}^{ji}}{ \rho_{L} ^{ii}}\bar{F}_{RL}(x^{\U}_i,\xi_{\U})\right)\, ,
\label{bLambdaU}
\eea
and for the {\it down} quark sector
\bea
\dfrac{1}{(\bar{\Lambda}^{\D}_L)_{ij}}&=&\dfrac{m^{\D}_i}{\overline{m}^2_{\D}} \left( e\,
e_i^{\D}\dfrac{ 
\eta_{R}^{ji}}{\eta_{R}^{ii}}\bar{F}_{LR}(x^D_i,\xi_{\D}) 
\right) \, ,
\\
\dfrac{1}{(\bar{\Lambda}^{\D}_R)_{ij}}&=&\dfrac{m^{\D}_i}{\overline{m}^2_{\D}} \left(e\,e_i^{\D} \dfrac{\eta_{L}^{ji}}{ \eta_{L}^{ii}}\bar{F}_{RL}(x^{\D}_i,\xi_{\D}) \right)\, ,
\label{bLambdaD}
\eea
where $e_i^{\U(\D)}$ are the  EM charges of SM fermions in the {\it up} ({\it down}) sector,  in unit of the EM charge $e$. The  loop function $\bar{F}_{RL}(x,\xi)$ is given by
\bea
\bar{F}_{LR}(x,\xi)=\bar{F}_{RL}(x,\xi)&=&\dfrac{\bar{f}_{2}(x,\xi)}{f_1(x,\xi)}\, ,
\label{fbarLR}
\eea
where $f_1(x,\xi)$ is defined in Eq.(\ref{f1}), and $\bar{f}_2(x,\xi)$
is given by
\bea
\bar{f}_{2}(x,\xi)&=&\dfrac{1}{2\,\xi}\left[ 
\dfrac{(1+\xi)^2-x^2  +2x(1+\xi) \log\left( \frac{x}{1+\xi}\right) }{2(x-1-\xi)^3}
-\Big\{\xi\to -\xi\Big\}\right]\, .
\label{effeduebarra}
\eea
In particular, the limits at small and large  mixing  are, respectively,
\bea
\lim_{\xi\to 0} \bar{F}_{LR}(x,\xi)&=&\lim_{\xi\to 0} \bar{F}_{RL}(x,\xi)=
\frac{1+4x-5x^2+2x(2+x)\log{x}}{4(1-x)^2(1-x+x\log{x})}, \\
\lim_{\xi\to 1} \bar{F}_{LR}(x,\xi)&=&\lim_{\xi\to 1} \bar{F}_{RL}(x,\xi)=
\frac{(1-x)^2\left(4-8x+3x^2-2x^2\log{\frac{x}{2}}\right)}{4x(2-x)^3(1-x+x\log{x})}\, .
\label{FbarLR}
\eea
Contrary to the dark-photon loop function 
$F_{LR}(x,\xi)$  in Eq.(\ref{FLR}),  the  
  $\bar{F}_{LR}(x,\xi)$ expansion at $\xi\sim 1$  in Eq.~(\ref{FbarLR}) 
does not present  $\log(1-\xi)$ singularities at the denominator.


\end{document}